\documentclass[prc,amsmath,showpacs,twocolumn,floatfix,unsortedaddress]{revtex4}
\usepackage{graphicx}
\usepackage{dcolumn}
\usepackage[pdftex,dvipsnames,usenames]{color}		
\usepackage{epstopdf}
\usepackage{slashed}
\usepackage{verbatim}
\usepackage{mathtools}
\usepackage{longtable}
\usepackage{amssymb}
\usepackage{bbold}
\usepackage{soul}   

\long\def\Omit#1{}

\allowdisplaybreaks[1]

\renewcommand*{\eqref}[1]{(\ref{eq:#1})}
\newcommand*{\eqlab}[1]{\label{eq:#1}}

\newcommand{\SP}[3]{\left<{#1}\right|{#2}\left|{#3}\right>}

\newcommand{\beq}{\begin{equation}}
\newcommand{\eeq}{\end{equation}}
\newcommand{\nn}{\nonumber}

\newcommand{\ghat}{\hat g}

\newcommand{\rbot}{{r_{^{\tiny \rfloor \! \!\lfloor} }\!}}
\newcommand{\rbarbot}{{\bar r_{^{\tiny \rfloor \! \!\lfloor} }\!}}
\newcommand{\wRbot}{{w_{^{\tiny \lfloor} }\! }}
\newcommand{\wLbot}{{w_{^{\tiny \rfloor} }\!}}

\newcommand{\wRbarbot}{{\bar w_{^{\tiny \lfloor} }\! }}
\newcommand{\wLbarbot}{{\bar w_{^{\tiny \rfloor} }\!}}

\newcommand{\tsp}[2]{\left<\mbox{\small{${#1}$}}\right|T \left|\mbox{\small{${#2}$}}\right>}

\def\rescale{\fontsize{7}{1.9}}

\begin{document}

\title{On kinematical constraints in boson-boson systems}
\author{M.F.M.\ Lutz}
\email{m.lutz@gsi.de}
\affiliation{GSI Helmholtzzentrum f\"ur Schwerionenforschung GmbH,\\
Planck Str. 1, 64291 Darmstadt, Germany}
\author{I.\ Vida\~na }
\email{ividana@fis.uc.pt}
\affiliation{Centro de F\'{i}sica Computacional. Department of Physics, University of Coimbra, PT-3004-516 Coimbra Portugal}

\date{\today}

\begin{abstract}

We consider the scattering of two-bosons  with negative parity and spin $0$ or $1$.
Starting from helicity partial-wave scattering amplitudes we derive transformations
that eliminate all kinematical constraints. Such amplitudes are expected to satisfy
partial-wave dispersion relations and therefore provide a suitable basis for data analysis and
the construction of effective field theories. Our derivation relies on a decomposition
of the various scattering amplitudes into suitable sets of invariant functions.
A novel algebra was developed that permits the efficient computation of such functions
in terms of computer algebra codes.
\end{abstract}

\maketitle


\section{Introduction}

In a strongly interacting quantum field theory like QCD an important challenge is the reliably
and predictive treatment of final-state interactions. Given some effective degrees of freedom
micro-causality and coupled-channel unitarity are crucial constraints that help
to establish coupled-channel reaction amplitudes from a suitable effective Lagrangian (see
e.g. \cite{Gasparyan:2010xz,Danilkin:2010xd,Danilkin:2011fz}).

Though it is straight forward to introduce partial-wave scattering amplitudes in the helicity formalism
of Jacob and Wick \cite{Jacob:1959at}, it is a nontrivial task to derive transformations that lead to
amplitudes that are kinematically unconstrained. Such amplitudes are useful for partial-wave analysis or
effective field theory approaches which consider the consequences of micro causality in terms of partial-wave
dispersion-integral representations~\cite{Chew:1957tf,Nakanishi:1962,Berends:1967vi,Lutz:1999yr,Lutz:2001mi,Lutz:2001yb,Lutz:2001dq,Lutz:2007bh,Korpa:2008ut,Gasparyan:2010xz}. It is the purpose of the present work to derive such amplitudes by suitable transformations of the helicity
partial-wave scattering amplitudes for two-body systems with $J^P=0^-$ or $1^-$ particles. In a previous
work one of the authors studied the scattering of $0^-$ off $1^-$ particles \cite{Lutz:2003fm} and
fermion-antifermion annihilation processes with $\frac{1}{2}^+$ and $\frac{3}{2}^+$
particles \cite{Stoica:2011cy}. So far reactions involving two body states with two $1^-$ particles have not been dealt with.
The technique applied in this work has been used previously in studies of two-body scattering systems with
photons, pions and nucleons \cite{Chew:1957tf,Ball:1961zza,Barut:1963zz,Hara:1964zza,PhysRev.169.1248,PhysRev.142.1187,PhysRev.170.1606,Scadron:1969rw,CohenTannoudji:1968,Bardeen:1969aw}.
A possibly related approach is by Chung and Friedrich \cite{PhysRevD.48.1225,Chung:2007nn}.
Our results will be relevant for the PANDA experiment at FAIR,
where protons and antiprotons may be annihilated into systems of spin $0$ or $1$ states \cite{Lutz:2009ff}.

We consider partial-wave projections of the scattering amplitude. Our goal is to establish
partial-wave amplitudes with convenient analytic properties that justify the use of uncorrelated
integral-dispersion relations. We consider all two-body reactions possible with spin $0$ and $1$ bosons.
In an initial step we decompose the scattering amplitude into invariant
functions that are free of kinematical constraints. Such amplitudes are expected to satisfy a Mandelstam
dispersion-integral representation~\cite{Mandelstam:1958xc,Ball:1961zza}. A given choice of basis
is free of kinematical constraints if any additional structure can be decomposed into
the basis with coefficients that are regular. The identification of such a basis is a nontrivial task as the spins of the
involved particles increase. Helicity partial-wave amplitudes are correlated at various kinematical conditions. 
The derivation of such constraints is based on an application of the previously constructed
basis of kinematically unconstrained invariant amplitudes. The kinematical constraints in the helicity partial-wave 
amplitudes are eliminated by means of non-unitary transformation matrices that map the initial, respectively final 
helicity sates to new covariant states.

The work is organized as follows. Section II introduces the conventions used for the kinematics and the
spin-$1$ helicity wave functions. The scattering amplitudes are decomposed into sets of invariant amplitudes
free of kinematical constraints. In the following section  the helicity partial-wave amplitudes are constructed
within the given convention. The central results are presented in section IV, where the transformation to
partial-wave amplitudes free of kinematical constraints are derived and discussed.

\section{On-shell scattering amplitudes}

We consider two-body reactions involving pseudo-scalar and vector particles. All derivations will be completely generic.
We introduce the 4-momenta $p_{1}$ and $\bar{p}_{1}$ of the incoming and outgoing first particle and those of the
second particle, $p_{2}$ and $\bar{p}_{2}$. In the center of mass frame we write
\begin{eqnarray}
&& p_{1}^\mu= \left(\omega_{1},0,0,+p \right)\,, \quad  \bar{p}_{1}^\mu= \left(\bar{\omega}_1, +\bar{p}\sin{\theta},0 , +\bar{p}\cos\theta \right)\,,
\nonumber\\
&& \omega_{1}= \sqrt{m_{1}^{2} + p^2}   \,,  \qquad \qquad \bar{\omega}_{1}= \sqrt{\bar{m}_{1}^{2} + \bar{p}^2}\,,
\nonumber\\
&& p_{2}^\mu= \left(\omega_{2},0,0,-p \right)\,, \quad  \bar{p}_{2}^\mu= \left(\bar{\omega}_2, -\bar{p}\sin{\theta},0 , -\bar{p}\cos\theta \right)\,,
\nonumber\\
&& \omega_{2}= \sqrt{m_{2}^{2} + p^2}   \,,  \qquad \qquad \bar{\omega}_{2}= \sqrt{\bar{m}_{2}^{2} + \bar{p}^2}\,,
\label{eq:momenta}
\end{eqnarray}
where $\theta$ is the scattering angle, $p$ and $\bar{p}$ are the magnitudes of the initial and final three-momenta.
The relative momenta $p$ and $\bar{p}$ can be expressed in terms of the total energy $\sqrt{s}$ of the system
\begin{eqnarray}
&& p^2=\frac{1}{4\,s}\,\big(s-(m_1 +m_2)^2\big)\,\big(s-(m_1 -m_2)^2\big)
\nonumber\\
&& w^\mu = p_{1}^\mu+p_{2}^\mu = \bar{p}_{1}^\mu + \bar{p}_{2}^\mu\,, \qquad s = w^2\,,
\nonumber\\
&& \bar{p}^2=\frac{1}{4\,s}\,\big(s-(\bar{m}_1 +\bar{m}_2)^2\big)\,\big(s-(\bar{m}_1 - \bar{m}_2)^2\big)\,.
\label{p-sqr}
\end{eqnarray}
It is convenient to introduce some further notation
\begin{eqnarray}
&& k^\mu=\frac{1}{2}\,(p^\mu_1-p^\mu_2)\,, \qquad \quad  r_\mu = k_\mu - \frac{1}{2}\, \frac{p_1^2-p_2^2}{s}\,w_\mu  \,,
\nonumber\\
&& \bar k^\mu =\frac{1}{2}\,(\bar p^\mu_1-\bar p^\mu_2)  \,, \qquad \quad
\bar r_\mu = \bar k_\mu-\frac{1}{2}\, \frac{\bar p_1^2-\bar p_2^2}{s}\,w_\mu \,,
\nonumber\\
&& \ghat_{ \mu  \nu} = g_{ \mu  \nu}- \frac{1}{s}\,w_{\mu}\,w_{\nu}\,,\quad \;\,    \bar r \cdot r = -\bar p\,p\,\cos \theta \,,
\label{def-k}
\end{eqnarray}
where the two 4-vectors $r_\mu$ and $\bar r_\mu $ have a transparent relation to the
center-of-mass momenta $p$ and $\bar p$.

We specify the spin-one wave functions
\begin{eqnarray}
&& \epsilon^\mu(\bar{p}_1, \pm 1)=\left(
                           \begin{array}{c}
                             0 \\
                              \frac{\mp\cos \theta }{\sqrt{2}} \\
                            \frac{- i}{\sqrt{2}} \\
                              \frac{ \pm \sin \theta }{\sqrt{2}} \\
                           \end{array}
                         \right)\,,\;
\epsilon^\mu(\bar{p}_1,  0)=      \left(
                           \begin{array}{c}
                             \frac{\bar p}{\bar m_1} \\
                             \frac{\bar{\omega}_1}{\bar m_1} \sin \theta \\
                             0 \\
                             \frac{\bar{\omega}_1}{\bar m_1} \cos \theta \\
                           \end{array}
                         \right)\,,       \nonumber
\nonumber\\
&& \epsilon^\mu(\bar{p}_2, \pm 1) = \left(
                           \begin{array}{c}
                             0 \\
                            \frac{\pm \cos \theta }{\sqrt{2}} \\
                            \frac{-i}{\sqrt{2}} \\
                            \frac{\mp \sin \theta }{\sqrt{2}} \\
                           \end{array}
                         \right) \,, \;
\epsilon^\mu(\bar{p}_2, 0) = \left(
                           \begin{array}{c}
                             \frac{\bar p}{\bar m_2} \\
                             -\frac{\bar{\omega}_2}{\bar m_2} \sin \theta \\
                             0 \\
                             -\frac{\bar{\omega}_2}{\bar m_2} \cos \theta \\
                           \end{array}
                         \right) \,,\nonumber
\end{eqnarray}
where the wave function of the corresponding initial states is recovered with $\theta =0$.

The on-shell production and scattering amplitudes are defined in terms
of plane-wave matrix elements of the scattering operator $T$. We represent the scattering amplitudes in terms of a complete set
of invariant functions $F_n(s,t)$. The merit of the decomposition lies in the transparent analytic properties of
such functions $F_n(s,t)$, which are expected to satisfy Mandelstam's dispersion integral
representation ~\cite{Mandelstam:1958xc,Ball:1961zza}. For reactions involving spin-one particles it is not straight
forward to identify such amplitudes.

We begin with the elastic scattering of two pseudoscalar particles
\begin{eqnarray}
&& T_{0\,0 \to 0\,0}(\bar k,\, k,\, w) = F_1 (s,t)\,,
\label{def-00to00}
\end{eqnarray}
which is characterized by one scalar function $F_1(s,t)$ depending on
two Mandelstam variables, e.g. $s$ and $t$ with
\begin{eqnarray}
s+t+u = m_1^2+m_2^2+\bar m_1^2+\bar m_2^2\,.
\label{def-Mandelstam}
\end{eqnarray}
We suppress internal degrees of freedom like isospin or strangeness quantum
numbers for simplicity.

A slightly more complicated process involves one vector particle in the final state
\begin{eqnarray}
&& T_{0\,0 \to 0\,1}(\bar k,\, k,\, w) = F_1 (s,t)\,\langle
T^{(1)}_{\bar \nu } \rangle^{\bar \nu }_{0\,0 \to 0\,1} \,,
\nonumber\\ \nonumber\\
&& \langle \,T^{(1)}_{\bar \nu } \,\rangle^{\bar \nu}_{0\,0 \to 0\,1} =
\epsilon^{\dagger, \bar \nu}(\bar p_2,\, \bar \lambda_2)\,
T^{(1)}_{\bar \nu }\,,
\nonumber\\
&& T^{(1)}_{\bar \nu} =i\,\epsilon_{\bar \nu \tau \alpha \beta}\,
w^{\tau}\,\bar p_2^\alpha \,p_2^\beta\,,
\label{def-00to01}
\end{eqnarray}
where we use a notation analogous to the one introduced in \cite{Stoica:2011cy}. For notational
simplicity we do not introduce different notations for the invariant amplitudes
$F_1(s,t)$ in the two reactions (\ref{def-00to00}, \ref{def-00to01}).
Further processes related  to (\ref{def-00to01}) are obtained by the exchange of the out or
ingoing momenta.

The structure of the on-shell reaction amplitudes turns more complicated with
increasing number of spin-1 particles involved. Consider the production of two vector
particles
\begin{eqnarray}
&&T_{0\,0 \to 1\,1}(\bar k, \,k,\, w) = \sum_{n=1}^5 \,F_n (s,t)\,
\langle T^{(n)}_{\bar \mu \bar \nu} \rangle^{\bar \mu \bar \nu}_{0\,0 \to 1\,1} \,,
\nonumber\\
&& \langle \,T^{(n)}_{\bar \mu \bar \nu} \,\rangle^{\bar \mu \bar \nu}_{0\,0 \to 1\,1} =
\epsilon^{\dagger\,\bar \mu}(\bar p_1,\, \bar \lambda_1)\,
\epsilon^{\dagger\,\bar \nu}(\bar p_2,\, \bar \lambda_2)\,
T^{(n)}_{\bar \mu \bar \nu} \,,
\nonumber\\ \nonumber\\
&& \begin{array}{ll}
T^{(1)}_{\bar \mu \bar \nu} = \ghat_{\bar \mu \bar \nu}\,, \qquad \qquad \qquad  &
T^{(2)}_{\bar \mu \bar \nu} = w_{\bar \mu}\, w_{\bar \nu}\,,\\
T^{(3)}_{\bar \mu \bar \nu} = w_{\bar \mu}\, r_{\bar \nu}\,, &
T^{(4)}_{\bar \mu \bar \nu} = r_{\bar \mu}\, w_{\bar \nu}\,,\\
T^{(5)}_{\bar \mu \bar \nu} = r_{\bar \mu}\,  r_{\bar \nu}\,, &
\\
\end{array}
\label{def-00to11}
\end{eqnarray}
which is characterized by five invariant amplitudes, $F_n(s,t)$. The choice of Lorentz tensors in (\ref{def-00to11}) is not unambiguous. Various linear combinations of the given tensors may be used. For instance we could have used the 5 tensors which follow from  (\ref{def-00to11}) by the replacements $r_\mu \to k_\mu$ and $\bar r_\mu \to \bar k_\mu$. The suggested form proves most convenient when calculating helicity  matrix elements.

The number of invariant amplitudes is easily determined for the reaction $0\,0 \to 1\,1$. The task is to construct all rank two tensor in terms of the four vectors  $\bar k, k, w$ and
\begin{eqnarray}
v_{\mu} = \epsilon_{\mu \alpha \tau \beta}\,
\bar k^\alpha \,w^{\tau}\, k^\beta
\label{def-eps-vector}
\end{eqnarray}
At first there are $4 \times 4 + 1=17$ distinct
tensors that one may construct. Parity conservation requires the pairwise occurrence of the
vector $v_{\mu}$ introduced in (\ref{def-eps-vector}). This eliminates 6 structures.
The transversality of the spin-one wave functions with
\begin{eqnarray}
&& 2\,\epsilon_\mu(\bar p_1)\,\bar k^{\mu}=  + \epsilon_\mu(\bar p_1)\,w^{\mu} \,,
\nonumber\\
&& 2\,\epsilon_\mu(\bar p_2)\,\bar k^{\mu}=  - \epsilon_\mu(\bar p_2)\,w^{\mu} \,,
\end{eqnarray}
eliminates additional 5 structures for on-shell conditions. Altogether there are
6 structures left. The five terms displayed in (\ref{def-00to11}) and the Lorentz tensor
$v_{\bar \mu}\,v_{\bar \nu}$. To show the on-shell redundance of the extra term requires
an explicit computation of on-shell matrix elements.

In a practical application it is important to derive explicit expressions for the invariant
amplitudes $F_n(s,t)$ (see e.g. \cite{Lutz:2007sk}). In the general case this is may be a tedious exercise, which is considerably
streamlined by the derivation and application of a set of projection tensors $P^{(n)}_{\mu  \nu}$ with the
following properties
\begin{eqnarray}
&& P^{(n)}_{\mu  \nu} \, g^{\bar \mu \mu}\,g^{\bar \nu \nu}\,T^{(m)}_{\bar \mu \bar \nu} = \delta_{nm} \,,
\nonumber\\
&& P^{(n)}_{\bar \mu  \bar \nu} \,\bar p_1^{\bar \mu } = 0 \, ,\qquad \quad P^{(n)}_{\bar \mu  \bar \nu} \,\bar p_2^{\bar \nu } =0 \,.
\end{eqnarray}
Given any off-shell production amplitude the invariant function $F_n(s,t)$ is obtained by the contraction with the $n$th projection tensor.
We decompose the projection tensors into a basis
\begin{eqnarray}
&& P^{(n)}_{\bar \mu  \bar \nu} = \sum_{k=1}^5\,c_k^{(n)} \,Q^{(k)}_{\bar \mu  \bar \nu} \,,
\nonumber\\
&& Q_{1}^{\bar \mu  \bar \nu} = \,v^{\bar \mu }\,v^{\bar \nu} /v^2\,, \qquad
\nonumber\\
&& Q_{2}^{\bar \mu  \bar \nu} = \wLbot^{\bar \mu}\,\wRbot^{\bar \nu}\,,\qquad
 Q_{3}^{\bar \mu  \bar \nu} = \wLbot^{\bar \mu }\,\,\rbot^{\bar \nu} \,,
\nonumber\\
&&Q_{4}^{\bar \mu  \bar \nu} = \,\rbot^{\bar \mu}\,\wRbot^{\bar \nu }\,,
 \qquad Q_{5}^{\bar \mu  \bar \nu} =\, \rbot^{\bar \mu}\,\,\rbot^{\bar \nu }\,,
\label{res-Q:00to11}
\end{eqnarray}
where the 4-vectors $\rbot, \wLbot$ and $\wRbot$ are suitable linear combinations of $\bar r, r$ and $w$ as
to have the convenient properties
\begin{eqnarray}
&& \,\rbot \cdot r \,= 1\,, \qquad \,\,\rbot \cdot \bar r = 0 = \,\rbot \cdot w \,,
\nonumber\\
&& \wLbot  \cdot w = 1\,,\qquad \wLbot  \cdot r = 0 = \wLbot  \cdot \bar p_1 \,,
\nonumber\\
&& \wRbot  \cdot w = 1\,,\qquad \wRbot  \cdot r = 0 = \wRbot  \cdot \bar p_2 \,.
\label{bot-properties}
\end{eqnarray}
The index ${\small \rfloor}$ and ${\small \lfloor}$ of a vector indicates whether it is orthogonal
to the 4 momentum of the first or second particle respectively. The patched symbol ${\small \rfloor \! \!\lfloor}$ implies
the orthogonality to both 4 momenta. Given such vectors the
coefficients $c_k^{(n)} $ are readily determined. We find
\begin{eqnarray}
&& c^{(n)}_n = 1 \,, \quad  c^{(2)}_1 = -\wLbot \cdot \wRbot + 1/s\, ,
\label{}\\
&&   c^{(3)}_1 = -\wLbot \cdot \rbot \, ,\quad c^{(4)}_1 = -\rbot \cdot \wRbot \, , \quad
  c^{(5)}_1 = -\rbot \cdot \rbot \, , \nonumber
\end{eqnarray}
where we display non-vanishing elements only.

We construct the auxiliary vectors $\rbot, \wLbot$ and $\wRbot$. For this purpose we introduce an intermediate notation.
Given three 4-vectors $a_\mu, b_\mu$ and $c_\mu$
we introduce a vector,
$a^\mu_{b\, c}= a^\mu_{c\, b}$, as follows
\begin{eqnarray}
&& \frac{a^\mu_{b \,c}}{a_{b \,c} \cdot a_{b\, c}} = a^\mu  - \frac{a \cdot c}{c \cdot c}\,c^\mu
\nonumber\\
&& \qquad \qquad \,-\, \frac{a \cdot (b- \frac{c \cdot b }{c \cdot c}\,c )}{(b- \frac{c \cdot b }{c \cdot c}\,c )^2}
\left( b^\mu - \frac{c \cdot b }{c \cdot c}\,c^\mu  \right) \,,
\nonumber\\
&& a^\mu_{b\, c} \,a_\mu =1 \,, \qquad a^\mu_{b\, c} \,b_\mu =0 \,,\qquad a^\mu_{b\, c} \,c_\mu =0\,.
\label{def-abc}
\end{eqnarray}
In the notation of (\ref{def-abc}) the desired vectors are identified with
\begin{eqnarray}
&&  \rbot^\mu = r^\mu_{\bar r \,w} \,, \qquad \wLbot^\mu = w^\mu_{r \,\bar p_1}\,, \qquad  \wRbot^\mu = w^\mu_{r \,\bar p_2}\,,
\nonumber\\
&&  \rbarbot^\mu = \bar r^\mu_{r\, w} \,,\qquad \wLbarbot^\mu = w^\mu_{\bar r\, p_1}\,, \qquad  \wRbarbot^\mu = w^\mu_{\bar r \,p_2} \,,
\label{def-rbot}
\end{eqnarray}
where we introduced the additional vectors $\rbarbot, \wLbarbot$ and $\wRbarbot$ that will turn useful below.

It remains the question why did we select the five tensors in (\ref{def-00to11}) and did not
include the extra structure $v_{\bar \mu}\,v_{\bar \nu}$ into our basis? The reason is our
request that the invariant amplitudes should be free of kinematical constraints. The issue is
nicely illustrated at hand of the on-shell identity
\begin{eqnarray}
&&v_{\bar \mu}\, v_{\bar \nu} = v^2\,\Big[ g_{\bar \mu \bar \nu}
 -\,(\wLbot \cdot \wRbot) \,w_{\bar \mu}\, w_{\bar \nu} -(\wLbot \cdot \rbot )\, w_{\bar \mu}\, r_{\bar \nu}
\nonumber\\
&& \qquad \qquad -\, (\wLbot \cdot \rbot)\,r_{\bar \mu}\, w_{\bar \nu}- (\rbot \cdot \rbot )\,r_{\bar \mu}\,  r_{\bar \nu} \Big]\,,
\nonumber\\
&& v^2 = s\,\Big[( \bar r \cdot r)^2 - \bar r^2 \,r^2\Big] \,,
\label{res-vv}
\end{eqnarray}
with the tensor basis introduced in (\ref{def-00to11}). Eliminating any of the five tensors in
favor of the structure $v_{\bar \mu}\, v_{\bar \nu}$ leads to invariant functions
singular at various kinematical conditions. This is evident from the regularity of the expressions
\begin{eqnarray}
&& v^2\,(\,\rbot \cdot \,\rbot ) = - s\, \bar r^2 \,, \quad
\nonumber\\
&&v^2\,(\wLbot \cdot \wRbot) = (\bar r \cdot r)^2 + r^2 \,\bar p_1\cdot \bar p_2\,,
\nonumber\\
&& v^2\,(\,\rbot \cdot \wLbot ) = {\textstyle {1 \over 2}}\,(\bar m_2^2-\bar m_1^2 -s)\,(\bar r \cdot r) \,, \qquad
\nonumber\\
&& v^2\,(\,\rbot \cdot \wRbot) = {\textstyle {1 \over 2}}\,(\bar m_2^2-\bar m_1^2 +s)\, (\bar r \cdot r)\,.
\end{eqnarray}

We continue with the scattering of pseudo-scalar  off vector particles. There are
again five invariant amplitudes needed to characterize the scattering amplitude
\begin{eqnarray}
&&T_{0\,1 \to 0\,1}(\bar k, \,k,\, w) = \sum_{n=1}^5 \,F_n (s,t)\,
\langle T^{(n)}_{\bar \nu \nu} \rangle^{\bar \nu \nu}_{0\,1 \to 0\,1} \,,
\nonumber\\
&& \langle \,T^{(n)}_{\bar \nu \nu} \,\rangle^{\bar \nu \nu}_{0\,1 \to 0\,1} =
\epsilon^{\dagger\,\bar \nu}(\bar p_2,\, \bar \lambda_2)\,
T^{(n)}_{\bar \nu \nu}\,\epsilon^\nu(p_2,\,\lambda_2) \,,
\nonumber\\ \nonumber\\
&& \begin{array}{ll}
T^{(1)}_{\bar \nu \nu} = \ghat_{\bar \nu \nu}\,, \qquad \qquad \qquad  &
T^{(2)}_{\bar \nu \nu} = w_{\bar \nu}\, w_\nu\,,\\
T^{(3)}_{\bar \nu \nu} = w_{\bar \nu}\, \bar r_\nu\,, &
T^{(4)}_{\bar \nu \nu} = r_{\bar \nu}\, w_\nu\,,\\
T^{(5)}_{\bar \nu \nu} = r_{\bar \nu}\, \bar r_\nu\,, & \\
\end{array}
\label{def-01to01}
\end{eqnarray}
where it suffices to assume the second particles with momenta $p_2$ and $\bar p_2$ to carry
the spin. The type arguments that lead to the given choice of tensors in (\ref{def-01to01}) are
identical to those given for the two vector production process (\ref{def-00to11}).
The construction of the associated projection tensors
\begin{eqnarray}
&& P_{\bar \mu  \mu}^{(n)} \, g^{\mu \nu}\,g^{\bar \mu \bar \nu}\,T^{(m)}_{\bar \nu \nu} = \delta_{nm} \,,
\nonumber\\
&&  P_{\bar \nu  \nu}^{(m)} \,\bar p_2^{\bar \nu } = 0 \, ,\qquad \quad  P_{\bar \nu  \nu}^{(m)} \,p_2^{\nu } =0 \,,
\end{eqnarray}
is analogous to (\ref{res-Q:00to11}). We find
\begin{eqnarray}
&& P^{(n)}_{\bar \nu  \nu} = \sum_{k=1}^5\,c_k^{(n)} \,Q^{(k)}_{\bar \nu  \nu} \,,
\label{res-Q:01to01}\\
&& Q_{1}^{\bar \nu  \nu} = \,v^{\bar \nu }\,v^{\nu}/v^2 \,, \qquad
\nonumber\\
&& Q_{2}^{\bar \nu  \nu} = \wRbot^{\bar \nu}\,\wRbarbot^{\nu}\,,\qquad
 Q_{3}^{\bar \nu  \nu} = \wRbot^{\bar \nu }\,\,\rbarbot^{\nu} \,,
\nonumber\\
&&Q_{4}^{\bar \nu  \nu} = \,\rbot^{\bar \nu}\,\wRbarbot^{\nu }\,,
 \qquad Q_{5}^{\bar \nu  \nu} =\, \rbot^{\bar \nu}\,\,\rbarbot^{\nu }\,,
\nonumber\\ \nonumber\\
&& c^{(n)}_n = 1 \,, \quad  c^{(2)}_1 = -\wRbot \cdot \wRbarbot + 1/s\, ,
\nonumber\\
&& c^{(3)}_1 = -\wRbot \cdot \rbarbot \, ,\quad
  c^{(4)}_1 = -\rbot \cdot \wRbarbot \, , \quad c^{(5)}_1 = -\rbot \cdot \rbarbot  \,,
\nonumber
\end{eqnarray}
where we display non-vanishing elements only and use the 4-vectors introduced in (\ref{def-rbot}).

\begin{widetext}

\begin{table}[t]
\rescale
\renewcommand{\arraystretch}{1.1}
\begin{tabular}{|ll|c |ll|c |ll|c  }
$k$ & $n$ & $c_{k}^{(n)}$ \phantom{xxxxxxxxxxxxxxxxxxxxxxxx}&
$k$ & $n$ & $c_{k}^{(n)}$ \phantom{xxxxxxxxxxxxxxxxxxxxxxxx}&
$k$ & $n$ & $c_{k}^{(n)}$ \phantom{xxxxxxxxxxxxxxxxxxxxxxxx}\\ \hline

$1$&$ 11$&$ 1$&$ 2$&$ 2$&$ \bar r^2\,s$&$ 2$&$ 3$&$ -(\bar r \cdot r)\,s$\\
  $2$&$ 4$&$ (\bar r \cdot r)$&$ 2$&$ 5$&$ -(\bar r \cdot r)$&$ 2$&$ 7$&$ \bar r^2$\\
  $2$&$ 8$&$ -\bar r^2$&$ 2$&$ 13$&$ \bar \alpha_+$&$ 3$&$ 2$&$ -{\textstyle{1\over 2}}\,\bar \alpha_+\,\bar r^2\,
    s$\\ $3$&$ 3$&$ {\textstyle{1\over 2}}\,\bar \alpha_+\,(\bar r \cdot r)\,s$&$ 3$&$ 4$&$ {\textstyle{1\over 2}}\,\bar \alpha_-\,(\bar r \cdot r)$&$ 3$&$
   5$&$ {\textstyle{1\over 2}}\,\bar \alpha_+\,(\bar r \cdot r)$\\ $3$&$ 7$&$ {\textstyle{1\over 2}}\,\bar \alpha_-\,\bar r^2$&$ 3$&$ 8$&$
  {\textstyle{1\over 2}}\, \bar \alpha_+\,\bar r^2$&$ 3$&$ 10$&$ -{\textstyle{1\over 2}}\,\bar \alpha_+\,s$\\
  $3$&$ 13$&$ -{\textstyle{1\over 2}}\,\bar \alpha_+^2$&$ 4$&$ 12$&$ 1$&$ 5$&$ 2$&$
   -{\textstyle{1\over 4}}\,\alpha_-\,\bar \alpha_+\,\bar r^2\,s$\\
  $5$&$ 3$&$ {\textstyle{1\over 4}}\,\alpha_-\,\bar \alpha_+\,(\bar r \cdot r)\,s$&$ 5$&$ 4$&$
   -{\textstyle{1\over 4}}\,\alpha_-\,\bar \alpha_+\,(\bar r \cdot r)$&$ 5$&$ 5$&$
   {\textstyle{1\over 4}}\,\bar \alpha_+\,(\alpha_-\,(\bar r \cdot r) + 2\,r^2)$\\
  $5$&$ 6$&$ (\bar r \cdot r)$&$ 5$&$ 7$&$ -{\textstyle{1\over 4}}\,\alpha_-\,\bar \alpha_+\,\bar r^2$&$ 5$&$ 8$&$
   {\textstyle{1\over 4}}\,\bar \alpha_+\,(\alpha_-\,\bar r^2 + 2\,(\bar r \cdot r))$\\
  $5$&$ 9$&$ \bar r^2$&$ 5$&$ 13$&${\textstyle{1\over 4}}\, \bar \alpha_-\,\alpha_-\,\bar \alpha_+$&$ 6$&$ 2$&$ \bar r^2\,s$\\
  $6$&$ 3$&$ -(\bar r \cdot r)\,s$&$ 6$&$ 4$&$ (\bar r \cdot r)$&$ 6$&$ 5$&$ -(\bar r \cdot r)$\\
  $6$&$ 7$&$ \bar r^2$&$ 6$&$ 8$&$ -\bar r^2$&$ 6$&$ 13$&$ -\bar \alpha_-$\\
  $7$&$ 2$&$ {\textstyle{1\over 2}}\,\bar \alpha_-\,\bar r^2\,s$&$ 7$&$ 3$&$ -{\textstyle{1\over 2}}\,\bar \alpha_-\,(\bar r \cdot r)\,s$&$ 7$&$
   4$&$ {\textstyle{1\over 2}}\,\bar \alpha_-\,(\bar r \cdot r)$\\ $7$&$ 5$&$ {\textstyle{1\over 2}}\,\bar \alpha_+\,(\bar r \cdot r)$&$ 7$&$ 7$&$
   {\textstyle{1\over 2}}\,\bar \alpha_-\,\bar r^2$&$ 7$&$ 8$&$ {\textstyle{1\over 2}}\,\bar \alpha_+\,\bar r^2$\\
  $7$&$ 10$&$ {\textstyle{1\over 2}}\,\bar \alpha_-\,s$&$ 7$&$ 13$&$ {\textstyle{1\over 2}}\,\bar \alpha_-\,\bar \alpha_+$&$ 8$&$ 2$&$
   -(\bar r \cdot r)\,s$\\ $8$&$ 3$&$ r^2\,s$&$ 8$&$ 4$&$ -r^2$&$ 8$&$ 5$&$
   r^2$\\ $8$&$ 7$&$ -(\bar r \cdot r)$&$ 8$&$ 8$&$ (\bar r \cdot r)$&$ 8$&$ 12$&$ 1$\\
  $9$&$ 1$&$ (\bar r \cdot r)^2 - \bar r^2\,r^2$&$ 9$&$ 2$&$
   -{\textstyle{1\over 4}}\,\alpha_-\,\bar \alpha_+\,\bar r^2\,s$&$ 9$&$ 3$&$ {\textstyle{1\over 4}}\,\alpha_-\,\bar \alpha_+\,(\bar r \cdot r)\,
    s$\\ $9$&$ 4$&$ {\textstyle{1\over 4}}\,(-\alpha_-\,\bar \alpha_+\,(\bar r \cdot r) - 2\,\bar \alpha_-\,r^2)$&$
   9$&$ 5$&$ {\textstyle{1\over 4}}\,\alpha_-\,\bar \alpha_+\,(\bar r \cdot r)$&$ 9$&$ 6$&$ (\bar r \cdot r)$\\
  $9$&$ 7$&$ {\textstyle{1\over 4}}\,(-\alpha_-\,\bar \alpha_+\,\bar r^2 - 2\,\bar \alpha_-\,(\bar r \cdot r))$&$ 9$&$ 8$&$
   {\textstyle{1\over 4}}\,\alpha_-\,\bar \alpha_+\,\bar r^2$&$ 9$&$ 9$&$ \bar r^2$\\
  $9$&$ 13$&$ {\textstyle{1\over 4}}\,\bar \alpha_-\,\alpha_-\,\bar \alpha_+$&$ 10$&$ 2$&$ \bar r^2\,s$&$ 10$&$ 3$&$
   -(\bar r \cdot r)\,s$\\ $10$&$ 4$&$ (\bar r \cdot r)$&$ 10$&$ 5$&$ -(\bar r \cdot r)$&$ 10$&$ 7$&$
   \bar r^2$\\ $10$&$ 8$&$ -\bar r^2$&$ 10$&$ 10$&$ s$&$ 10$&$ 13$&$ \bar \alpha_+$\\
  $11$&$ 2$&$ {\textstyle{1\over 2}}\,\alpha_-\,\bar r^2\,s$&$ 11$&$ 3$&$ -{\textstyle{1\over 2}}\,\alpha_-\,(\bar r \cdot r)\,s$&$
   11$&$ 4$&$ {\textstyle{1\over 2}}\,\alpha_-\,(\bar r \cdot r)$\\
  $11$&$ 5$&$ -{\textstyle{1\over 2}}\,\alpha_-\,(\bar r \cdot r) - r^2$&$ 11$&$ 7$&$
   {\textstyle{1\over 2}}\,\alpha_-\,\bar r^2$&$ 11$&$ 8$&$ -{\textstyle{1\over 2}}\,\alpha_-\,\bar r^2 -(\bar r \cdot r)$\\
  $11$&$ 13$&$ -{\textstyle{1\over 2}}\,\bar \alpha_-\,\alpha_-$&$ 12$&$ 2$&$ {\textstyle{1\over 2}}\,\alpha_-\,\bar r^2\,s$&$ 12$&$
   3$&$ -{\textstyle{1\over 2}}\,\alpha_-\,(\bar r \cdot r)\,s$\\
  $12$&$ 4$&$ {\textstyle{1\over 2}}\,\alpha_-\,(\bar r \cdot r) - r^2$&$ 12$&$ 5$&$ -{\textstyle{1\over 2}}\,\alpha_-\,(\bar r \cdot r)$&$
   12$&$ 7$&$ {\textstyle{1\over 2}}\,\alpha_-\,\bar r^2 - (\bar r \cdot r)$\\
  $12$&$ 8$&$ -{\textstyle{1\over 2}}\,\alpha_-\,\bar r^2$&$ 12$&$ 13$&$ {\textstyle{1\over 2}}\,\alpha_-\,\bar \alpha_+$&$ 13$&$ 6$&$
   -r^2$\\ $13$&$ 9$&$ -(\bar r \cdot r)$&$ 13$&$ 12$&$ {\textstyle{1\over 2}}\,\alpha_-$&&&

\end{tabular}
\label{tab-1}
\caption{The non-vanishing coefficients $c_k^{(n)}$ in the expansion (\ref{res-Q:01to11}). Here is
$ \alpha_\pm = 1 \pm \frac{m_1^2-m_2^2}{s} $ and $ \bar \alpha_\pm = 1 \pm \frac{\bar m_1^2-\bar m_2^2}{s}\,.$}
\end{table}
\end{widetext}

While the construction of a suitable basis
was almost trivially implied for the reactions $0\,0 \to 1\,1$ and $0\,1 \to 0\,1$
by choosing the tensors that involve the minimal number
of momenta, the task is considerably more complicated once there are
three vector particles involved. It suffices to construct tensors composed out of
the metric tensor $g_{\mu \nu}$ and the three 4 momenta $\bar r_\mu, r_\mu$ and $w_\mu$. Owing to the
relation (\ref{res-vv}) any structure involving an even number of  $v_\mu$ vectors is redundant.
Altogether there are 61 Lorentz structures with the proper parity transformation. Due to
the Schouten identity \cite{West:1991xv,Bondarev:1994ti}
\begin{eqnarray}
&& g_{\sigma\tau}\epsilon_{\alpha\beta\gamma\delta}= g_{\alpha\tau}\epsilon_{\sigma\beta\gamma\delta}
+g_{\beta\tau}\epsilon_{\alpha\sigma\gamma\delta} +g_{\gamma\tau}\epsilon_{\alpha\beta\sigma\delta}
\nonumber\\
&& \qquad +\,g_{\delta\tau}\epsilon_{\alpha\beta\gamma\sigma}\,,
\label{eps-identity}
\end{eqnarray}
only a subset of 28 structures are off-shell independent. This is in contrast to the number of
independent helicity amplitudes, which there are 13. Thus using on-shell conditions out of
the 28 tensors only 13 are linear independent. The task is to find a subset which is free of
kinematical constraints. The construction of such a set is quite tedious and to
the best knowledge of the authors such amplitudes did not exist for the considered reaction.
The on-shell scattering amplitude may be parameterized in terms of the following
13 scalar amplitudes $F_{1,...,13}$ with
\begin{eqnarray}
&&T_{0\,1 \to 1\,1}(\bar k, \,k,\, w) = \sum_{n=1}^{13} \,F_n (s,t)\,
\langle \,i\,T^{(n)}_{\bar \mu \bar \nu ,\nu} \rangle^{\bar \mu \bar \nu,\nu}_{0\,1 \to 1\,1} \,,
\label{def-01to11}\\
&& \langle \,T^{(n)}_{\bar \mu \bar \nu,\nu} \,\rangle^{\bar \mu \bar \nu,\nu}_{0\,1 \to 1\,1} =
\epsilon^{\dagger\,\bar \mu}(\bar p_1, \, \bar \lambda_1)\,
\epsilon^{\dagger\,\bar \nu}(\bar p_2, \, \bar \lambda_2)\,
\nonumber\\
&& \qquad \qquad \quad \quad \;\times\,T^{(n)}_{\bar \mu \bar \nu, \nu}\,\epsilon^\nu(p_2,\,\lambda_2) \,,
\nonumber\\ \nonumber\\
&& \begin{array}{ll}
T^{(1)}_{\bar \mu \bar \nu, \nu} =  \epsilon_{\bar \mu \bar \nu \nu \alpha}\, w^\alpha\,,  &
T^{(2)}_{\bar \mu \bar \nu, \nu} = \epsilon_{\bar \mu \bar \nu \nu \alpha}\, r^\alpha\,,\\
T^{(3)}_{\bar \mu \bar \nu, \nu} = \epsilon_{\bar \mu \bar \nu \nu \alpha}\, \bar r^\alpha\,, &
T^{(4)}_{\bar \mu \bar \nu, \nu} =w_{\bar \nu}\,\epsilon_{  \bar \mu  \nu \alpha  \beta} \,\bar r^\alpha\, w^\beta\,,\\
T^{(5)}_{\bar \mu \bar \nu, \nu} = w_{\bar \mu }\,\epsilon_{ \bar \nu  \nu  \alpha  \beta} \,\bar r^\alpha \, w^\beta\,, &
T^{(6)}_{\bar \mu \bar \nu, \nu} = r_{\bar \nu}\,\epsilon_{  \bar \mu  \nu \alpha  \beta} \,\bar r^\alpha\, w^\beta\,,\\
T^{(7)}_{\bar \mu \bar \nu, \nu} =  w_{\bar \nu}\,\epsilon_{  \bar \mu  \nu \alpha  \beta} \,w^\alpha\, r^\beta\,, &
T^{(8)}_{\bar \mu \bar \nu, \nu} = w_{\bar \mu }\,\epsilon_{ \bar \nu  \nu  \alpha  \beta} \,w^\alpha \, r^\beta \,,\\
T^{(9)}_{\bar \mu \bar \nu, \nu} = r_{\bar \nu}\,\epsilon^{  \bar \mu  \nu \alpha  \beta} \,w^\alpha\, r^\beta\,,&
T^{(10)}_{\bar \mu \bar \nu, \nu} = w_{\bar \nu}\,\epsilon_{  \bar \mu  \nu \alpha  \beta} \,\bar r^\alpha\, r^\beta\,,\\
T^{(11)}_{\bar \mu \bar \nu, \nu} = \ghat_{\bar \nu \nu}\,\epsilon_{  \bar \mu   \alpha \tau \beta} \,\bar  r^\alpha\,w^\tau \, r^\beta\,,
\!\!\!\!\!\!\!\!\! \!\!\!& \\
T^{(12)}_{\bar \mu \bar \nu, \nu} = r_{\bar \nu}\,w_\nu\,\epsilon_{  \bar \mu   \alpha \tau \beta} \,\bar r^\alpha\, w^\tau
\!\!\!\! \!\!\!\!\!\! \!\!\! \!\!\!\!\!\!& r^\beta\,, \\
T^{(13)}_{\bar \mu \bar \nu, \nu} = {\textstyle{1\over 2}}\,\big( w_{\bar \nu }\,w_\nu \,\epsilon_{  \bar \mu   \alpha \tau \beta}
\;- &w_{\bar \mu }\,w_\nu \, \epsilon_{  \bar \nu   \alpha \tau \beta} \big)\, \bar r^\alpha\, w^\tau \, r^\beta \,.
\\
\end{array}
\nonumber
\end{eqnarray}
We assure that our choice of amplitudes in (\ref{def-01to11}) excludes the occurrence of kinematical constraints with the possible exception at $s=0$.
Using slightly modified amplitudes as implied by the replacement $r_\mu \to k_\mu $ and $\bar r_\mu \to \bar k_\mu$ removes the constraints
at $s=0$.

Again we provide the convenient projection tensors that streamline the computation of the invariant amplitudes $F_n(s,t)$ by means of
algebraic computer codes. We find
\begin{eqnarray}
&& P^{(n)}_{\alpha  \beta ,\tau} \, g^{\alpha \bar \mu}\,g^{\beta \bar \nu}\,g^{\tau \nu}\,T^{(m)}_{\bar \mu  \bar \nu ,\nu} = \delta_{nm} \,,
\label{}\\
&&  P^{(n)}_{\bar \mu  \bar \nu, \nu } \,\bar p_1^{\bar \mu } = 0 \, ,\qquad
P^{(n)}_{\bar \mu  \bar \nu, \nu } \,\bar p_2^{\bar \nu } = 0 \, ,\qquad  P^{(n)}_{\bar \mu  \bar \nu, \nu } \,p_2^{ \nu } =0 \,.
\nonumber\\
&& P^{(n)}_{\bar \mu  \bar \nu, \nu} = \sum_{k=1}^{13}\,c_k^{(n)} \,Q^{(k)}_{\bar \mu  \bar \nu, \nu} \,,
\label{res-Q:01to11}\\
&& \begin{array}{ll}
  Q_{1}^{\bar \mu  \bar \nu, \nu} = v^{\bar \mu }\,v^{\bar \nu}\,v^{\nu} /v^2 /v^2\,, \qquad & \\

  Q_{2}^{\bar \mu \bar \nu , \nu} = v^{\bar \mu }\,\wRbot^{\bar \nu}\,\wRbarbot^{\nu}/v^2  \;- \!\!&(\wRbot \cdot \wRbarbot)\,Q_{1}^{\bar \mu  \bar \nu, \nu} \,, \\
  Q_{3}^{\bar \mu \bar \nu , \nu}  = v^{\bar \mu }\, \wRbot^{\bar \nu }\,\,\rbarbot^{\nu}/v^2  \;- \!\!& (\wRbot\cdot\,\rbarbot)\,Q_{1}^{\bar \mu  \bar \nu, \nu} \,, \\
  Q_{4}^{\bar \mu\bar \nu , \nu}  = v^{\bar \mu }\,\rbot^{\bar \nu}\,\wRbarbot^{\nu }/v^2  \;\,- \!\!&(\,\rbot\cdot\wRbarbot)\,Q_{1}^{\bar \mu  \bar \nu, \nu} \,, \\
  Q_{5}^{\bar \mu \bar \nu,  \nu} = v^{\bar \mu }\, \rbot^{\bar \nu}\,\,\rbarbot^{\nu }/v^2  \;\,- \!\!&(\,\rbot \cdot\,\rbarbot)\,Q_{1}^{\bar \mu  \bar \nu, \nu} \,, \\

  Q_{6}^{\bar \mu \bar \nu , \nu} = \wLbot^{\bar \mu}\,v^{\bar \nu }\,\wRbarbot^{\nu}/v^2\,,
& Q_{7}^{\bar \mu \bar \nu , \nu} = \wLbot^{\bar \mu }\,v^{\bar \mu }\,\,\rbarbot^{\nu}/v^2 \,, \\
  Q_{8}^{\bar \mu\bar \nu , \nu}  = \,\rbot^{\bar \mu}\,v^{\bar \nu }\,\,\wRbarbot^{\nu }/v^2\,,
& Q_{9}^{\bar \mu \bar \nu,  \nu} = \,\rbot^{\bar \mu}\,v^{\bar \nu }\, \,\,\rbarbot^{\nu }/v^2\,, \\
  Q_{10}^{\bar \mu  \bar \nu, \nu} = \wLbot^{\bar \mu}\,\wRbot^{\bar \nu}\,v^\nu/v^2\,,
& Q_{11}^{\bar \mu  \bar \nu, \nu} = \wLbot^{\bar \mu }\,\,\rbot^{\bar \nu}\,v^\nu/v^2 \,, \\
  Q_{12}^{\bar \mu  \bar \nu, \nu} = \,\rbot^{\bar \mu}\,\wRbot^{\bar \nu }\,v^\nu/v^2\,,
& Q_{13}^{\bar \mu  \bar \nu, \nu} = \,\rbot^{\bar \mu}\,\,\rbot^{\bar \nu } \,v^\nu/v^2\,,
\end{array}
\nonumber
\end{eqnarray}
where the explicit form of the coefficients $c_k^{(n)}$ can be found in Tab. I.

We turn to the most complicated reaction $1 \,1 \to 1\,1$. Excluding Lorentz
structures involving an even number of  $v_\mu$ vectors there are 138 structures
with the proper parity transformation that one may write down.
A subset of 136 structures are off-shell independent. Using on-shell conditions out of
the 136 tensors only 41 are linear independent. We constructed a subset which is free of kinematical constraints
with the possible exception of $s=0$. To the best knowledge of the authors such amplitudes
did not exist for the considered scattering process. The on-shell scattering amplitude may be parameterized
in terms of the following 41  scalar amplitudes $F_{1,...,41}$ with
\allowdisplaybreaks[1]
\begin{eqnarray}
&&T_{1\,1 \to 1\,1}(\bar k, \,k,\, w) = \sum_{n=1}^{41} \,F_n (s,t)\,
\langle T^{(n)}_{\bar \mu \bar \nu ,\mu \nu} \rangle^{\bar \mu \bar \nu,\mu \nu}_{1\,1 \to 1\,1} \,,
\label{def-11to11}\\
&& \langle \,T^{(n)}_{\bar \mu \bar \nu,\mu \nu} \,\rangle^{\bar \mu \bar \nu,\mu \nu}_{1\,1 \to 1\,1} =
\epsilon^{\dagger\,\bar \mu}(\bar p_1, \, \bar \lambda_1)\,
\epsilon^{\dagger\,\bar \nu}(\bar p_2, \, \bar \lambda_2)\,
\nonumber\\
&& \qquad \qquad \quad \quad \;\;\,\times\,T^{(n)}_{\bar \mu \bar \nu, \mu \nu}
\,\epsilon^\nu(p_1,\,\lambda_1)\,\epsilon^\nu(p_2,\,\lambda_2)  \,,
\nonumber\\ \nonumber\\
&& \begin{array}{ll}
T^{(1)}_{\bar \mu \bar \nu, \mu \nu} = \ghat_{\bar \mu \mu}\,\ghat_{\bar \nu \nu}  \,,&
T^{(2)}_{\bar \mu \bar \nu, \mu \nu} = \ghat_{\bar \mu \nu}\,\ghat_{\bar \nu \mu}\,,\\
T^{(3)}_{\bar \mu \bar \nu, \mu \nu} = \ghat_{\bar \mu \bar \nu}\,\ghat_{\mu \nu } \,,&

T^{(4)}_{\bar \mu \bar \nu, \mu \nu}  = \ghat_{\bar \nu \nu}\, w_{\bar \mu}\,w_\mu\,,\\
T^{(5)}_{\bar \mu \bar \nu, \mu \nu}  = \ghat_{\bar \nu \nu}\, r_{\bar \mu}\,w_\mu  \,,&
T^{(6)}_{\bar \mu \bar \nu, \mu \nu}  = \ghat_{\bar \nu \nu}\, w_{\bar \mu}\,\bar r_\mu\,,\\
T^{(7)}_{\bar \mu \bar \nu, \mu \nu}  = \ghat_{\bar \nu \nu}\, r_{\bar \mu}\,\bar r_\mu  \,,&

T^{(8)}_{\bar \mu \bar \nu, \mu \nu}  = \ghat_{\bar \mu \mu}\, w_{\bar \nu}\,w_{\nu }\,,\\
T^{(9)}_{\bar \mu \bar \nu, \mu \nu}  = \ghat_{\bar \mu \mu}\, r_{\bar \nu}\,w_\nu   \,,&
T^{(10)}_{\bar \mu \bar \nu, \mu \nu} = \ghat_{\bar \mu \mu}\, w_{\bar \nu}\,\bar r_\nu \,,\\
T^{(11)}_{\bar \mu \bar \nu, \mu \nu} = \ghat_{\bar \mu \mu}\, r_{\bar \nu}\,\bar r_\nu \,,&

T^{(12)}_{\bar \mu \bar \nu, \mu \nu} = \ghat_{\bar \nu \mu}\, w_{\bar \mu}\,w_\nu\,,\\
T^{(13)}_{\bar \mu \bar \nu, \mu \nu} = \ghat_{\bar \nu \mu}\, r_{\bar \mu}\,w_\nu  \,,&
T^{(14)}_{\bar \mu \bar \nu, \mu \nu} = \ghat_{\bar \nu \mu}\, w_{\bar \mu}\,\bar r_\nu\,,\\
T^{(15)}_{\bar \mu \bar \nu, \mu \nu} = \ghat_{\bar \nu \mu}\, r_{\bar \mu}\,\bar r_\nu \,,&

T^{(16)}_{\bar \mu \bar \nu, \mu \nu} = \ghat_{\bar \mu \nu}\, w_{\bar \nu}\,w_\mu\,,\\
T^{(17)}_{\bar \mu \bar \nu, \mu \nu} = \ghat_{\bar \mu \nu}\, r_{\bar \nu}\,w_\mu  \,,&
T^{(18)}_{\bar \mu \bar \nu, \mu \nu} = \ghat_{\bar \mu \nu}\, w_{\bar \nu}\,\bar r_\mu\,,\\
T^{(19)}_{\bar \mu \bar \nu, \mu \nu} = \ghat_{\bar \mu \nu}\, r_{\bar \nu}\,\bar r_\mu \,,&

T^{(20)}_{\bar \mu \bar \nu, \mu \nu} = \ghat_{\bar \mu \bar \nu}\,w_\mu\, w_{\nu}\,,\\
T^{(21)}_{\bar \mu \bar \nu, \mu \nu} = \ghat_{\bar \mu \bar \nu}\,\bar r_\mu\, w_{\nu}\,,&
T^{(22)}_{\bar \mu \bar \nu, \mu \nu} = \ghat_{\bar \mu \bar \nu}\,w_\mu\,\bar r_{\nu} \,, \\
T^{(23)}_{\bar \mu \bar \nu, \mu \nu} = \ghat_{\bar \mu \bar \nu}\,\bar r_\mu \, \bar r_{\nu}\,,&

T^{(24)}_{\bar \mu \bar \nu, \mu \nu} = \ghat_{\mu \nu}\,w_{\bar \mu }\, w_{\bar \nu}\,,\\
T^{(25)}_{\bar \mu \bar \nu, \mu \nu} = \ghat_{\mu \nu}\,r_{\bar \mu}\, w_{\bar \nu}\,,&
T^{(26)}_{\bar \mu \bar \nu, \mu \nu} = \ghat_{\mu \nu}\,w_{\bar \mu}\,r_{\bar \nu} \,, \\
T^{(27)}_{\bar \mu \bar \nu, \mu \nu} = \ghat_{\mu \nu}\,r_{\bar \mu}\, r_{\bar \nu} \,,&

T^{(28)}_{\bar \mu \bar \nu, \mu \nu} =  w_{\bar \mu }\,w_{\bar \nu}\,w_\mu\,w_\nu\,,\\
T^{(29)}_{\bar \mu \bar \nu, \mu \nu} =  r_{\bar \mu }\,r_{\bar \nu}\,w_\mu\,w_\nu\,,&
T^{(30)}_{\bar \mu \bar \nu, \mu \nu} =  w_{\bar \mu }\,w_{\bar \nu}\,\bar r_\mu\,\bar r_\nu\,,\\

T^{(31)}_{\bar \mu \bar \nu, \mu \nu} =  r_{\bar \mu }\,w_{\bar \nu}\,w_\mu\,w_\nu\,,&
T^{(32)}_{\bar \mu \bar \nu, \mu \nu} =  w_{\bar \mu }\,r_{\bar \nu}\,w_\mu\,w_\nu\,,\\
T^{(33)}_{\bar \mu \bar \nu, \mu \nu} =  r_{\bar \mu }\,w_{\bar \nu}\,\bar r_\mu\,\bar r_\nu\,,&
T^{(34)}_{\bar \mu \bar \nu, \mu \nu} =  w_{\bar \mu }\,r_{\bar \nu}\,\bar r_\mu\,\bar r_\nu\,,\\

T^{(35)}_{\bar \mu \bar \nu, \mu \nu} =  w_{\bar \mu }\,w_{\bar \nu}\,\bar r_\mu\,w_\nu\,,&
T^{(36)}_{\bar \mu \bar \nu, \mu \nu} =  r_{\bar \mu }\,r_{\bar \nu}\,\bar r_\mu\,w_\nu\,,\\
T^{(37)}_{\bar \mu \bar \nu, \mu \nu} =  w_{\bar \mu }\,w_{\bar \nu}\,w_\mu\,\bar r_\nu\,,&
T^{(38)}_{\bar \mu \bar \nu, \mu \nu} =  r_{\bar \mu }\,r_{\bar \nu}\,w_\mu\,\bar r_\nu\,,\\

T^{(39)}_{\bar \mu \bar \nu, \mu \nu} =  {\textstyle{1\over 4}}\,\big( r_{\bar \mu}  \,w_{\bar \nu} +w_{\bar \mu}  \,r_{\bar \nu} \big)&
\big( \bar r_{\mu}  \,w_{\nu} - w_{\mu}  \,\bar r_{\nu} \big)\,,\\

T^{(40)}_{\bar \mu \bar \nu, \mu \nu} = {\textstyle{1\over 4}}\,\big( r_{\bar \mu}  \,w_{\bar \nu} - w_{\bar \mu}  \,r_{\bar \nu} \big) &
\big( \bar r_{\mu}  \,w_{\nu} + w_{\mu}  \,\bar r_{\nu} \big)\,,\\

T^{(41)}_{\bar \mu \bar \nu, \mu \nu} = {\textstyle{1\over 4}}\,\big( r_{\bar \mu}  \,w_{\bar \nu} - w_{\bar \mu}  \,r_{\bar \nu} \big) &
\big( \bar r_{\mu}  \,w_{\nu} - w_{\mu}  \,\bar r_{\nu} \big)\,,\\

\end{array}
\nonumber
\end{eqnarray}
where our invariant functions are kinematically correlated at $s=0$ only. The latter constraint can be
eliminated by the use of the modified vectors $r_\mu \to k_\mu$ and $\bar r_\mu \to \bar k_\mu $
in (\ref{def-11to11}). An algebra to project onto the invariant amplitudes $F_n(s,t)$ is developed in Appendix A.

We emphasize that all amplitudes $F_n(s,t)$ introduced in this section are truly uncorrelated and satisfy
Mandelstam's dispersion integral representation ~\cite{Mandelstam:1958xc,Ball:1961zza}.

\section{Partial-wave decomposition}

The helicity matrix elements of the scattering operator, $T$, are decomposed into
partial-wave amplitudes characterized by the total angular momentum $J$.
Given a specific process together with our convention of the helicity wave functions it suffices
to specify the helicity projection  $\lambda_{1},\lambda_2$ and
$\bar \lambda_{1}, \bar \lambda_2$ as introduced in the previous section.
We write
\begin{eqnarray}
&& \SP{\bar{\lambda}_1\bar{\lambda}_2}{T}{\lambda_1\lambda_2}= \sum_{J}
(2\, J + \!1) \,\langle \bar{\lambda}_1 \bar{\lambda}_2 | \,T_J | \lambda_1 \lambda_2 \rangle \,
d^{(J)}_{\lambda,\bar{\lambda}} (\theta)\,,
\nonumber\\
&& d^{(J)}_{\lambda, \bar \lambda} (\theta )=(-)^{\lambda-\bar \lambda}\, d^{(J)}_{-\lambda, -\bar \lambda} (\theta )
\nonumber\\
&& \qquad \quad \;= (-)^{\lambda-\bar \lambda}\, d^{(J)}_{\bar \lambda, \lambda} (\theta ) =
d^{(J)}_{-\bar \lambda, -\lambda} (\theta )\,,
\label{def-Wigner} \\
&& \langle \bar{\lambda}_1 \bar{\lambda}_2 | \,T_J | \lambda_1 \lambda_2 \rangle
= \int_{-1}^{-1} \frac{\mathrm{d}\cos\theta}{2} \tsp{\bar{\lambda}_1\bar{\lambda}_2}{\lambda_1 \lambda_2} \,
d^{(J)}_{\lambda,\bar{\lambda}} (\theta)\,, \nonumber
\end{eqnarray}
with $\lambda=\lambda_{1}-\lambda_{2}$ and $\bar{\lambda}=\bar{\lambda}_{1}-\bar{\lambda}_{2}$.
Wigner's rotation functions, $d^{(J)}_{\lambda,\bar \lambda}(\theta)$, are used in a convention as
characterized by (\ref{def-Wigner}). The phase conventions assumed in this work imply the parity relations
\begin{eqnarray}
&& \langle -\bar \lambda_1, -\bar \lambda_2 | \,T\, | -\lambda_1 ,-\lambda_2 \rangle
 = (-)^{\Delta  } \,
\langle \bar \lambda_1, \bar \lambda_2 | \,T\, | \lambda_1, \lambda_2 \rangle\,,
\nonumber\\
&& \qquad \qquad \Delta = S_{1}-S_{2} +\bar S_1 -\bar S_2
\nonumber\\
&& \qquad \qquad \quad \,+\,\lambda_1-\lambda_2 -\bar \lambda_1 +\bar \lambda_2\,.
\label{def-helicity-flip}
\end{eqnarray}
It is useful to decouple the two parity sectors by
introducing parity eigenstates of good total angular momentum $J$,
formed in terms of the helicity states \cite{Jacob:1959at}.
Following (\ref{def-Wigner}) we introduce the angular momentum
projection, $| \lambda_1, \lambda_2 \,\rangle_J$, of the helicity
state $|\lambda_1, \lambda_2 \,\rangle$. We write
\begin{eqnarray}
| \lambda_1, \lambda_2  \,\rangle_J \,, \quad {\rm with}\quad
T\,| \lambda_1, \lambda_2 \,\rangle_J = T_J|\lambda_1, \lambda_2 \,\rangle \,.
\label{def-angular-projection}
\end{eqnarray}

\begin{widetext}

\begin{table}[t]

\rescale

\renewcommand{\arraystretch}{1.1}
\setlength{\tabcolsep}{1.2mm}
\setlength{\arraycolsep}{1.4mm}

\begin{tabular}{rl|c|rl|c|rl|c  }

$a$ & $b$ & $\big[U^J_{+,0\,1}\big]_{ab} \phantom{\Big]}$ \phantom{xxxxxxxxxxxxxxxx} &
$a$ & $b$ & $\big[U^J_{+,0\,1}\big]_{ab} \phantom{\Big]}$ \phantom{xxxxxxxxxxxxxxxx} &
$a$ & $b$ & $\big[U^J_{+,0\,1}\big]_{ab} \phantom{\Big]}$ \phantom{xxxxxxxxxxxxxxxx}\\ \hline

$1$&$ 1$&$ (M_+ - M_-)\,\frac{\sqrt{s}}{2\,p}$&$ 2$&$ 1$&$ \alpha_-\,\frac{s}{2\,p}\,
     \sqrt{\frac{J}{J+1}}
     $&$ 2$&$ 2$&$ -p$

\\ \\

$a$ & $b$ & $\big[U^J_{+,1\,1}\big]_{ab} \phantom{\Big]}$ \phantom{xxxxxxxxxxxxxxxx} &
$a$ & $b$ & $\big[U^J_{+,1\,1}\big]_{ab} \phantom{\Big]}$ \phantom{xxxxxxxxxxxxxxxx} &
$a$ & $b$ & $\big[U^J_{+,1\,1}\big]_{ab} \phantom{\Big]}$ \phantom{xxxxxxxxxxxxxxxx}\\ \hline

$1$&$ 1$&$ -\sqrt{2}\,\frac{s}{p}$&$ 2$&$ 1$&$ -(M_+ + M_-)\,\frac{\sqrt{s}}{p}\,\sqrt{\frac{J}{J+1}}
     $&$ 2$&$ 2$&$ \alpha_-\,(M_+ + M_-)\,\frac{\sqrt{s}}{4\, p}$\\
     $2$&$ 3$&$ (M_+ + M_-)\,\frac{p}{2\,\sqrt{s}}$&$ 3$&$ 1$&$
   - (M_+ - M_-)\,\frac{\sqrt{s}}{p}\,\sqrt{\frac{J}{J+1}}$&$ 3$&$ 2$&$ -\alpha_+\,(M_+ - M_-)\,
     \frac{ \sqrt{s}}{4\, p}$\\
     $3$&$ 3$&$ (M_+ - M_-)\,\frac{p}{2\,\sqrt{s}}$&$ 4$&$ 1$&$ M_+\,M_-\,\frac{1}{p}\,
    \sqrt{ \frac{2\, (J-1)\,J}{(J+1)\,(J+2)}}
     $&$ 4$&$ 2$&$ -\alpha_-\,\alpha_+\,\frac{s}{4\, p}\,\sqrt{\frac{2\,(J-1)}{J+2}}$\\$4$&$ 3$&$ -M_+\,M_-\,\frac{p}{2\,s}\,
     \sqrt{\frac{2\,(J-1)}{J+2}}
     $&$ 4$&$ 4$&$ \frac{p}{\sqrt{2}}$&$  $&&$ $

\\ \\

$a$ & $b$ & $\big[U^J_{-,1\,1}\big]_{ab} \phantom{\Big]}$ \phantom{xxxxxxxxxxxxxxxx} &
$a$ & $b$ & $\big[U^J_{-,1\,1}\big]_{ab} \phantom{\Big]}$ \phantom{xxxxxxxxxxxxxxxx} &
$a$ & $b$ & $\big[U^J_{-,1\,1}\big]_{ab} \phantom{\Big]}$ \phantom{xxxxxxxxxxxxxxxx}\\ \hline

$1$&$ 1$&$ (M_-^2\,- M_+^2)\,\frac{s}{2\,p^2}\,\sqrt{\frac{J+1}{J}}$&$ 2$&$ 1$&$ \alpha_-\,\alpha_+\,
     \frac{s^2}{4\, p^2} \,\sqrt{\frac{2 \,(J+1)}{J}}$&$ 2$&$ 2$&$ -s\,\sqrt{\frac{2\,(J+1)}{J}}$\\$
   3$&$ 1$&$ -\alpha_-\,(M_- + M_+)\,\frac{\sqrt{s}\,s}{2\,p^2}$&$ 3$&$ 3$&$ (M_- + M_+)\,\frac{\sqrt{s}}{2}$&$ 4$&$ 1$&$
    \alpha_+\,(M_+ - M_-)\,\frac{\sqrt{s}\,s}{2\,p^2}$\\$4$&$ 3$&$(M_+ - M_-)\,\frac{ \sqrt{s}}{2}$&$ 4$&$ 4$&$
    \sqrt{s}\,(M_- - M_+)$&$ 5$&$ 1$&$ \alpha_-\,\alpha_+\,\frac{s^2}{4 \,p^2}\,\sqrt{\frac{2\,(J-1)}{J+2}}$\\$5$&$ 2$&$ s\,
     \sqrt{\frac{2\,(J-1)}{J+2}}
     $&$ 5$&$ 3$&$ -M_+\,M_-\,\sqrt{\frac{2\,(J-1)}{J+2}}$&$ 5$&$ 4$&$ -\alpha_-\,s\,\sqrt{\frac{2\,(J-1)}{J+2}}$\\$5$&$ 5$&$ -p^2\,
     \frac{\sqrt{2} \,(2\,J+1)}{\sqrt{J-1}\, \sqrt{J+2}}
     $&&&&&&

\end{tabular}
\label{tab-2}
\caption{Non-zero elements of the transformation matrices  $U^J_{+,0\,1}$ and $U^J_{\pm,1\,1}$. We use the notation of (\ref{def-alpha-delta}).}
\end{table}

\end{widetext}

We introduce parity eigenstates states, $|n_\pm ,J\, \rangle$, that are
eigenstates of the total angular momentum. We will be applying the
following state convention
\begin{eqnarray}
&&|1_- ,J \,\rangle_{0\,0} = | \,0\,,\,0\, \rangle_J\,,
\label{def-states-plus}\\
\nonumber\\
&&|1_- ,J \,\rangle_{0\,1} = {\textstyle{1\over \sqrt{2}}}\,\Big( | \,0\,,- \rangle_J - |\,0\,,+ \rangle_J\Big)\,,
\nonumber\\ \nonumber\\
&&|1_- ,J \,\rangle_{1\,1} = | \,0\,,\,0\, \rangle_J\,,
\nonumber\\
&&|2_- ,J \,\rangle_{1\,1} = {\textstyle{1\over \sqrt{2}}}\,\Big(
| +,+ \rangle_J +| -,- \rangle_J\Big)\,,
\nonumber\\
&&|3_- ,J \,\rangle_{1\,1} = {\textstyle{1\over \sqrt{2}}}\,\Big(
| \,0\,, - \rangle_J +| \,0\,,+ \rangle_J\Big)\,,
\nonumber\\
&&|4_- ,J \,\rangle_{1\,1} = {\textstyle{1\over \sqrt{2}}}\,\Big(
|+, \,0\, \rangle_J+| -, \,0\, \rangle_J\Big)\,,
\nonumber\\
&&|5_- ,J \,\rangle_{1\,1} = {\textstyle{1\over \sqrt{2}}}\,\Big(
| +,- \rangle_J + | -,+ \rangle_J \Big)\,,
\nonumber
\end{eqnarray}
and
\begin{eqnarray}
&&|1_+ ,J \,\rangle_{0\,1} = | \,0\,,\,0\, \rangle_J\,,
\label{def-states-minus}\\
&&|2_+ ,J \,\rangle_{0\,1} ={\textstyle{1\over \sqrt{2}}}\,\Big( | \,0\,,- \rangle_J + |\,0\,,+ \rangle_J\Big)\,,
\nonumber\\ \nonumber\\
&&|1_+ ,J \,\rangle_{1\,1} = {\textstyle{1\over \sqrt{2}}}\,\Big(
| +,+ \rangle_J -| -,- \rangle_J\Big)\,,
\nonumber\\
&&|2_+ ,J \,\rangle_{1\,1} = {\textstyle{1\over \sqrt{2}}}\,\Big(
| \,0\,, - \rangle_J -| \,0\,,+ \rangle_J\Big)\,,
\nonumber\\
&&|3_+ ,J \,\rangle_{1\,1} = {\textstyle{1\over \sqrt{2}}}\,\Big(
|+, \,0\, \rangle_J-| -, \,0\, \rangle_J\Big)\,,
\nonumber\\
&&|4_+ ,J \,\rangle_{1\,1} = {\textstyle{1\over \sqrt{2}}}\,\Big(
| +,- \rangle_J - | -,+ \rangle_J \Big)\,,
\nonumber
\end{eqnarray}
where we suppress the sector index $0\,0$, $0\,1$ or $1\,1$ on the right hand sides and use the short-hand notation
$\pm \equiv \pm 1$. The states have the following property
\begin{eqnarray}
P\,|n_\pm ,J \,\rangle = \pm (-1)^{J+1} \,|n_\pm ,J\,\rangle  \,.
\label{parity-eigenstates}
\end{eqnarray}

The partial-wave helicity amplitudes $t^{J}_{\pm,ab}$ that carry good angular momentum $J$ and
good parity are defined with
\begin{eqnarray}
t^{J}_{\pm,ab} =  \SP{a_{\pm},J\,}{T}{b_{\pm},J}\,,
\eqlab{def-tij}
\end{eqnarray}
where $a$ and $b$ label the states. For sufficiently large $s$ the unitarity condition takes the simple form
\begin{eqnarray}
\Im \big[ t^{(J)}\big]^{-1}_{ab} =
-\frac{p_a}{8\pi\,\sqrt{s}}\,\delta_{ab} \,,
\label{unitarity-appendix}
\end{eqnarray}
where the index $a$ and $b$ spans the basis of two-particle helicity states in
the $(0,0),(0,1)$ and $(1,1)$ sectors.

Helicity-partial-wave amplitudes are correlated at specific kinematical conditions.
This is seen once the amplitudes $t^{J}_{\pm,ab}(s)$ are expressed in terms of the invariant functions
$ F_{n}(s,t)$. This is a well know problem related
to the use of the helicity basis in  covariant models, see for example the review \cite{CohenTannoudji:1968}.
In contrast covariant-partial wave amplitudes $T^{J}_\pm(s) $ are free of kinematical constraints and can therefore
be used  efficiently in partial-wave dispersion relation. They are associated to covariant states and a
covariant projector algebra which diagonalizes the Bethe-Salpeter two-body scattering equation for
local interactions \cite{Lutz:2001yb,Lutz:2001mi,Lutz:2003fm,Stoica:2011cy}. We introduce
\begin{align}
T^{J}_\pm (s) &= \left( \frac{s}{  \bar{p} \,p} \right)^J \big[\bar{U}_{\pm}^{J}(s) \big]^T\, t^J_\pm(s) \, U^J_{\pm}(s) \;,
\label{defTJ}
\end{align}
with nontrivial matrices $U^J_{\pm}(s)$ and $\bar{U}^J_{\pm}(s)$ characterizing the transformation
for the initial and final states from the helicity basis to the new kinematic-free basis.
The matrices $U^J_{\pm}(s)$ are given in Tab. \ref{tab-2}, where we use the conventions
\begin{eqnarray}
&& M_{\pm } = m_1 \pm  m_2 \,, \qquad \bar M_{\pm } =\bar m_1 \pm \bar m_2 \,,
\nonumber\\
&& \delta = \frac{M_+\,M_-}{s} \,, \qquad \quad \;\;\,\bar \delta = \frac{\bar M_+\,\bar M_-}{s}\,,
\nonumber\\
&& \alpha_\pm = 1\pm \delta \,, \qquad  \qquad \;\bar \alpha_\pm =1 \pm \bar \delta \,.
\label{def-alpha-delta}
\end{eqnarray}

The transformation (\ref{defTJ}) implies a change in the phase-space distribution:
\begin{align}
    \nn
    \rho_{\pm}^{J}(s ) &= - \Im\Big[T_{\pm}^{J}(s)\Big]^{-1} \\
        & = \frac{1 }{8 \,\pi} \left(\frac{p}{\sqrt{s}}\right)^{2\,J+1}
    \Big[ U^J_{\pm}(s)\Big]^{-1} \Big[U_{\pm}^{J}(s)\Big]^{T,-1}\,.
    \label{def-rho}
\end{align}
We adapt a convention for the transformation matrices that lead to an asymptotically bounded
phase-space matrix, i.e. we require
\begin{eqnarray}
\lim_{s\to \infty}\det \rho _{\pm}^{J}( s ) =  {\rm const} \neq 0\,.
\label{def-rho-constraint}
\end{eqnarray}
In contrast to the helicity states the phase-space matrix in the covariant states does have off-diagonal elements. The quest for
the elimination of kinematical constraints leads necessarily to off-diagonal elements. To the best knowledge of the authors transformation
matrices for the '$1\,1$' case in Tab. \ref{tab-2} are novel and not presented in the literature before.

We provide  particularly detailed results for the $ 0\, 1 \to 0\,1 $, $ 0\, 1 \to 1\,1 $ and $ 0\, 0 \to 1\,1 $  reactions, but refrain from
giving the tedious details for the remaining cases. There are two one-dimensional cases with
\begin{eqnarray}
U^J_{-,0\,0} = 1\,, \qquad U^J_{-,0\,1}= 1\,.
\label{def-trivial-U}
\end{eqnarray}
It follows
\begin{eqnarray}
&& T^J_{-,0\,0 \to 0\,0 }(s) =  A^J_1(s) \,, \qquad
\nonumber\\
&& T^J_{-,0\,0 \to 0\,1 }(s) = \frac{\sqrt{J\,(J+1)}}{(2\,J+1)\,\sqrt{s}}\,\Big[ s^2\, A^{J-1}_1(s)
\nonumber\\
&& \qquad \qquad \quad \;\;-\,\bar p^2 \,p^2 A^{J+1}_1(s)  \Big] \,, \qquad
\nonumber\\
&&T^J_{-,0\,1 \to 0\,1 }(s) = - A^J_1(s) +  \frac{\bar p^2 \,p^2}{(2\,J+1)\,s}\,A^{J+1}_5(s)
\nonumber\\
&&\qquad \qquad \quad \;\;-\, \frac{s}{2\,J+1}\,A^{J-1}_5(s)\,,
\label{trivial-cases}
\end{eqnarray}
with
\begin{eqnarray}
&& A^{J}_{n}(s) = \left( \frac{s}{\bar p\,p} \right)^J \int_{-1}^{1} \!
\frac{\mathrm{d} \cos\theta}{2} \, F_n(s,t) \, P_J(\cos \theta) \,.
\label{def-As}
\end{eqnarray}
The important merit of (\ref{trivial-cases}) is the absence of kinematical constraints, with the repeatedly discussed exception 
at $s=0$. A potential singularity at $\bar p\,p=0$ in (\ref{def-As}) is not realized due to the properties
of the Legendre polynomials $P_J(\cos \theta)$.

The corresponding decomposition of the partial-wave scattering amplitudes $T^J(s)$ into integrals over the invariant
amplitude $F_n(s,t)$ for the $ 0\, 1 \to 0\,1 $,  $ 0\, 1 \to 1\,1 $,
$ 0\, 0 \to 1\,1 $ and $ 1\, 1 \to 1\,1 $  reactions is considerably more involved. We write
\begin{eqnarray}
T^J_\pm(s) = \sum_{k,n}\,a_{\pm n}^{J+k}(s)\,A_n^{J+k}(s)\,,
\label{def-decomposition-TJ}
\end{eqnarray}
with coefficients $a_{\pm n}^{J+k}(s)$.

In  Tab. \ref{tab-3} we detail those results for the coefficients $a_{\pm n}^{J+k}(s)$ which demonstrate
the absence of kinematical singularities in all partial-wave amplitudes considered in this work.  We note that the transformation
matrices in Tab. \ref{tab-2} are derived from a study of the three reactions $ 0\, 1 \to 0\,1 $,  $ 0\, 1 \to 1\,1 $ and 
$ 0\, 0 \to 1\,1 $ with appropriate parity selection only.
From the regularity of the $a_{\pm n}^{J+k}(s)$ in Tab. \ref{tab-3} it follows the absence of kinematical singularities. Additional 
and tedious computations reveal that there are also no correlations of the various partial-wave amplitudes $T^J_\pm (s)$ 
at any kinematical point but at $s=0$. The covariant amplitudes are associated to a projector algebra for the Bethe-Salpeter 
scattering equation  \cite{Lutz:2001yb,Lutz:2001mi,Lutz:2003fm,Stoica:2011cy}. A consistency check was performed that confirm the 
claimed properties for the remaining cases in particular the most tedious $1\,1\to 1\,1$ reaction.

\begin{widetext}

\begin{table}[t]
\rescale
\setlength{\tabcolsep}{0.5mm}
\renewcommand{\arraystretch}{1.4}
\begin{tabular}{rc|c|rl|c|rl|c  }

\phantom{-}$\,k$ & $n\;$ & $\big[a_{+n}^{J+k}\big]_{11}$ \phantom{xxxxxxx} $0\,1\to 0\,1 $\phantom{xxxxxxx}&
\phantom{-}$\,k$ & $n\;$ & $\big[a_{+n}^{J+k}\big]_{11}$ \phantom{xxxxxxx} $0\,1\to 0\,1 $\phantom{xxxxxxx}&
\phantom{-}$\,k$ & $n\;$ & $\big[a_{+n}^{J+k}\big]_{11}$ \phantom{xxxxxxx} $0\,1\to 0\,1 $\phantom{xxxxxxx}\\ \hline

$0$&$ 2$&$ s^2$&$ 0$&$ 5$&$ \bar \alpha_-\,\alpha_-\,s^2\,\frac{1}{4}\,\frac{1}{(J+1) (2\,J+3)}$&$ 1$&$ 1$&$ -\bar \alpha_-\,\alpha_-\,s\,\frac{1}{4}\,
     \frac{2\,J+1}{J+1}
     $\\$1$&$ 3$&$ \alpha_-\,\bar p^2\,s\,\frac{1}{2}$&$ 1$&$ 4$&$ \bar \alpha_-\,p^2\,s\,\frac{1}{2}$&$ 2$&$ 5$&$ \bar \alpha_-\,\alpha_-\,\bar p^2\,p^2\,\frac{1}{4}\,
     \frac{(J+2) (2\,J+1)}{(J+1) (2\,J+3)}
     $

\\ \\

\phantom{-}$\,k$ & $n\;$ & $\big[a_{+n}^{J+k}\big]_{12}$ \phantom{xxxxxxx} $0\,1\to 0\,1 $\phantom{xxxxxxx}&
\phantom{-}$\,k$ & $n\;$ & $\big[a_{+n}^{J+k}\big]_{12}$ \phantom{xxxxxxx} $0\,1\to 0\,1 $\phantom{xxxxxxx}&
\phantom{-}$\,k$ & $n\;$ & $\big[a_{+n}^{J+k}\big]_{12}$ \phantom{xxxxxxx} $0\,1\to 0\,1 $\phantom{xxxxxxx}\\ \hline

$-1$&$ 3$&$ s^2\,\frac{\sqrt{J}\,\sqrt{J+1}}{2\,J+1}$&$ 0$&$ 5$&$ \bar \alpha_-\,p^2\,s\,\frac{1}{2}\,
     \frac{\sqrt{J}\,(J+2)}{\sqrt{J+1}\,(2\,J+3)}
     $&$ 1$&$ 1$&$ \bar \alpha_-\,p^2\,\frac{1}{2}\,\frac{\sqrt{J}}{\sqrt{J+1}}$\\$1$&$ 3$&$ -\bar p^2\,p^2\,\frac{\sqrt{J}\,\sqrt{J+1}}{2\,J+1}$&$ 2$&$
    5$&$ -\bar \alpha_-\,\bar p^2\,p^4\,\frac{1}{2\,s}\,\frac{\sqrt{J}\,(J+2)}{\sqrt{J+1}\,(2\,J+3)}$&$  $&&$ $

\\ \\

\phantom{-}$\,k$ & $n\;$ & $\big[a_{+n}^{J+k}\big]_{21}$ \phantom{xxxxxxx} $0\,1\to 0\,1 $\phantom{xxxxxxx}&
\phantom{-}$\,k$ & $n\;$ & $\big[a_{+n}^{J+k}\big]_{21}$ \phantom{xxxxxxx} $0\,1\to 0\,1 $\phantom{xxxxxxx}&
\phantom{-}$\,k$ & $n\;$ & $\big[a_{+n}^{J+k}\big]_{21}$ \phantom{xxxxxxx} $0\,1\to 0\,1 $\phantom{xxxxxxx}\\ \hline

$-1$&$ 4$&$ s^2\,\frac{\sqrt{J}\,\sqrt{J+1}}{2\,J+1}$&$ 0$&$ 5$&$ \alpha_-\,\bar p^2\,s\,\frac{1}{2}\,
     \frac{\sqrt{J}\,(J+2)}{\sqrt{J+1}\,(2\,J+3)}
     $&$ 1$&$ 1$&$ \alpha_-\,\bar p^2\,\frac{1}{2}\,\frac{\sqrt{J}}{\sqrt{J+1}}$\\$1$&$ 4$&$ -\bar p^2\,p^2\,\frac{\sqrt{J}\,\sqrt{J+1}}{2\,J+1}$&$ 2$&$
    5$&$ -\alpha_-\,\bar p^4\,p^2\,\frac{1}{2\,s}\,\frac{\sqrt{J}\,(J+2)}{\sqrt{J+1}\,(2\,J+3)}$&$  $&&$ $

\\ \\

\phantom{-}$\,k$ & $n\;$ & $\big[a_{+n}^{J+k}\big]_{22}$ \phantom{xxxxxxx} $0\,1\to 0\,1 $\phantom{xxxxxxx}&
\phantom{-}$\,k$ & $n\;$ & $\big[a_{+n}^{J+k}\big]_{22}$ \phantom{xxxxxxx} $0\,1\to 0\,1 $\phantom{xxxxxxx}&
\phantom{-}$\,k$ & $n\;$ & $\big[a_{+n}^{J+k}\big]_{22}$ \phantom{xxxxxxx} $0\,1\to 0\,1 $\phantom{xxxxxxx}\\ \hline

$-2$&$ 5$&$ s^2\,\frac{J^2-1}{4 J^2-1}$&$ -1$&$ 1$&$ -s\,\frac{J+1}{2\,J+1}$&$ 0$&$ 5$&$ -\bar p^2\,p^2\,
     \frac{2\,J\,(J+1)-3}{4 J\,(J+1)-3}
     $\\$1$&$ 1$&$ -\bar p^2\,p^2\,\frac{1}{s}\,\frac{J}{2\,J+1}$&$ 2$&$ 5$&$ \bar p^4\,p^4\,\frac{1}{s^2}\,\frac{J\,(J+2)}{4 J\,(J+2)+3}$&&&

\\ \\

\phantom{-}$\,k$ & $n\;$ & $\big[a_{+n}^{J+k}\big]_{11}$ \phantom{xxxxxxx} $0\,1\to 1\,1 $\phantom{xxxxxxx}&
\phantom{-}$\,k$ & $n\;$ & $\big[a_{+n}^{J+k}\big]_{11}$ \phantom{xxxxxxx} $0\,1\to 1\,1 $\phantom{xxxxxxx}&
\phantom{-}$\,k$ & $n\;$ & $\big[a_{+n}^{J+k}\big]_{11}$ \phantom{xxxxxxx} $0\,1\to 1\,1 $\phantom{xxxxxxx}\\ \hline

$-1$&$ 10$&$ 2\,\sqrt{s}\,s^2\,\frac{J}{2\,J+1}$&$ -1$&$ 11$&$ -2\,\sqrt{s}\,s^2\,\frac{J}{2\,J+1}$&$ 0$&$ 3$&$ -2\,\sqrt{s}\,s$\\
   $0$&$ 6$&$ \alpha_-\,\sqrt{s}\,s^2\,\frac{\bar \alpha_+\, J+1}{(J+1) (2\,J+3)}$&$ 0$&$ 7$&$ \alpha_-\,\sqrt{s}\,s^2\,\frac{J}{J+1}$&$ 0$&$ 8$&$ \alpha_-\,\sqrt{s}\,s^2\,
     \frac{J}{J+1}
     $\\$0$&$ 11$&$ \alpha_-\,\sqrt{s}\,s^2\,\frac{1}{2}\,\frac{(2\,J+1) (\bar \alpha_-\, J+2)}{(J+1) (2\,J+3)}$&$ 1$&$ 1$&$ -\alpha_-\,\sqrt{s}\,s\,
     \frac{2\,J+1}{J+1}
     $&$ 1$&$ 2$&$ -2\,p^2\,\sqrt{s}\,$\\$1$&$ 4$&$ -\alpha_-\,\bar p^2\,\sqrt{s}\,s\,\frac{J}{J+1}$&$ 1$&$ 5$&$ -\alpha_-\,\bar p^2\,\sqrt{s}\,s\,\frac{J}{J+1}
     $&$ 1$&$ 9$&$ \bar \alpha_-\,\alpha_-\,p^2\,\sqrt{s}\,s\,\frac{1}{2}\,\frac{J}{J+1}$\\$1$&$ 10$&$ -2\,\bar p^2\,p^2\,\sqrt{s}\,\,\frac{J}{2\,J+1}$&$ 1$&$ 11$&$
    2\,\bar p^2\,p^2\,\sqrt{s}\,\,\frac{J}{2\,J+1}$&$ 2$&$ 6$&$ -\alpha_-\,\bar p^2\,p^2\,\sqrt{s}\,\,\frac{1}{2}\,
     \frac{(2\,J+1) (\bar \alpha_-\, J+2)}{(J+1) (2\,J+3)}
     $\\$2$&$ 11$&$ -\alpha_-\,\bar p^2\,p^2\,\sqrt{s}\,\frac{1}{2}\,\frac{(2\,J+1) (\bar \alpha_-\, J+2)}{(J+1) (2\,J+3)}$&&&&&&

\\ \\

\phantom{-}$\,k$ & $n\;$ & $\big[a_{+n}^{J+k}\big]_{12}$ \phantom{xxxxxxx} $0\,1\to 1\,1 $\phantom{xxxxxxx}&
\phantom{-}$\,k$ & $n\;$ & $\big[a_{+n}^{J+k}\big]_{12}$ \phantom{xxxxxxx} $0\,1\to 1\,1 $\phantom{xxxxxxx}&
\phantom{-}$\,k$ & $n\;$ & $\big[a_{+n}^{J+k}\big]_{12}$ \phantom{xxxxxxx} $0\,1\to 1\,1 $\phantom{xxxxxxx}\\ \hline

$-1$&$ 4$&$ 2\,\sqrt{s}\,s^2\,\frac{\sqrt{J}\,\sqrt{J+1}}{2\,J+1}$&$ -1$&$ 5$&$ 2\,\sqrt{s}\,s^2\,\frac{\sqrt{J}\,\sqrt{J+1}}{2\,J+1}$&$ 0$&$ 6$&$
    p^2\,\sqrt{s}\,s\,\frac{\sqrt{J}\,(\bar \alpha_-\, J-3 \,\bar \delta+1)}{\sqrt{J+1}\,(2\,J+3)}$\\$0$&$ 7$&$ -2\,p^2\,\sqrt{s}\,s\,
     \frac{\sqrt{J}}{\sqrt{J+1}}
     $&$ 0$&$ 8$&$ -2\,p^2\,\sqrt{s}\,s\,\frac{\sqrt{J}}{\sqrt{J+1}}$&$ 0$&$ 11$&$ -p^2\,\sqrt{s}\,s\,
     \frac{\sqrt{J}\,(\bar \alpha_-\, J+2)}{\sqrt{J+1}\,(2\,J+3)}
     $\\$1$&$ 1$&$ 2\,p^2\,\sqrt{s}\,\,\frac{\sqrt{J}}{\sqrt{J+1}}$&$ 1$&$ 4$&$ 2\,\bar p^2\,p^2\,\sqrt{s}\,\,
     \frac{\sqrt{J}\,J\,}{\sqrt{J+1}\,(2\,J+1)}
     $&$ 1$&$ 5$&$ 2\,\bar p^2\,p^2\,\sqrt{s}\,\,\frac{\sqrt{J}\,J\,}{\sqrt{J+1}\,(2\,J+1)}$\\$1$&$ 9$&$ -\bar \alpha_-\,p^4\,\sqrt{s}\,\,
     \frac{\sqrt{J}}{\sqrt{J+1}}
     $&$ 2$&$ 6$&$ \bar p^2\,p^4\,\frac{1}{\sqrt{s}}\,\frac{\sqrt{J}\,(\bar \alpha_-\, J+2)}{\sqrt{J+1}\,(2\,J+3)}$&$ 2$&$ 11$&$ \bar p^2\,p^4\,
     \frac{1}{\sqrt{s}}
     \,\frac{\sqrt{J}\,(\bar \alpha_-\, J+2)}{\sqrt{J+1}\,(2\,J+3)}$
\\ \\

\phantom{-}$\,k$ & $n\;$ & $\big[a_{+n}^{J+k}\big]_{21}$ \phantom{xxxxxxx} $0\,1\to 1\,1 $\phantom{xxxxxxx}&
\phantom{-}$\,k$ & $n\;$ & $\big[a_{+n}^{J+k}\big]_{21}$ \phantom{xxxxxxx} $0\,1\to 1\,1 $\phantom{xxxxxxx}&
\phantom{-}$\,k$ & $n\;$ & $\big[a_{+n}^{J+k}\big]_{21}$ \phantom{xxxxxxx} $0\,1\to 1\,1 $\phantom{xxxxxxx}\\ \hline

$-1$&$ 10$&$ \bar \alpha_+\,\sqrt{s}\,s^2\,\frac{1}{2}\,\frac{\sqrt{J}\,\sqrt{J+1}}{2\,J+1}$&$ -1$&$ 11$&$ -\bar \alpha_+\,\sqrt{s}\,s^2\,\frac{1}{2}\,
     \frac{\sqrt{J}\,\sqrt{J+1}}{2\,J+1}
     $&$ 0$&$ 2$&$ -\alpha_-\,\sqrt{s}\,s\,\frac{1}{2}\,\frac{\sqrt{J}}{\sqrt{J+1}}$\\$0$&$ 6$&$ -\bar \alpha_-\,\bar \alpha_+\,\alpha_-\,\sqrt{s}\,s^2\,\frac{1}{4}\,\frac{\sqrt{J}}{\sqrt{J+1}\,(2\,J+3)}
     $&$ 0$&$ 7$&$ \bar \alpha_+\,\alpha_-\,\sqrt{s}\,s^2\,\frac{1}{4}\,\frac{\sqrt{J}}{\sqrt{J+1}}$&$ 0$&$ 8$&$ -\bar \alpha_-\,\alpha_-\,\sqrt{s}\,s^2\,\frac{1}{4}\,
     \frac{\sqrt{J}}{\sqrt{J+1}}
     $\\$0$&$ 11$&$ \bar \alpha_-\,\bar \alpha_+\,\alpha_-\,\sqrt{s}\,s^2\,\frac{1}{8}\,\frac{\sqrt{J}\,J\, (2\,J+1)}{\sqrt{J+1}\,(2\,J+3)}$&$ 1$&$ 3$&$ -\alpha_-\,\bar p^2\,\sqrt{s}\,\,
     \frac{1}{2}
     \,\frac{\sqrt{J}}{\sqrt{J+1}}$&$ 1$&$ 4$&$ -\bar \alpha_+\,\alpha_-\,\bar p^2\,\sqrt{s}\,s\,\frac{1}{4}\,\frac{\sqrt{J}}{\sqrt{J+1}}$\\$
   1$&$ 5$&$ \bar \alpha_-\,\alpha_-\,\bar p^2\,\sqrt{s}\,s\,\frac{1}{4}\,\frac{\sqrt{J}}{\sqrt{J+1}}$&$ 1$&$ 9$&$ \bar \alpha_-\,\bar \alpha_+\,\alpha_-\,p^2\,\sqrt{s}\,s\,\frac{1}{8}\,
     \frac{\sqrt{J}\,J\,}{\sqrt{J+1}}
     $&$ 1$&$ 10$&$ -\bar \alpha_+\,\bar p^2\,p^2\,\sqrt{s}\,\,\frac{1}{2}\,\frac{\sqrt{J}\,\sqrt{J+1}}{2\,J+1}$\\$
   1$&$ 11$&$ \bar \alpha_+\,\bar p^2\,p^2\,\sqrt{s}\,\,\frac{1}{2}\,\frac{\sqrt{J}\,\sqrt{J+1}}{2\,J+1}$&$ 2$&$ 6$&$
    -\bar \alpha_-\,\bar \alpha_+\,\alpha_-\,\bar p^2\,p^2\,\sqrt{s}\,\,\frac{1}{8}\,\frac{\sqrt{J}\, (2\,J+1)}{\sqrt{J+1}\,(2\,J+3)}$&$ 2$&$ 11$&$
    -\bar \alpha_-\,\bar \alpha_+\,\alpha_-\,\bar p^2\,p^2\,\sqrt{s}\,\,\frac{1}{8}\,\frac{\sqrt{J}\, (2\,J+1)}{\sqrt{J+1}\,(2\,J+3)}$

\\ \\

\phantom{-}$\,k$ & $n\;$ & $\big[a_{+n}^{J+k}\big]_{22}$ \phantom{xxxxxxx} $0\,1\to 1\,1 $\phantom{xxxxxxx}&
\phantom{-}$\,k$ & $n\;$ & $\big[a_{+n}^{J+k}\big]_{22}$ \phantom{xxxxxxx} $0\,1\to 1\,1 $\phantom{xxxxxxx}&
\phantom{-}$\,k$ & $n\;$ & $\big[a_{+n}^{J+k}\big]_{22}$ \phantom{xxxxxxx} $0\,1\to 1\,1 $\phantom{xxxxxxx}\\ \hline

$-1$&$ 3$&$ \sqrt{s}\,s\,\frac{J+1}{2\,J+1}$&$ -1$&$ 4$&$ \bar \alpha_+\,\sqrt{s}\,s^2\,\frac{1}{2}\,\frac{J+1}{2\,J+1}$&$ -1$&$ 5$&$ -\bar \alpha_-\,\sqrt{s}\,s^2\,
     \frac{1}{2}
     \,\frac{J+1}{2\,J+1}$\\$0$&$ 2$&$ p^2\,\sqrt{s}\,$&$ 0$&$ 6$&$ \bar \alpha_-\,\bar \alpha_+\,p^2\,\sqrt{s}\,s\,\frac{1}{4}\,\frac{J+3}{2\,J+3}$&$ 0$&$ 7$&$
    -\bar \alpha_+\,p^2\,\sqrt{s}\,s\,\frac{1}{2}$\\$0$&$ 8$&$ \bar \alpha_-\,p^2\,\sqrt{s}\,s\,\frac{1}{2}$&$ 0$&$ 11$&$ -\bar \alpha_-\,\bar \alpha_+\,p^2\,\sqrt{s}\,s\,
     \frac{1}{4}
     \,\frac{J}{2\,J+3}$&$ 1$&$ 3$&$ \bar p^2\,p^2\,\frac{1}{\sqrt{s}}\,\frac{J}{2\,J+1}$\\$1$&$ 4$&$ \bar \alpha_+\,\bar p^2\,p^2\,\sqrt{s}\,\,
     \frac{1}{2}
     \,\frac{J}{2\,J+1}$&$ 1$&$ 5$&$ -\bar \alpha_-\,\bar p^2\,p^2\,\sqrt{s}\,\,\frac{1}{2}\,\frac{J}{2\,J+1}$&$ 1$&$ 9$&$ -\bar \alpha_-\,\bar \alpha_+\,p^4\,\sqrt{s}\,\,
     \frac{1}{4}
     $\\$2$&$ 6$&$ \bar \alpha_-\,\bar \alpha_+\,\bar p^2\,p^4\,\frac{1}{4 \sqrt{s}}\,\frac{J^2}{2\,J+3}$&$ 2$&$ 11$&$ \bar \alpha_-\,\bar \alpha_+\,\bar p^2\,p^4\,
     \frac{1}{4 \sqrt{s}}
     \,\frac{J}{2\,J+3}$&&&

\end{tabular}
\label{tab-3}
\caption{Non-vanishing coefficients $a_{\pm n}^{J+k}$  in the expansion (\ref{def-decomposition-TJ}) for the reactions $0\,1 \to 0\,1$,
$0\,1\to 1\,1$ and $0\,0\to 1\,1$. }
\end{table}

\begin{table}[t]
\rescale
\setlength{\tabcolsep}{0.5mm}
\renewcommand{\arraystretch}{1.4}
\begin{tabular}{rc|c|rl|c|rl|c  }

\phantom{-}$\,k$ & $n\;$ & $\big[a_{+n}^{J+k}\big]_{31}$ \phantom{xxxxxxx} $0\,1\to 1\,1 $\phantom{xxxxxxx}&
\phantom{-}$\,k$ & $n\;$ & $\big[a_{+n}^{J+k}\big]_{31}$ \phantom{xxxxxxx} $0\,1\to 1\,1 $\phantom{xxxxxxx}&
\phantom{-}$\,k$ & $n\;$ & $\big[a_{+n}^{J+k}\big]_{31}$ \phantom{xxxxxxx} $0\,1\to 1\,1 $\phantom{xxxxxxx}\\ \hline

$-1$&$ 2$&$ -\sqrt{s}\,s\,\frac{\sqrt{J}\,\sqrt{J+1}}{2\,J+1}$&$ -1$&$ 9$&$ -\alpha_-\,\sqrt{s}\,s^2\,\frac{1}{4}\,
     \frac{\sqrt{J}\,\sqrt{J+1}}{2\,J+1}
     $&$ -1$&$ 10$&$ -\bar p^2\,\sqrt{s}\,s\,\frac{\sqrt{J}\,\sqrt{J+1}}{2\,J+1}$\\$-1$&$ 11$&$ \bar p^2\,\sqrt{s}\,s\,\frac{\sqrt{J}\,\sqrt{J+1}}{2\,J+1}$&$
    0$&$ 6$&$ -\alpha_-\,\bar p^2\,\sqrt{s}\,s\,\frac{1}{4}\,-\frac{\sqrt{J}\,(-2\,\bar \delta+J+1)}{\sqrt{J+1}\,(2\,J+3)}$&$ 0$&$ 7$&$ -\alpha_-\,\bar p^2\,\sqrt{s}\,s
    \,\frac{1}{2}\,\frac{\sqrt{J}}{\sqrt{J+1}}$\\$0$&$ 8$&$ -\alpha_-\,\bar p^2\,\sqrt{s}\,s\,\frac{1}{2}\,\frac{\sqrt{J}}{\sqrt{J+1}}$&$ 0$&$ 11$&$
    \alpha_-\,\bar p^2\,\sqrt{s}\,s\,\frac{1}{4}\,\frac{\sqrt{J}\,(2\,\bar \delta \, J+\bar \delta-J-2)}{\sqrt{J+1}\,(2\,J+3)}$&$ 1$&$ 1$&$ \alpha_-\,\bar p^2\,\sqrt{s}\,\,
     \frac{1}{2}
     \,\frac{\sqrt{J}}{\sqrt{J+1}}$\\$1$&$ 2$&$ \bar p^2\,p^2\,\frac{1}{\sqrt{s}}\,\frac{\sqrt{J}\,\sqrt{J+1}}{2\,J+1}$&$ 1$&$ 4$&$
    \alpha_-\,\bar p^4\,\sqrt{s}\,\,\frac{1}{2}\,\frac{\sqrt{J}}{\sqrt{J+1}}$&$ 1$&$ 5$&$ \alpha_-\,\bar p^4\,\sqrt{s}\,\,\frac{1}{2}\,\frac{\sqrt{J}}{\sqrt{J+1}}
     $\\$1$&$ 9$&$ \alpha_-\,\bar p^2\,p^2\,\sqrt{s}\,\,\frac{1}{4}\,\frac{\sqrt{J}\,(2\,\bar \delta \, J+\bar \delta-J)}{\sqrt{J+1}\,(2\,J+1)}$&$ 1$&$ 10$&$
    \bar p^4\,p^2\,\frac{1}{\sqrt{s}}\,\frac{\sqrt{J}\,\sqrt{J+1}}{2\,J+1}$&$ 1$&$ 11$&$ -\bar p^4\,p^2\,\frac{1}{\sqrt{s}}\,
     \frac{\sqrt{J}\,\sqrt{J+1}}{2\,J+1}
     $\\$2$&$ 6$&$ -\alpha_-\,\bar p^4\,p^2\,\frac{1}{4\,\sqrt{s}}\,\frac{\sqrt{J}\,(2\,\bar \delta \, J+\bar \delta-J-2)}{\sqrt{J+1}\,(2\,J+3)}$&$ 2$&$ 11$&$
    -\alpha_-\,\bar p^4\,p^2\,\frac{1}{4\,\sqrt{s}}\,\frac{\sqrt{J}\,(2\,\bar \delta \, J+\bar \delta-J-2)}{\sqrt{J+1}\,(2\,J+3)}$&$  $&&$ $

\\ \\

\phantom{-}$\,k$ & $n\;$ & $\big[a_{+n}^{J+k}\big]_{32}$ \phantom{xxxxxxx} $0\,1\to 1\,1 $\phantom{xxxxxxx}&
\phantom{-}$\,k$ & $n\;$ & $\big[a_{+n}^{J+k}\big]_{32}$ \phantom{xxxxxxx} $0\,1\to 1\,1 $\phantom{xxxxxxx}&
\phantom{-}$\,k$ & $n\;$ & $\big[a_{+n}^{J+k}\big]_{32}$ \phantom{xxxxxxx} $0\,1\to 1\,1 $\phantom{xxxxxxx}\\ \hline

$-2$&$ 6$&$ -\sqrt{s}\,s^2\,\frac{1}{2}\,\frac{J^2-1}{4\,J^2-1}$&$ -2$&$ 11$&$ -\sqrt{s}\,s^2\,\frac{1}{2}\,\frac{J^2-1}{4\,J^2-1}$&$ -1$&$ 1$&$
    -\sqrt{s}\,s\,\frac{J+1}{2\,J+1}$\\$-1$&$ 4$&$ -\bar p^2\,\sqrt{s}\,s\,\frac{J+1}{2\,J+1}$&$ -1$&$ 5$&$ -\bar p^2\,\sqrt{s}\,s\,\frac{J+1}{2\,J+1}$&$
    -1$&$ 9$&$ p^2\,\sqrt{s}\,s\,\frac{1}{2}\,\frac{J+1}{2\,J+1}$\\$0$&$ 6$&$ \bar p^2\,p^2\,\sqrt{s}\,\,\frac{1}{2}\,
     \frac{\bar \delta (J+3) (2\,J-1)-2\,J\,(J+1)}{4\,J\,(J+1)-3}
     $&$ 0$&$ 7$&$ \bar p^2\,p^2\,\sqrt{s}\,$&$ 0$&$ 8$&$ \bar p^2\,p^2\,\sqrt{s}\,$\\$0$&$ 11$&$ -\bar p^2\,p^2\,\sqrt{s}\,\,\frac{1}{2}\,
     \frac{2\,(\bar \delta-1) J^2-(\bar \delta+2) J+3}{4\,J\,(J+1)-3}
     $&$ 1$&$ 1$&$ -\bar p^2\,p^2\,\frac{1}{\sqrt{s}}\,\frac{J}{2\,J+1}$&$ 1$&$ 4$&$ -\bar p^4\,p^2\,\frac{1}{\sqrt{s}}\,\frac{J}{2\,J+1}$\\$
   1$&$ 5$&$ -\bar p^4\,p^2\,\frac{1}{\sqrt{s}}\,\frac{J}{2\,J+1}$&$ 1$&$ 9$&$ -\bar p^2\,p^4\,\frac{1}{2\,\sqrt{s}}\,
     \frac{2\,\bar \delta \, J+\bar \delta-J}{2\,J+1}
     $&$ 2$&$ 6$&$ \bar p^4\,p^4\,\frac{1}{2\,\sqrt{s}\,s}\,\frac{J\,(2\,\bar \delta \, J+\bar \delta-J-2)}{4\,J\,(J+2)+3}$\\$2$&$ 11$&$ \bar p^4\,p^4\,
     \frac{1}{2\,\sqrt{s}\,s}
     \,\frac{J\,(2\,\bar \delta \, J+\bar \delta-J-2)}{4\,J\,(J+2)+3}$&$  $&&&$ $&$ $&$ $

\\ \\

\phantom{-}$\,k$ & $n\;$ & $\big[a_{+n}^{J+k}\big]_{41}$ \phantom{xxxxxxx} $0\,1\to 1\,1 $\phantom{xxxxxxx}&
\phantom{-}$\,k$ & $n\;$ & $\big[a_{+n}^{J+k}\big]_{41}$ \phantom{xxxxxxx} $0\,1\to 1\,1 $\phantom{xxxxxxx}&
\phantom{-}$\,k$ & $n\;$ & $\big[a_{+n}^{J+k}\big]_{41}$ \phantom{xxxxxxx} $0\,1\to 1\,1 $\phantom{xxxxxxx}\\ \hline

$-1$&$ 9$&$ \alpha_-\,\sqrt{s}\,s^2\,\frac{1}{4}\,\frac{\sqrt{J}\,\sqrt{J\,(J+1)-2}}{\sqrt{J+1}\,(2\,J+1)}$&$ 0$&$ 6$&$ -\alpha_-\,\bar p^2\,\sqrt{s}\,s\,\frac{1}{4}\,\frac{\sqrt{J}\,\sqrt{J\,(J+1)-2}}{\sqrt{J+1}\,(2\,J+3)}$&$ 0$&$ 11$&$ -\alpha_-\,\bar p^2\,\sqrt{s}\,s\,\frac{1}{4}\,
     \frac{\sqrt{J}\,\sqrt{J\,(J+1)-2}}{\sqrt{J+1}\,(2\,J+3)}
     $\\$1$&$ 9$&$ -\alpha_-\,\bar p^2\,p^2\,\sqrt{s}\,\,\frac{1}{4}\,\frac{\sqrt{J}\,\sqrt{J\,(J+1)-2}}{\sqrt{J+1}\,(2\,J+1)}$&$ 2$&$ 6$&$
    \alpha_-\,\bar p^4\,p^2\,\frac{1}{4\,\sqrt{s}}\,\frac{\sqrt{J}\,\sqrt{J\,(J+1)-2}}{\sqrt{J+1}\,(2\,J+3)}$&$ 2$&$ 11$&$ \alpha_-\,\bar p^4\,p^2\,
     \frac{1}{4\,\sqrt{s}}
     \,\frac{\sqrt{J}\,\sqrt{J\,(J+1)-2}}{\sqrt{J+1}\,(2\,J+3)}$

\\ \\

\phantom{-}$\,k$ & $n\;$ & $\big[a_{+n}^{J+k}\big]_{42}$ \phantom{xxxxxxx} $0\,1\to 1\,1 $\phantom{xxxxxxx}&
\phantom{-}$\,k$ & $n\;$ & $\big[a_{+n}^{J+k}\big]_{42}$ \phantom{xxxxxxx} $0\,1\to 1\,1 $\phantom{xxxxxxx}&
\phantom{-}$\,k$ & $n\;$ & $\big[a_{+n}^{J+k}\big]_{42}$ \phantom{xxxxxxx} $0\,1\to 1\,1 $\phantom{xxxxxxx}\\ \hline

$-2$&$ 6$&$ \sqrt{s}\,s^2\,\frac{1}{2}\,\frac{(J+1) \sqrt{J\,(J+1)-2}}{(2\,J-1) (2\,J+1)}$&$ -2$&$ 11$&$ \sqrt{s}\,s^2\,\frac{1}{2}\,
     \frac{(J+1) \sqrt{J\,(J+1)-2}}{(2\,J-1) (2\,J+1)}
     $&$ -1$&$ 9$&$ -p^2\,\sqrt{s}\,s\,\frac{1}{2}\,\frac{\sqrt{J\,(J+1)-2}}{2\,J+1}$\\$0$&$ 6$&$ -3\,\bar p^2\,p^2\,\sqrt{s}\,\,\frac{1}{2}\,
     \frac{\sqrt{J\,(J+1)-2}}{(2\,J-1) (2\,J+3)}
     $&$ 0$&$ 11$&$ -3\,\bar p^2\,p^2\,\sqrt{s}\,\,\frac{1}{2}\,\frac{\sqrt{J\,(J+1)-2}}{(2\,J-1) (2\,J+3)}$&$ 1$&$ 9$&$ \bar p^2\,p^4\,
     \frac{1}{2 \sqrt{s}}
     \,\frac{\sqrt{J\,(J+1)-2}}{2\,J+1}$\\$2$&$ 6$&$ -\bar p^4\,p^4\,\frac{1}{2\,\sqrt{s}\,s}\,\frac{J \sqrt{J\,(J+1)-2}}{(2\,J+1) (2\,J+3)}$&$ 2$&$
    11$&$ -\bar p^4\,p^4\,\frac{1}{2\,\sqrt{s}\,s}\,\frac{J \sqrt{J\,(J+1)-2}}{(2\,J+1) (2\,J+3)}$&&&

\\ \\

\phantom{-}$\,k$ & $n\;$ & $\big[a_{-n}^{J+k}\big]_{11}$ \phantom{xxxxxxx} $0\,0\to 1\,1 $\phantom{xxxxxxx}&
\phantom{-}$\,k$ & $n\;$ & $\big[a_{-n}^{J+k}\big]_{11}$ \phantom{xxxxxxx} $0\,0\to 1\,1 $\phantom{xxxxxxx}&
\phantom{-}$\,k$ & $n\;$ & $\big[a_{-n}^{J+k}\big]_{11}$ \phantom{xxxxxxx} $0\,0\to 1\,1 $\phantom{xxxxxxx}\\ \hline

$0$&$ 2$&$ -2\,s^2\,\frac{\sqrt{J+1}}{\sqrt{J}}$&$ 1$&$ 3$&$ -\bar \alpha_-\,p^2\,s\,\frac{\sqrt{J+1}}{\sqrt{J}}$&$ 1$&$ 4$&$ \bar \alpha_+\,p^2\,s
    \,\frac{\sqrt{J+1}}{\sqrt{J}}$\\$2$&$ 5$&$ \bar \alpha_-\,\bar \alpha_+\,p^4\,\frac{1}{2}\,\frac{\sqrt{J+1}}{\sqrt{J}}$&$  $&&&&&

\\ \\

\phantom{-}$\,k$ & $n\;$ & $\big[a_{-n}^{J+k}\big]_{21}$ \phantom{xxxxxxx} $0\,0\to 1\,1 $\phantom{xxxxxxx}&
\phantom{-}$\,k$ & $n\;$ & $\big[a_{-n}^{J+k}\big]_{21}$ \phantom{xxxxxxx} $0\,0\to 1\,1 $\phantom{xxxxxxx}&
\phantom{-}$\,k$ & $n\;$ & $\big[a_{-n}^{J+k}\big]_{21}$ \phantom{xxxxxxx} $0\,0\to 1\,1 $\phantom{xxxxxxx}\\ \hline

$0$&$ 1$&$ -2\,s\,\frac{\sqrt{J+1}}{\sqrt{J}}$&$ 0$&$ 5$&$ 2\,p^2\,s\,\frac{\sqrt{J+1}}{\sqrt{J}\,(2\,J+3)}$&$ 2$&$ 5$&$
    -2\,\bar p^2\,p^4\,\frac{1}{s}\,\frac{\sqrt{J+1}}{\sqrt{J}\,(2\,J+3)}$

\\ \\
\phantom{-}$\,k$ & $n\;$ & $\big[a_{-n}^{J+k}\big]_{31}$ \phantom{xxxxxxx} $0\,0\to 1\,1 $\phantom{xxxxxxx}&
\phantom{-}$\,k$ & $n\;$ & $\big[a_{-n}^{J+k}\big]_{31}$ \phantom{xxxxxxx} $0\,0\to 1\,1 $\phantom{xxxxxxx}&
\phantom{-}$\,k$ & $n\;$ & $\big[a_{-n}^{J+k}\big]_{31}$ \phantom{xxxxxxx} $0\,0\to 1\,1 $\phantom{xxxxxxx}\\ \hline

$-1$&$ 3$&$ -s^2\,\frac{\sqrt{J}\,\sqrt{J+1}\,(2\,J+3)}{4\,J\,(J+2)+3}$&$ -1$&$ 4$&$ -s^2\,
     \frac{\sqrt{J}\,\sqrt{J+1}\,(2\,J+3)}{4\,J\,(J+2)+3}
     $&$ 0$&$ 5$&$ \bar M_-\,\bar M_+\,p^2\,\frac{\sqrt{J}\,\sqrt{J+1}\,(2\,J+1)}{4\,J\,(J+2)+3}$\\$1$&$ 3$&$ \bar p^2\,p^2\,
     \frac{\sqrt{J}\,\sqrt{J+1}\,(2\,J+3)}{4\,J\,(J+2)+3}
     $&$ 1$&$ 4$&$ \bar p^2\,p^2\,\frac{\sqrt{J}\,\sqrt{J+1}\,(2\,J+3)}{4\,J\,(J+2)+3}$&$ 2$&$ 5$&$ -\bar M_-\,\bar M_+\,\bar p^2\,p^4\,\frac{1}{s^2}\,
     \frac{\sqrt{J}\,\sqrt{J+1}\,(2\,J+1)}{4\,J\,(J+2)+3}
     $

\\ \\

\phantom{-}$\,k$ & $n\;$ & $\big[a_{-n}^{J+k}\big]_{41}$ \phantom{xxxxxxx} $0\,0\to 1\,1 $\phantom{xxxxxxx}&
\phantom{-}$\,k$ & $n\;$ & $\big[a_{-n}^{J+k}\big]_{41}$ \phantom{xxxxxxx} $0\,0\to 1\,1 $\phantom{xxxxxxx}&
\phantom{-}$\,k$ & $n\;$ & $\big[a_{-n}^{J+k}\big]_{41}$ \phantom{xxxxxxx} $0\,0\to 1\,1 $\phantom{xxxxxxx}\\ \hline

$-1$&$ 4$&$ 2\,s^2\,\frac{\sqrt{J}\,\sqrt{J+1}\,(2\,J+3)}{4\,J\,(J+2)+3}$&$ 0$&$ 5$&$ \bar \alpha_-\,p^2\,s\,
     \frac{\sqrt{J}\,\sqrt{J+1}\,(2\,J+1)}{4\,J\,(J+2)+3}
     $&$ 1$&$ 4$&$ -2\,\bar p^2\,p^2\,\frac{\sqrt{J}\,\sqrt{J+1}\,(2\,J+3)}{4\,J\,(J+2)+3}$\\$2$&$ 5$&$ -\bar \alpha_-\,\bar p^2\,p^4\,\frac{1}{s}\,
     \frac{\sqrt{J}\,\sqrt{J+1}\,(2\,J+1)}{4\,J\,(J+2)+3}
     $&&&&&&

\\ \\

\phantom{-}$\,k$ & $n\;$ & $\big[a_{-n}^{J+k}\big]_{51}$ \phantom{xxxxxxx} $0\,0\to 1\,1 $\phantom{xxxxxxx}&
\phantom{-}$\,k$ & $n\;$ & $\big[a_{-n}^{J+k}\big]_{51}$ \phantom{xxxxxxx} $0\,0\to 1\,1 $\phantom{xxxxxxx}&
\phantom{-}$\,k$ & $n\;$ & $\big[a_{-n}^{J+k}\big]_{51}$ \phantom{xxxxxxx} $0\,0\to 1\,1 $\phantom{xxxxxxx}\\ \hline

$-2$&$ 5$&$ -s^2\,\frac{\sqrt{J}\,\sqrt{J+1}\,(2\,J+3)}{4\,J\,(J+1)-3}$&$ 0$&$ 5$&$ 2\,\bar p^2\,p^2\,
     \frac{\sqrt{J}\,\sqrt{J+1}\,(2\,J+1)}{4\,J\,(J+1)-3}
     $&$ 2$&$ 5$&$ -\bar p^4\,p^4\,\frac{1}{s^2}\,\frac{\sqrt{J}\,\sqrt{J+1}\,(2\,J-1)}{4\,J\,(J+1)-3}$

\end{tabular}
\label{tab-3}
\caption{Continuation of Tab. \ref{tab-3}. We use the notation of (\ref{def-alpha-delta}).}
\end{table}
\end{widetext}

\clearpage

\section{Conclusions}

We have constructed partial-wave amplitudes for two-body reactions involving $J^P=0^-$ and $J^P=1^-$
particles which are free from kinematical constraints and frame independent. Those covariant partial-wave amplitudes are well suited to be
used in partial-wave dispersion relations and data analysis. Explicit transformations from the conventional helicity
states to the covariant states were derived and presented in this work. In an initial step we identified complete sets
of invariant functions that parameterize the scattering amplitudes of the various processes and are kinematically unconstrained.
The latter are expected to satisfy Mandelstam's dispersion integral representation. Explicit expressions for the covariant
partial-wave scattering amplitudes in terms of integrals over the invariant amplitudes were derived and partially presented in this work.
A convenient projection algebra was constructed that streamlines the derivation of the invariant amplitudes  by means of computer
algebra codes significantly.

The present paper thus offers an efficient starting point for analyzing
boson-boson scattering in a covariant coupled-channel approach that takes into account the constraints
set by micro-causality and coupled-channel unitarity.

\vskip0.5cm
{\bfseries{Acknowledgments}}
\vskip1.cm
M.F.M.L. thanks the Centro de F\'{i}sica Computacional at the University of Coimbra for kind hospitalilty.


\appendix

\section{}

We provide projection tensors for the $1 \,1 \to 1\,1$ reaction. They streamline the computation of the 41 invariant amplitudes
$F_n(s,t)$ introduced in (\ref{def-11to11}) by means of algebraic computer codes. We find
\allowdisplaybreaks[2]
\begin{eqnarray}
&& P^{(n)}_{\alpha  \beta ,\tau \sigma } \, g^{\alpha \bar \mu}\,g^{\beta \bar \nu}\,g^{\tau \mu}\,g^{\sigma \nu}\,T^{(m)}_{\bar \mu  \bar \nu ,\mu \nu} = \delta_{nm} \,,
\nonumber\\
&&  P^{(n)}_{\bar \mu  \bar \nu, \mu \nu } \,\bar p_1^{\bar \mu } = 0 \, ,\qquad
P^{(n)}_{\bar \mu  \bar \nu, \mu \nu } \,\bar p_2^{\bar \nu } = 0 \, ,\qquad
\nonumber\\
&& P^{(n)}_{\bar \mu  \bar \nu, \mu \nu } \,p_1^{\mu } =0  \,, \qquad P^{(n)}_{\bar \mu  \bar \nu, \mu \nu } \,p_2^{ \nu } =0 \,.
\nonumber\\
&& P^{(n)}_{\bar \mu  \bar \nu, \mu \nu} = \sum_{k=1}^{41}\,c_k^{(n)} \,Q^{(k)}_{\bar \mu  \bar \nu, \nu \mu } \,,
\label{res-Q:11to11}\\
&& Q_n^{\bar \mu \bar \nu, \mu \nu} =
\bar L^{\bar \mu \bar \nu}_{\lceil  n/5 \rceil}\,L^{\mu \nu}_{n+5 -5\,\lceil  n/5 \rceil}  \quad  {\rm for} \quad  n\leq 25 \,,
\nonumber\\
&& \!\!\begin{array}{ll}
Q_{26}^{\bar \mu \bar \nu, \mu \nu} = \hat v^{\bar \mu}\,\hat v^\mu\, \wRbot^{\bar \nu}\,\wRbarbot^{\nu} \,, \;\;\;\;&
Q_{27}^{\bar \mu \bar \nu, \mu \nu} = \hat v^{\bar \mu}\,\hat v^\mu\,\wRbot^{\bar \nu }\,\,\rbarbot^{\nu} \,, \\
Q_{28}^{\bar \mu \bar \nu, \mu \nu} = \hat v^{\bar \mu}\,\hat v^\mu\, \,\rbot^{\bar \nu}\,\wRbarbot^{\nu } \,, &
Q_{29}^{\bar \mu \bar \nu, \mu \nu} = \hat v^{\bar \mu}\,\hat v^\mu\,\, \rbot^{\bar \nu}\,\,\rbarbot^{\nu } \,, \\

Q_{30}^{\bar \mu \bar \nu, \mu \nu} = \hat v^{\bar \nu}\,\hat v^\nu\, \wLbot^{\bar \mu}\,\wLbarbot^{\mu} \,, &
Q_{31}^{\bar \mu \bar \nu, \mu \nu} = \hat v^{\bar \nu}\,\hat v^\nu\,\wLbot^{\bar \mu }\,\,\rbarbot^{\mu}\,, \\
Q_{32}^{\bar \mu \bar \nu, \mu \nu} = \hat v^{\bar \nu}\,\hat v^\nu\, \,\rbot^{\bar \mu}\,\wLbarbot^{\mu } \,, &
Q_{33}^{\bar \mu \bar \nu, \mu \nu} = \hat v^{\bar \nu}\,\hat v^\nu\,\, \rbot^{\bar \mu}\,\,\rbarbot^{\mu }\,,
\end{array}
\nonumber\\
&& \!\!\begin{array}{ll}
Q_{34}^{\bar \mu \bar \nu, \mu \nu} = \hat v^{\bar \nu}\,\hat v^\mu\, \wLbot^{\bar \mu}\,\wRbarbot^{\nu}\,, &
Q_{35}^{\bar \mu \bar \nu, \mu \nu} = \hat v^{\bar \nu}\,\hat v^\mu\,\wLbot^{\bar \mu }\,\,\rbarbot^{\nu} \,, \\
Q_{36}^{\bar \mu \bar \nu, \mu \nu} = \hat v^{\bar \nu}\,\hat v^\mu\, \,\rbot^{\bar \mu}\,\wRbarbot^{\nu } \,, &
Q_{37}^{\bar \mu \bar \nu, \mu \nu} = \hat v^{\bar \nu}\,\hat v^\mu\,\, \rbot^{\bar \mu}\,\,\rbarbot^{\nu }\,, \\

Q_{38}^{\bar \mu \bar \nu, \mu \nu} = \hat v^{\bar \mu}\,\hat v^\nu\, \wRbot^{\bar \nu}\,\wLbarbot^{\mu}\,, &
Q_{39}^{\bar \mu \bar \nu, \mu \nu} = \hat v^{\bar \mu}\,\hat v^\nu\,\wRbot^{\bar \nu }\,\,\rbarbot^{\mu} \,, \\
Q_{40}^{\bar \mu \bar \nu, \mu \nu} = \hat v^{\bar \mu}\,\hat v^\nu\, \,\rbot^{\bar \nu}\,\wLbarbot^{\mu }\,, &
Q_{41}^{\bar \mu \bar \nu, \mu \nu} = \hat v^{\bar \mu}\,\hat v^\nu\,\, \rbot^{\bar \nu}\,\,\rbarbot^{\mu }\,,
\end{array}
\nonumber
\end{eqnarray}
with $\hat v_\mu = v_\mu/v^2$. The ceiling function,
\begin{eqnarray}
\lceil x \rceil -1\leq x \leq \lceil x \rceil \,,
\label{def-ceiling}
\end{eqnarray}
maps a real number $x$ onto an integer number $\lceil x \rceil $  as defined by (\ref{def-ceiling}). The tensors $L_{n}^{ \mu  \nu} $ and
$\bar L_{n}^{\bar \mu  \bar \nu} $ used in (\ref{res-Q:11to11}) are
\begin{eqnarray}
&&L_{1}^{ \mu  \nu} = v^{\mu}\,v^{\nu}/v^2\,,
\nonumber\\
&& L_{2}^{\mu  \nu} = \wLbarbot^{\mu}\,\wRbarbot^{\nu}
\nonumber\\
&& \qquad -\,\big[ (\wLbarbot\cdot\wRbarbot)- (\wLbarbot\cdot w)\,(w\cdot\wRbarbot)/w^2\big] \,L_{1}^{ \mu  \nu} \,,\qquad
\nonumber\\
&& L_{3}^{\mu  \nu} = \wLbarbot^{\mu }\,\,\rbarbot^{\nu} -(\wLbarbot\cdot\rbarbot)\,L_{1}^{ \mu  \nu} \,,
\nonumber\\
&& L_{4}^{\mu  \nu} = \,\rbarbot^{\mu}\,\wRbarbot^{\nu }-(\rbarbot\cdot\wRbarbot)\,L_{1}^{ \mu  \nu} \,,
\nonumber\\
&& L_{5}^{\mu  \nu} =\, \rbarbot^{\mu}\,\,\rbarbot^{\nu }-(\rbarbot \cdot\rbarbot)\,L_{1}^{ \mu  \nu} \,,
\nonumber\\
&&\bar L_{1}^{\bar \mu  \bar \nu} = v^{\bar \mu}\,v^{\bar \nu}/v^2\,,
\nonumber\\
&& \bar L_{2}^{\bar \mu  \bar \nu} = \wLbot^{\bar \mu}\,\wRbot^{\bar \nu}
 \nonumber\\
&& \qquad -\,\big[(\wLbot \cdot \wRbot)-(\wLbot \cdot w)\,(w \cdot \wRbot)/w^2 \big]\,\bar L_{1}^{\bar \mu  \bar \nu} \,,\qquad
\nonumber\\
&& \bar L_{3}^{\bar \mu  \bar \nu} = \wLbot^{\bar \mu }\,\,\rbot^{\bar \nu}-(\wLbot\cdot\rbot)\,\bar L_{1}^{\bar \mu  \bar \nu} \,,
\nonumber\\
&&\bar L_{4}^{\bar \mu  \bar \nu} = \,\rbot^{\bar \mu}\,\wRbot^{\bar \nu }-(\rbot\cdot\wRbot)\,\bar L_{1}^{\bar \mu  \bar \nu}\,,
\nonumber\\
&& \bar L_{5}^{\bar \mu  \bar \nu} =\, \rbot^{\bar \mu}\,\,\rbot^{\bar \nu }-( \rbot\cdot\rbot)\,\bar L_{1}^{\bar \mu  \bar \nu}\,.
\end{eqnarray}

The derivation of the coefficient matrix $c_k^{(n)}$ appears prohibitively cumbersome at first. A $41\times 41 $ matrix needs to be inverted.
The  merit of the algebra developed in this work are concise and manageable expressions for the coefficients $c_k^{(n)}$. They are presented in terms of the building blocks
\begin{eqnarray}
&& \delta = \frac{m_1^2-m_2^2}{s}\,, \qquad \quad \alpha_\pm =1 \pm \delta \,,
\label{def-kinematics-ckn}\\
&& \bar \delta = \frac{\bar m_1^2-\bar m_2^2}{s}\,, \qquad \quad \bar \alpha_\pm = 1 \pm \bar \delta \,,
\nonumber\\
&& \beta_\pm = \pm \,3\,\bar \delta \,\delta + \bar \delta -\delta\,,\qquad
 \gamma_\pm = \pm \,3\,\bar \delta \,\delta + \bar \delta +\delta\,.\qquad \nonumber
\end{eqnarray}
In Tab. V we specify $k$ and $n$ for all coefficients with $c_k^{(n)}=1$. In Tab. VI we detail all remaining non-vanishing elements in the expansion (\ref{res-Q:11to11}).

\begin{table}[b]
\setlength{\tabcolsep}{2.0mm}
\setlength{\arraycolsep}{5mm}
\renewcommand{\arraystretch}{1.1}
\begin{tabular}{c|cccccccccc  }
\hline
$k$ & 1&  2&  3&  4&  5&  6&  7&  8& 9& 10 \\
$n$ & 3& 20& 22& 21& 23& 24& 28&37& 35& 30 \\
\hline

$k$ &11& 12& 15& 16& 17& 20& 21& 22& 23& 24 \\
$n$ &26& 32& 34& 25& 31& 33& 27& 29& 38& 36        \\ \hline
\end{tabular}
\label{tab-0}
\caption{The non-vanishing coefficients  $c_k^{(n)}=1$.}
\end{table}

\begin{widetext}

\begin{table}[t]
\rescale
\renewcommand{\arraystretch}{1.04}
\begin{tabular}{|ll|c |ll|c |ll|c  }
$k$ & $n$ & $c_{k}^{(n)}$ \phantom{xxxxxxxxxxxxxxxxxxxxxxxx}&
$k$ & $n$ & $c_{k}^{(n)}$ \phantom{xxxxxxxxxxxxxxxxxxxxxxxx}&
$k$ & $n$ & $c_{k}^{(n)}$ \phantom{xxxxxxxxxxxxxxxxxxxxxxxx}\\ \hline

$13$&$ 1$&$ -v^2/s$&$ 13$&$ 2$&$ v^2/s$&$ 13$&$ 4$&$
   {\textstyle{1\over 4}}\, \bar \alpha_+\,\alpha_+\,(\bar r \cdot r)$\\
  $13$&$ 5$&$ {\textstyle{1\over 2}}\, \alpha_+\,\bar r^2$&$ 13$&$ 6$&$
   {\textstyle{1\over 2}}\, \bar \alpha_+\,r^2$&$ 13$&$ 7$&$ (\bar r \cdot r)$\\
  $13$&$ 8$&$ {\textstyle{1\over 4}}\, \bar \alpha_-\,\alpha_-\,(\bar r \cdot r)$&$ 13$&$ 9$&$
  - {\textstyle{1\over 2}}\, \alpha_-\,\bar r^2$&$ 13$&$ 10$&$
  - {\textstyle{1\over 2}}\, \bar \alpha_-\,r^2$\\ $13$&$ 11$&$ (\bar r \cdot r)$&$ 13$&$ 12$&$
   {\textstyle{1\over 4}}\, \alpha_-\,\bar \alpha_+\,(\bar r \cdot r)$&$ 13$&$ 13$&$
   {\textstyle{1\over 2}}\, \alpha_-\,\bar r^2$\\
  $13$&$ 14$&$ -{\textstyle{1\over 2}}\, \bar \alpha_+\,r^2$&$ 13$&$ 15$&$ -(\bar r \cdot r)$&$
   13$&$ 16$&$ {\textstyle{1\over 4}}\, \bar \alpha_-\,\alpha_+\,(\bar r \cdot r)$\\
  $13$&$ 17$&$ -{\textstyle{1\over 2}}\, \alpha_+\,\bar r^2$&$ 13$&$ 18$&$
   {\textstyle{1\over 2}}\, \bar \alpha_-\,r^2$&$ 13$&$ 19$&$ -(\bar r \cdot r)$\\
  $13$&$ 39$&$ -\alpha_-$&$ 13$&$ 40$&$ -\bar \alpha_-$&$ 13$&$ 41$&$ 1 - \bar \delta\,\delta$\\
  $14$&$ 1$&$ -v^2/s$&$ 14$&$ 2$&$ v^2/s$&$ 14$&$ 4$&$
   {\textstyle{1\over 4}}\, \bar \alpha_+\,\alpha_+\,(\bar r \cdot r)$\\
  $14$&$ 5$&$ {\textstyle{1\over 2}}\, \alpha_+\,\bar r^2$&$ 14$&$ 6$&$
   {\textstyle{1\over 2}}\, \bar \alpha_+\,r^2$&$ 14$&$ 7$&$ (\bar r \cdot r)$\\
  $14$&$ 8$&$ {\textstyle{1\over 4}}\, \bar \alpha_-\,\alpha_-\,(\bar r \cdot r)$&$ 14$&$ 9$&$
  - {\textstyle{1\over 2}}\, \alpha_-\,\bar r^2$&$ 14$&$ 10$&$
  - {\textstyle{1\over 2}}\, \bar \alpha_-\,r^2$\\ $14$&$ 11$&$ (\bar r \cdot r)$&$ 14$&$ 12$&$
   {\textstyle{1\over 4}}\, \alpha_-\,\bar \alpha_+\,(\bar r \cdot r)$&$ 14$&$ 13$&$
   {\textstyle{1\over 2}}\, \alpha_-\,\bar r^2$\\
  $14$&$ 14$&$ {\textstyle{1\over 2}}\,- \bar \alpha_+\,r^2$&$ 14$&$ 15$&$ -(\bar r \cdot r)$&$
   14$&$ 16$&$ {\textstyle{1\over 4}}\, \bar \alpha_-\,\alpha_+\,(\bar r \cdot r)$\\
  $14$&$ 17$&$ {\textstyle{1\over 2}}\,- \alpha_+\,\bar r^2$&$ 14$&$ 18$&$
   {\textstyle{1\over 2}}\, \bar \alpha_-\,r^2$&$ 14$&$ 19$&$ -(\bar r \cdot r)$\\
  $14$&$ 39$&$ \alpha_+$&$ 14$&$ 40$&$ -\bar \alpha_-$&$ 14$&$ 41$&$ -\bar \delta\,\delta - 1$\\
  $18$&$ 1$&$ -v^2/s$&$ 18$&$ 2$&$ v^2/s$&$ 18$&$ 4$&$
   {\textstyle{1\over 4}}\, \bar \alpha_+\,\alpha_+\,(\bar r \cdot r)$\\
  $18$&$ 5$&$ {\textstyle{1\over 2}}\, \alpha_+\,\bar r^2$&$ 18$&$ 6$&$
   {\textstyle{1\over 2}}\, \bar \alpha_+\,r^2$&$ 18$&$ 7$&$ (\bar r \cdot r)$\\
  $18$&$ 8$&$ {\textstyle{1\over 4}}\, \bar \alpha_-\,\alpha_-\,(\bar r \cdot r)$&$ 18$&$ 9$&$
   -{\textstyle{1\over 2}}\, \alpha_-\,\bar r^2$&$ 18$&$ 10$&$
   -{\textstyle{1\over 2}}\, \bar \alpha_-\,r^2$\\ $18$&$ 11$&$ (\bar r \cdot r)$&$ 18$&$ 12$&$
   {\textstyle{1\over 4}}\, \alpha_-\,\bar \alpha_+\,(\bar r \cdot r)$&$ 18$&$ 13$&$
   {\textstyle{1\over 2}}\, \alpha_-\,\bar r^2$\\
  $18$&$ 14$&$ -{\textstyle{1\over 2}}\, \bar \alpha_+\,r^2$&$ 18$&$ 15$&$ -(\bar r \cdot r)$&$
   18$&$ 16$&$ {\textstyle{1\over 4}}\, \bar \alpha_-\,\alpha_+\,(\bar r \cdot r)$\\
  $18$&$ 17$&$ -{\textstyle{1\over 2}}\, \alpha_+\,\bar r^2$&$ 18$&$ 18$&$
   {\textstyle{1\over 2}}\, \bar \alpha_-\,r^2$&$ 18$&$ 19$&$ -(\bar r \cdot r)$\\
  $18$&$ 39$&$ -\alpha_-$&$ 18$&$ 40$&$ \bar \alpha_+$&$ 18$&$ 41$&$ -\bar \delta\,\delta - 1$\\
  $19$&$ 1$&$ -v^2/s$&$ 19$&$ 2$&$ v^2/s$&$ 19$&$ 4$&$
   {\textstyle{1\over 4}}\, \bar \alpha_+\,\alpha_+\,(\bar r \cdot r)$\\
  $19$&$ 5$&$ {\textstyle{1\over 2}}\, \alpha_+\,\bar r^2$&$ 19$&$ 6$&$
   {\textstyle{1\over 2}}\, \bar \alpha_+\,r^2$&$ 19$&$ 7$&$ (\bar r \cdot r)$\\
  $19$&$ 8$&$ {\textstyle{1\over 4}}\, \bar \alpha_-\,\alpha_-\,(\bar r \cdot r)$&$ 19$&$ 9$&$
   -{\textstyle{1\over 2}}\, \alpha_-\,\bar r^2$&$ 19$&$ 10$&$
   -{\textstyle{1\over 2}}\, \bar \alpha_-\,r^2$\\ $19$&$ 11$&$ (\bar r \cdot r)$&$ 19$&$ 12$&$
   {\textstyle{1\over 4}}\, \alpha_-\,\bar \alpha_+\,(\bar r \cdot r)$&$ 19$&$ 13$&$
   {\textstyle{1\over 2}}\, \alpha_-\,\bar r^2$\\
  $19$&$ 14$&$ -{\textstyle{1\over 2}}\, \bar \alpha_+\,r^2$&$ 19$&$ 15$&$ -(\bar r \cdot r)$&$
   19$&$ 16$&$ {\textstyle{1\over 4}}\, \bar \alpha_-\,\alpha_+\,(\bar r \cdot r)$\\
  $19$&$ 17$&$ -{\textstyle{1\over 2}}\, \alpha_+\,\bar r^2$&$ 19$&$ 18$&$
   {\textstyle{1\over 2}}\, \bar \alpha_-\,r^2$&$ 19$&$ 19$&$ -(\bar r \cdot r)$\\
  $19$&$ 39$&$ \alpha_+$&$ 19$&$ 40$&$ \bar \alpha_+$&$ 19$&$ 41$&$ 1 - \bar \delta\,\delta$\\
  $25$&$ 1$&$ - {\textstyle{1\over 2}}\,(\bar \delta\,\delta + 1)\,v^2/s$&$ 25$&$ 2$&$
   {\textstyle{1\over 2}}\,(\bar \delta\,\delta - 1)\,v^2/s$&$ 25$&$ 3$&$ v^2/s$\\
  $25$&$ 4$&$ {\textstyle{1\over 8}}\,\bar \alpha_+\,\alpha_+\,(\bar \delta\,\delta + 1)\,(\bar r \cdot r)$&$ 25$&$ 5$&$
   {\textstyle{1\over 4}}\,\alpha_+\,(\bar \delta\,\delta + 1)\,\bar r^2$&$ 25$&$ 6$&$
   {\textstyle{1\over 4}}\,\bar \alpha_+\,(\bar \delta\,\delta + 1)\,r^2$\\
  $25$&$ 7$&$ {\textstyle{1\over 2}}\,(\bar \delta\,\delta + 1)\,(\bar r \cdot r)$&$ 25$&$ 8$&$
   {\textstyle{1\over 8}}\,\bar \alpha_-\,\alpha_-\,(\bar \delta\,\delta + 1)\,(\bar r \cdot r)$&$ 25$&$ 9$&$
   -{\textstyle{1\over 4}}\,\alpha_-\,(\bar \delta\,\delta + 1)\,\bar r^2$\\
  $25$&$ 10$&$ -{\textstyle{1\over 4}}\,\bar \alpha_-\,(\bar \delta\,\delta + 1)\,r^2$&$ 25$&$ 11$&$
   {\textstyle{1\over 2}}\,(\bar \delta\,\delta + 1)\,(\bar r \cdot r)$&$ 25$&$ 12$&$
   {\textstyle{1\over 8}}\,\alpha_-\,\bar \alpha_+\,(\bar \delta\,\delta - 1)\,(\bar r \cdot r)$\\
  $25$&$ 13$&$ {\textstyle{1\over 4}}\,\alpha_-\,(\bar \delta\,\delta - 1)\,\bar r^2$&$ 25$&$ 14$&$
   -{\textstyle{1\over 4}}\,\bar \alpha_+\,(\bar \delta\,\delta - 1)\,r^2$&$ 25$&$ 15$&$
   -{\textstyle{1\over 2}}\,(\bar \delta\,\delta - 1)\,(\bar r \cdot r)$\\
  $25$&$ 16$&$ {\textstyle{1\over 8}}\,\bar \alpha_-\,\alpha_+\,(\bar \delta\,\delta - 1)\,(\bar r \cdot r)$&$ 25$&$ 17$&$
   -{\textstyle{1\over 4}}\,\alpha_+\,(\bar \delta\,\delta - 1)\,\bar r^2$&$ 25$&$ 18$&$
   {\textstyle{1\over 4}}\,\bar \alpha_-\,(\bar \delta\,\delta - 1)\,r^2$\\
  $25$&$ 19$&$ -{\textstyle{1\over 2}}\,(\bar \delta\,\delta - 1)\,(\bar r \cdot r)$&$ 25$&$ 20$&$
   -{\textstyle{1\over 4}}\,\alpha_-\,\alpha_+\,\bar r^2$&$ 25$&$ 21$&$
   -{\textstyle{1\over 2}}\, \alpha_-\,(\bar r \cdot r)$\\
  $25$&$ 22$&$ {\textstyle{1\over 2}}\, \alpha_+\,(\bar r \cdot r)$&$ 25$&$ 23$&$ r^2$&$ 25$&$
   24$&$ -{\textstyle{1\over 4}}\,\bar \alpha_-\,\bar \alpha_+\,r^2$\\
  $25$&$ 25$&$ -{\textstyle{1\over 2}}\, \bar \alpha_-\,(\bar r \cdot r)$&$ 25$&$ 26$&$
   {\textstyle{1\over 2}}\, \bar \alpha_+\,(\bar r \cdot r)$&$ 25$&$ 27$&$ \bar r^2$\\
  $25$&$ 28$&$ -{\textstyle{1\over 16}}\,\bar \alpha_-\,\alpha_-\,\bar \alpha_+\,\alpha_+$&$ 25$&$ 39$&$
    -{\textstyle{1\over 2}}\,\alpha_-\,\alpha_+\,\bar \delta$&$ 25$&$ 40$&$ -{\textstyle{1\over 2}}\,\bar \alpha_-\,\bar \alpha_+\,\delta$\\
  $25$&$ 41$&$ -{\textstyle{1\over 2}}\,(\bar \delta^2\,\delta^2 - 1)$&$ 26$&$ 1$&$
   (\bar r \cdot r)\,v^2$&$ 26$&$ 2$&$ -(\bar r \cdot r)\,v^2$\\
  $26$&$ 4$&$ -{\textstyle{1\over 4}}\,\bar \alpha_+\,\alpha_+\,(\bar r \cdot r)^2\,s$&$ 26$&$ 5$&$
   -{\textstyle{1\over 2}}\,\alpha_+\,\bar r^2\,(\bar r \cdot r)\,s$&$ 26$&$ 6$&$
   -{\textstyle{1\over 2}}\,\bar \alpha_+\,(\bar r \cdot r)\,r^2\,s$\\ $26$&$ 7$&$ -(\bar r \cdot r)^2\,s$&$
   26$&$ 8$&$ {\textstyle{1\over 4}}\,(4\,v^2 - \bar \alpha_-\,\alpha_-\,(\bar r \cdot r)^2\,s)$&$ 26$&$ 9$&$
   {\textstyle{1\over 2}}\,\alpha_-\,\bar r^2\,(\bar r \cdot r)\,s$\\
  $26$&$ 10$&$ {\textstyle{1\over 2}}\,\bar \alpha_-\,(\bar r \cdot r)\,r^2\,s$&$ 26$&$ 11$&$
   (-(\bar r \cdot r)^2)\,s$&$ 26$&$ 12$&$ -{\textstyle{1\over 4}}\,\alpha_-\,\bar \alpha_+\,(\bar r \cdot r)^2\,s$\\
  $26$&$ 13$&$ -{\textstyle{1\over 2}}\,\alpha_-\,\bar r^2\,(\bar r \cdot r)\,s$&$ 26$&$ 14$&$
   {\textstyle{1\over 2}}\,\bar \alpha_+\,(\bar r \cdot r)\,r^2\,s$&$ 26$&$ 15$&$ (\bar r \cdot r)^2\,s$\\
  $26$&$ 16$&$ -{\textstyle{1\over 4}}\,\bar \alpha_-\,\alpha_+\,(\bar r \cdot r)^2\,s$&$ 26$&$ 17$&$
   {\textstyle{1\over 2}}\,\alpha_+\,\bar r^2\,(\bar r \cdot r)\,s$&$ 26$&$ 18$&$
   -{\textstyle{1\over 2}}\,\bar \alpha_-\,(\bar r \cdot r)\,r^2\,s$\\ $26$&$ 19$&$ (\bar r \cdot r)^2\,s$&$ 26$&$
   28$&$ -{\textstyle{1\over 4}}\,\bar \alpha_+\,\alpha_+\,(\bar r \cdot r)\,s$&$ 26$&$ 31$&$
   -{\textstyle{1\over 2}}\,\alpha_+\,\bar r^2\,s$\\ $26$&$ 35$&$
   -{\textstyle{1\over 2}}\,\bar \alpha_+\,r^2\,s$&$ 26$&$ 39$&$ -\alpha_+\,(\bar r \cdot r)\,s$&$ 26$&$
   40$&$ -\bar \alpha_+\,(\bar r \cdot r)\,s$\\ $26$&$ 41$&$ (\bar \delta\,\delta - 1)\,(\bar r \cdot r)\,s$&$
   27$&$ 1$&$ {\textstyle{1\over 2}}\, \alpha_+\,\bar r^2\,v^2)$&$ 27$&$ 2$&$
   -{\textstyle{1\over 2}}\,\alpha_+\,\bar r^2\,v^2$\\
  $27$&$ 4$&$ -{\textstyle{1\over 8}}\,\bar \alpha_+\,\alpha_+^2\,\bar r^2\,(\bar r \cdot r)\,s$&$ 27$&$ 5$&$
   -{\textstyle{1\over 4}}\,\alpha_+^2\,(\bar r^2)^2\,s$&$ 27$&$ 6$&$
   -{\textstyle{1\over 4}}\,\bar \alpha_+\,\alpha_+\,\bar r^2\,r^2\,s$\\
  $27$&$ 7$&$ -{\textstyle{1\over 2}}\,\alpha_+\,\bar r^2\,(\bar r \cdot r)\,s$&$ 27$&$ 8$&$
   -{\textstyle{1\over 8}}\,\bar \alpha_-\,\alpha_-\,\alpha_+\,\bar r^2\,(\bar r \cdot r)\,s$&$ 27$&$ 9$&$
   {\textstyle{1\over 4}}\,\alpha_-\,\alpha_+\,(\bar r^2)^2\,s$\\
  $27$&$ 10$&$ {\textstyle{1\over 4}}\,(\bar \alpha_-\,\alpha_+\,\bar r^2\,r^2\,s + 4\,v^2)$&$ 27$&$
   11$&$ -{\textstyle{1\over 2}}\,\alpha_+\,\bar r^2\,(\bar r \cdot r)\,s$&$ 27$&$ 12$&$
   -{\textstyle{1\over 8}}\,\alpha_-\,\bar \alpha_+\,\alpha_+\,\bar r^2\,(\bar r \cdot r)\,s$\\
  $27$&$ 13$&$ -{\textstyle{1\over 4}}\,\alpha_-\,\alpha_+\,(\bar r^2)^2\,s$&$ 27$&$ 14$&$
   {\textstyle{1\over 4}}\,\bar \alpha_+\,\alpha_+\,\bar r^2\,r^2\,s$&$ 27$&$ 15$&$
   {\textstyle{1\over 2}}\,\alpha_+\,\bar r^2\,(\bar r \cdot r)\,s$\\
  $27$&$ 16$&$ -{\textstyle{1\over 8}}\,\bar \alpha_-\,\alpha_+^2\,\bar r^2\,(\bar r \cdot r)\,s$&$ 27$&$ 17$&$
   {\textstyle{1\over 4}}\,\alpha_+^2\,(\bar r^2)^2\,s$&$ 27$&$ 18$&$ -{\textstyle{1\over 4}}\,\bar \alpha_-\,\alpha_+\,
    \bar r^2\,r^2\,s$\\ $27$&$ 19$&$ {\textstyle{1\over 2}}\,\alpha_+\,\bar r^2\,(\bar r \cdot r)\,s$&$ 27$&$
   30$&$ -{\textstyle{1\over 2}}\,\bar \alpha_+\,r^2\,s$&$ 27$&$ 33$&$ -(\bar r \cdot r)\,s$\\
  $27$&$ 37$&$ -{\textstyle{1\over 4}}\,\bar \alpha_+\,\alpha_+\,(\bar r \cdot r)\,s$&$ 27$&$ 39$&$
    {\textstyle{1\over 2}}\,\alpha_-\,\alpha_+\,\bar r^2\,s$&$ 27$&$ 40$&$ -{\textstyle{1\over 2}}\,\bar \alpha_+\,\alpha_+\,
    \bar r^2\,s$\\ $27$&$ 41$&$ {\textstyle{1\over 2}}\,\alpha_+\,(\bar \delta\,\delta + 1)\,\bar r^2\,s$&$
   28$&$ 1$&$ {\textstyle{1\over 2}}\, \bar \alpha_+\,r^2\,v^2$&$ 28$&$ 2$&$
   -{\textstyle{1\over 2}}\,\bar \alpha_+\,r^2\,v^2$\\
  $28$&$ 4$&$ -{\textstyle{1\over 8}}\,\bar \alpha_+^2\,\alpha_+\,(\bar r \cdot r)\,r^2\,s$&$ 28$&$ 5$&$
   -{\textstyle{1\over 4}}\,\bar \alpha_+\,\alpha_+\,\bar r^2\,r^2\,s$&$ 28$&$ 6$&$
   -{\textstyle{1\over 4}}\,\bar \alpha_+^2\,(r^2)^2\,s$\\
  $28$&$ 7$&$ -{\textstyle{1\over 2}}\,\bar \alpha_+\,(\bar r \cdot r)\,r^2\,s$&$ 28$&$ 8$&$
   -{\textstyle{1\over 8}}\,\bar \alpha_-\,\alpha_-\,\bar \alpha_+\,(\bar r \cdot r)\,r^2\,s$&$ 28$&$ 9$&$
   {\textstyle{1\over 4}}\,(\alpha_-\,\bar \alpha_+\,\bar r^2\,r^2\,s + 4\,v^2)$\\
  $28$&$ 10$&$ {\textstyle{1\over 4}}\,\bar \alpha_-\,\bar \alpha_+\,(r^2)^2\,s$&$ 28$&$ 11$&$
   -{\textstyle{1\over 2}}\,\bar \alpha_+\,(\bar r \cdot r)\,r^2\,s$&$ 28$&$ 12$&$
   -{\textstyle{1\over 8}}\,\alpha_-\,\bar \alpha_+^2\,(\bar r \cdot r)\,r^2\,s$\\
  $28$&$ 13$&$ -{\textstyle{1\over 4}}\,\alpha_-\,\bar \alpha_+\,\bar r^2\,r^2\,s$&$ 28$&$ 14$&$
    {\textstyle{1\over 4}}\,\bar \alpha_+^2\,(r^2)^2\,s$&$ 28$&$ 15$&$ {\textstyle{1\over 2}}\,\bar \alpha_+\,(\bar r \cdot r)\,r^2\,
    s$

\end{tabular}
\label{tab-A}
\caption{The non-vanishing coefficients $c_k^{(n)}$ in the expansion (\ref{res-Q:11to11}).  }
\end{table}

\begin{table}[t]
\rescale
\renewcommand{\arraystretch}{1.03}
\begin{tabular}{|ll|c |ll|c |ll|c  }
$k$ & $n$ & $c_{k}^{(n)}$ \phantom{xxxxxxxxxxxxxxxxxxxxxxxx}&
$k$ & $n$ & $c_{k}^{(n)}$ \phantom{xxxxxxxxxxxxxxxxxxxxxxxx}&
$k$ & $n$ & $c_{k}^{(n)}$ \phantom{xxxxxxxxxxxxxxxxxxxxxxxx}\\ \hline

$28$&$ 16$&$ -{\textstyle{1\over 8}}\,\bar \alpha_-\,\bar \alpha_+\,\alpha_+\,(\bar r \cdot r)\,r^2\,s$&$ 28$&$ 17$&$
   {\textstyle{1\over 4}}\,\bar \alpha_+\,\alpha_+\,\bar r^2\,r^2\,s$&$ 28$&$ 18$&$
   -{\textstyle{1\over 4}}\,\bar \alpha_-\,\bar \alpha_+\,(r^2)^2\,s$\\
  $28$&$ 19$&$ {\textstyle{1\over 2}}\,\bar \alpha_+\,(\bar r \cdot r)\,r^2\,s$&$ 28$&$ 29$&$
    -{\textstyle{1\over 2}}\,\alpha_+\,\bar r^2\,s$&$ 28$&$ 32$&$ -{\textstyle{1\over 4}}\,\bar \alpha_+\,\alpha_+\,(\bar r \cdot r)\,
    s$\\ $28$&$ 36$&$ -(\bar r \cdot r)\,s$&$ 28$&$ 39$&$
   -{\textstyle{1\over 2}}\,\bar \alpha_+\,\alpha_+\,r^2\,s$&$ 28$&$ 40$&$
   {\textstyle{1\over 2}}\,\bar \alpha_-\,\bar \alpha_+\,r^2\,s$\\
  $28$&$ 41$&$ {\textstyle{1\over 2}}\,\bar \alpha_+\,(\bar \delta\,\delta + 1)\,r^2\,s$&$ 29$&$ 1$&$
   {\textstyle{1\over 4}}\,(\gamma_+ + 3)\,(\bar r \cdot r)\,v^2$&$ 29$&$ 2$&$
   -{\textstyle{1\over 4}}\,(\gamma_+ - 1)\,(\bar r \cdot r)\,v^2$\\ $29$&$ 3$&$ -(\bar r \cdot r)\,v^2$&$
   29$&$ 4$&$ -{\textstyle{1\over 16}}\,\bar \alpha_+\,\alpha_+\,(\gamma_+ + 3)\,(\bar r \cdot r)^2\,s$&$ 29$&$ 5$&$
   -{\textstyle{1\over 8}}\,\alpha_+\,(\gamma_+ + 3)\,\bar r^2\,(\bar r \cdot r)\,s$\\
  $29$&$ 6$&$ -{\textstyle{1\over 8}}\,\bar \alpha_+\,(\gamma_+ + 3)\,(\bar r \cdot r)\,r^2\,s$&$ 29$&$ 7$&$
   -{\textstyle{1\over 4}}\,(\gamma_+ + 3)\,(\bar r \cdot r)^2\,s$&$ 29$&$ 8$&$
   -{\textstyle{1\over 16}}\,\bar \alpha_-\,\alpha_-\,(\gamma_+ + 3)\,(\bar r \cdot r)^2\,s$\\
  $29$&$ 9$&$ {\textstyle{1\over 8}}\,\alpha_-\,(\gamma_+ + 3)\,\bar r^2\,(\bar r \cdot r)\,s$&$ 29$&$ 10$&$
   {\textstyle{1\over 8}}\,\bar \alpha_-\,(\gamma_+ + 3)\,(\bar r \cdot r)\,r^2\,s$&$ 29$&$ 11$&$
   -{\textstyle{1\over 4}}\,((\gamma_+ - 1)\,(\bar r \cdot r)^2 + 4\,\bar r^2\,r^2)\,s$\\
  $29$&$ 12$&$ -{\textstyle{1\over 16}}\,\alpha_-\,\bar \alpha_+\,(\gamma_+ - 1)\,(\bar r \cdot r)^2\,s$&$ 29$&$
   13$&$ -{\textstyle{1\over 8}}\,\alpha_-\,(\gamma_+ - 1)\,\bar r^2\,(\bar r \cdot r)\,s$&$ 29$&$ 14$&$
   {\textstyle{1\over 8}}\,\bar \alpha_+\,(\gamma_+ - 1)\,(\bar r \cdot r)\,r^2\,s$\\
  $29$&$ 15$&$ {\textstyle{1\over 4}}\,(\gamma_+ - 1)\,(\bar r \cdot r)^2\,s$&$ 29$&$ 16$&$
   -{\textstyle{1\over 16}}\,\bar \alpha_-\,\alpha_+\,(\gamma_+ - 1)\,(\bar r \cdot r)^2\,s$&$ 29$&$ 17$&$
   {\textstyle{1\over 8}}\,\alpha_+\,(\gamma_+ - 1)\,\bar r^2\,(\bar r \cdot r)\,s$\\
  $29$&$ 18$&$ -{\textstyle{1\over 8}}\,\bar \alpha_-\,(\gamma_+ - 1)\,(\bar r \cdot r)\,r^2\,s$&$ 29$&$ 19$&$
   {\textstyle{1\over 4}}\,(\gamma_+ - 1)\,(\bar r \cdot r)^2\,s$&$ 29$&$ 20$&$
   {\textstyle{1\over 4}}\,\alpha_-\,\alpha_+\,\bar r^2\,(\bar r \cdot r)\,s$\\
  $29$&$ 21$&$ {\textstyle{1\over 2}}\,\alpha_-\,(\bar r \cdot r)^2\,s$&$ 29$&$ 22$&$
   -{\textstyle{1\over 2}}\,\alpha_+\,(\bar r \cdot r)^2\,s$&$ 29$&$ 23$&$ -(\bar r \cdot r)\,r^2\,s$\\
  $29$&$ 24$&$ {\textstyle{1\over 4}}\,\bar \alpha_-\,\bar \alpha_+\,(\bar r \cdot r)\,r^2\,s$&$ 29$&$ 25$&$
    {\textstyle{1\over 2}}\,\bar \alpha_-\,(\bar r \cdot r)^2\,s$&$ 29$&$ 26$&$ -{\textstyle{1\over 2}}\,\bar \alpha_+\,(\bar r \cdot r)^2\,
    s$\\ $29$&$ 27$&$ -\bar r^2\,(\bar r \cdot r)\,s$&$ 29$&$ 28$&$
   {\textstyle{1\over 16}}\,\bar \alpha_-\,\alpha_-\,\bar \alpha_+\,\alpha_+\,(\bar r \cdot r)\,s$&$ 29$&$ 34$&$
   -{\textstyle{1\over 2}}\,\bar \alpha_+\,r^2\,s$\\ $29$&$ 38$&$
   -{\textstyle{1\over 2}}\,\alpha_+\,\bar r^2\,s$&$ 29$&$ 39$&$ {\textstyle{1\over 4}}\,\alpha_-\,\alpha_+\,
    (3\,\bar \delta + 1)\,(\bar r \cdot r)\,s$&$ 29$&$ 40$&$ {\textstyle{1\over 4}}\,\bar \alpha_-\,\bar \alpha_+\,
    (3\,\delta + 1)\,(\bar r \cdot r)\,s$\\ $29$&$ 41$&$ {\textstyle{1\over 4}}\,(\bar \delta\,\delta - 1)\,
    (\gamma_+ + 3)\,(\bar r \cdot r)\,s$&$ 30$&$ 1$&$ (\bar r \cdot r)\,v^2$&$ 30$&$ 2$&$
   -(\bar r \cdot r)\,v^2$\\ $30$&$ 4$&$
   {\textstyle{1\over 4}}\,(4\,v^2 - \bar \alpha_+\,\alpha_+\,(\bar r \cdot r)^2\,s)$&$ 30$&$ 5$&$
   -{\textstyle{1\over 2}}\,\alpha_+\,\bar r^2\,(\bar r \cdot r)\,s$&$ 30$&$ 6$&$
   -{\textstyle{1\over 2}}\,\bar \alpha_+\,(\bar r \cdot r)\,r^2\,s$\\ $30$&$ 7$&$ -(\bar r \cdot r)^2\,s$&$
   30$&$ 8$&$ -{\textstyle{1\over 4}}\,\bar \alpha_-\,\alpha_-\,(\bar r \cdot r)^2\,s$&$ 30$&$ 9$&$
   {\textstyle{1\over 2}}\,\alpha_-\,\bar r^2\,(\bar r \cdot r)\,s$\\
  $30$&$ 10$&$ {\textstyle{1\over 2}}\,\bar \alpha_-\,(\bar r \cdot r)\,r^2\,s$&$ 30$&$ 11$&$
   -(\bar r \cdot r)^2\,s$&$ 30$&$ 12$&$ -{\textstyle{1\over 4}}\,\alpha_-\,\bar \alpha_+\,(\bar r \cdot r)^2\,s$\\
  $30$&$ 13$&$ -{\textstyle{1\over 2}}\,\alpha_-\,\bar r^2\,(\bar r \cdot r)\,s$&$ 30$&$ 14$&$
   {\textstyle{1\over 2}}\,\bar \alpha_+\,(\bar r \cdot r)\,r^2\,s$&$ 30$&$ 15$&$ (\bar r \cdot r)^2\,s$\\
  $30$&$ 16$&$ -{\textstyle{1\over 4}}\,\bar \alpha_-\,\alpha_+\,(\bar r \cdot r)^2\,s$&$ 30$&$ 17$&$
   {\textstyle{1\over 2}}\,\alpha_+\,\bar r^2\,(\bar r \cdot r)\,s$&$ 30$&$ 18$&$
   -{\textstyle{1\over 2}}\,\bar \alpha_-\,(\bar r \cdot r)\,r^2\,s$\\ $30$&$ 19$&$ (\bar r \cdot r)^2\,s$&$ 30$&$
   28$&$ -{\textstyle{1\over 4}}\,\bar \alpha_-\,\alpha_-\,(\bar r \cdot r)\,s$&$ 30$&$ 32$&$
   {\textstyle{1\over 2}}\,\alpha_-\,\bar r^2\,s $\\ $30$&$ 37$&$ {\textstyle{1\over 2}}\,\bar \alpha_-\,r^2\,s $&$ 30$&$ 39$&$
   \alpha_-\,(\bar r \cdot r)\,s$&$ 30$&$ 40$&$ \bar \alpha_-\,(\bar r \cdot r)\,s$\\
  $30$&$ 41$&$ (\bar \delta\,\delta - 1)\,(\bar r \cdot r)\,s$&$ 31$&$ 1$&$
   -{\textstyle{1\over 2}}\,\alpha_-\,\bar r^2\,v^2$&$ 31$&$ 2$&$ {\textstyle{1\over 2}}\,\alpha_-\,\bar r^2\,v^2 $\\
  $31$&$ 4$&$ {\textstyle{1\over 8}}\,\alpha_-\,\bar \alpha_+\,\alpha_+\,\bar r^2\,(\bar r \cdot r)\,s$&$ 31$&$ 5$&$
   {\textstyle{1\over 4}}\,\alpha_-\,\alpha_+\,(\bar r^2)^2\,s$&$ 31$&$ 6$&$
   {\textstyle{1\over 4}}\,(\alpha_-\,\bar \alpha_+\,\bar r^2\,r^2\,s + 4\,v^2)$\\
  $31$&$ 7$&$ {\textstyle{1\over 2}}\,\alpha_-\,\bar r^2\,(\bar r \cdot r)\,s$&$ 31$&$ 8$&$
   {\textstyle{1\over 8}}\,\bar \alpha_-\,\alpha_-^2\,\bar r^2\,(\bar r \cdot r)\,s$&$ 31$&$ 9$&$
   -{\textstyle{1\over 4}}\,\alpha_-^2\,(\bar r^2)^2\,s$\\
  $31$&$ 10$&$ -{\textstyle{1\over 4}}\,\bar \alpha_-\,\alpha_-\,\bar r^2\,r^2\,s$&$ 31$&$ 11$&$
   {\textstyle{1\over 2}}\,\alpha_-\,\bar r^2\,(\bar r \cdot r)\,s$&$ 31$&$ 12$&$
   {\textstyle{1\over 8}}\,\alpha_-^2\,\bar \alpha_+\,\bar r^2\,(\bar r \cdot r)\,s$\\
  $31$&$ 13$&$ {\textstyle{1\over 4}}\,\alpha_-^2\,(\bar r^2)^2\,s$&$ 31$&$ 14$&$
   -{\textstyle{1\over 4}}\,\alpha_-\,\bar \alpha_+\,\bar r^2\,r^2\,s$&$ 31$&$ 15$&$
   -{\textstyle{1\over 2}}\,\alpha_-\,\bar r^2\,(\bar r \cdot r)\,s$\\
  $31$&$ 16$&$ {\textstyle{1\over 8}}\,\bar \alpha_-\,\alpha_-\,\alpha_+\,\bar r^2\,(\bar r \cdot r)\,s$&$ 31$&$ 17$&$
   -{\textstyle{1\over 4}}\,\alpha_-\,\alpha_+\,(\bar r^2)^2\,s$&$ 31$&$ 18$&$
   {\textstyle{1\over 4}}\,\bar \alpha_-\,\alpha_-\,\bar r^2\,r^2\,s$\\
  $31$&$ 19$&$ -{\textstyle{1\over 2}}\,\alpha_-\,\bar r^2\,(\bar r \cdot r)\,s$&$ 31$&$ 30$&$
   (\bar \alpha_-\,r^2\,s /2$&$ 31$&$ 34$&$ -(\bar r \cdot r)\,s$\\
  $31$&$ 35$&$ -{\textstyle{1\over 4}}\,\bar \alpha_-\,\alpha_-\,(\bar r \cdot r)\,s$&$ 31$&$ 39$&$
    {\textstyle{1\over 2}}\,\alpha_-\,\alpha_+\,\bar r^2\,s$&$ 31$&$ 40$&$ -{\textstyle{1\over 2}}\,\bar \alpha_-\,\alpha_-\,
    \bar r^2\,s$\\ $31$&$ 41$&$ -{\textstyle{1\over 2}}\,\alpha_-\,(\bar \delta\,\delta + 1)\,\bar r^2\,
    s$&$ 32$&$ 1$&$ -{\textstyle{1\over 2}}\,\bar \alpha_-\,r^2\,v^2$&$ 32$&$ 2$&$
    (\bar \alpha_-\,r^2\,v^2 /2$\\ $32$&$ 4$&$ {\textstyle{1\over 8}}\,\bar \alpha_-\,\bar \alpha_+\,\alpha_+\,(\bar r \cdot r)\,
    r^2\,s$&$ 32$&$ 5$&$ {\textstyle{1\over 4}}\,(\bar \alpha_-\,\alpha_+\,\bar r^2\,r^2\,s +
     4\,v^2)$&$ 32$&$ 6$&$ {\textstyle{1\over 4}}\,\bar \alpha_-\,\bar \alpha_+\,(r^2)^2\,s$\\
  $32$&$ 7$&$ {\textstyle{1\over 2}}\,\bar \alpha_-\,(\bar r \cdot r)\,r^2\,s$&$ 32$&$ 8$&$
   {\textstyle{1\over 8}}\,\bar \alpha_-^2\,\alpha_-\,(\bar r \cdot r)\,r^2\,s$&$ 32$&$ 9$&$
   -{\textstyle{1\over 4}}\,\bar \alpha_-\,\alpha_-\,\bar r^2\,r^2\,s$\\
  $32$&$ 10$&$ -{\textstyle{1\over 4}}\,\bar \alpha_-^2\,(r^2)^2\,s$&$ 32$&$ 11$&$
   {\textstyle{1\over 2}}\,\bar \alpha_-\,(\bar r \cdot r)\,r^2\,s$&$ 32$&$ 12$&$
   {\textstyle{1\over 8}}\,\bar \alpha_-\,\alpha_-\,\bar \alpha_+\,(\bar r \cdot r)\,r^2\,s$\\
  $32$&$ 13$&$ {\textstyle{1\over 4}}\,\bar \alpha_-\,\alpha_-\,\bar r^2\,r^2\,s$&$ 32$&$ 14$&$
   -{\textstyle{1\over 4}}\,\bar \alpha_-\,\bar \alpha_+\,(r^2)^2\,s$&$ 32$&$ 15$&$
   -{\textstyle{1\over 2}}\,\bar \alpha_-\,(\bar r \cdot r)\,r^2\,s$\\
  $32$&$ 16$&$ {\textstyle{1\over 8}}\,\bar \alpha_-^2\,\alpha_+\,(\bar r \cdot r)\,r^2\,s$&$ 32$&$ 17$&$
   -{\textstyle{1\over 4}}\,\bar \alpha_-\,\alpha_+\,\bar r^2\,r^2\,s$&$ 32$&$ 18$&$
   {\textstyle{1\over 4}}\,\bar \alpha_-^2\,(r^2)^2\,s$\\
  $32$&$ 19$&$ -{\textstyle{1\over 2}}\,\bar \alpha_-\,(\bar r \cdot r)\,r^2\,s$&$ 32$&$ 29$&$
   (\alpha_-\,\bar r^2\,s /2$&$ 32$&$ 31$&$ -{\textstyle{1\over 4}}\,\bar \alpha_-\,\alpha_-\,(\bar r \cdot r)\,s$\\
  $32$&$ 38$&$ -(\bar r \cdot r)\,s$&$ 32$&$ 39$&$ -{\textstyle{1\over 2}}\,\bar \alpha_-\,\alpha_-\,r^2\,
    s$&$ 32$&$ 40$&$ {\textstyle{1\over 2}}\,\bar \alpha_-\,\bar \alpha_+\,r^2\,s$\\
  $32$&$ 41$&$ -{\textstyle{1\over 2}}\,\bar \alpha_-\,(\bar \delta\,\delta + 1)\,r^2\,s$&$ 33$&$ 1$&$
   -{\textstyle{1\over 4}}\,(\gamma_- - 3)\,(\bar r \cdot r)\,v^2$&$ 33$&$ 2$&$
   {\textstyle{1\over 4}}\,(\gamma_- + 1)\,(\bar r \cdot r)\,v^2$\\ $33$&$ 3$&$ -(\bar r \cdot r)\,v^2$&$ 33$&$
   4$&$ {\textstyle{1\over 16}}\,\bar \alpha_+\,\alpha_+\,(\gamma_- - 3)\,(\bar r \cdot r)^2\,s$&$ 33$&$ 5$&$
   {\textstyle{1\over 8}}\,\alpha_+\,(\gamma_- - 3)\,\bar r^2\,(\bar r \cdot r)\,s$\\
  $33$&$ 6$&$ {\textstyle{1\over 8}}\,\bar \alpha_+\,(\gamma_- - 3)\,(\bar r \cdot r)\,r^2\,s$&$ 33$&$ 7$&$
   {\textstyle{1\over 4}}\,((\gamma_- + 1)\,(\bar r \cdot r)^2 - 4\,\bar r^2\,r^2)\,s$&$ 33$&$ 8$&$
   {\textstyle{1\over 16}}\,\bar \alpha_-\,\alpha_-\,(\gamma_- - 3)\,(\bar r \cdot r)^2\,s$\\
  $33$&$ 9$&$ -{\textstyle{1\over 8}}\,\alpha_-\,(\gamma_- - 3)\,\bar r^2\,(\bar r \cdot r)\,s$&$ 33$&$ 10$&$
   -{\textstyle{1\over 8}}\,\bar \alpha_-\,(\gamma_- - 3)\,(\bar r \cdot r)\,r^2\,s$&$ 33$&$ 11$&$
   {\textstyle{1\over 4}}\,(\gamma_- - 3)\,(\bar r \cdot r)^2\,s$\\
  $33$&$ 12$&$ {\textstyle{1\over 16}}\,\alpha_-\,\bar \alpha_+\,(\gamma_- + 1)\,(\bar r \cdot r)^2\,s$&$ 33$&$ 13$&$
   {\textstyle{1\over 8}}\,\alpha_-\,(\gamma_- + 1)\,\bar r^2\,(\bar r \cdot r)\,s$&$ 33$&$ 14$&$
   -{\textstyle{1\over 8}}\,\bar \alpha_+\,(\gamma_- + 1)\,(\bar r \cdot r)\,r^2\,s$\\
  $33$&$ 15$&$ -{\textstyle{1\over 4}}\,(\gamma_- + 1)\,(\bar r \cdot r)^2\,s$&$ 33$&$ 16$&$
   {\textstyle{1\over 16}}\,\bar \alpha_-\,\alpha_+\,(\gamma_- + 1)\,(\bar r \cdot r)^2\,s$&$ 33$&$ 17$&$
   -{\textstyle{1\over 8}}\,\alpha_+\,(\gamma_- + 1)\,\bar r^2\,(\bar r \cdot r)\,s$\\
  $33$&$ 18$&$ {\textstyle{1\over 8}}\,\bar \alpha_-\,(\gamma_- + 1)\,(\bar r \cdot r)\,r^2\,s$&$ 33$&$ 19$&$
   -{\textstyle{1\over 4}}\,(\gamma_- + 1)\,(\bar r \cdot r)^2\,s$&$ 33$&$ 20$&$
   {\textstyle{1\over 4}}\,\alpha_-\,\alpha_+\,\bar r^2\,(\bar r \cdot r)\,s$\\
  $33$&$ 21$&$ {\textstyle{1\over 2}}\,\alpha_-\,(\bar r \cdot r)^2\,s$&$ 33$&$ 22$&$
   -{\textstyle{1\over 2}}\,\alpha_+\,(\bar r \cdot r)^2\,s$&$ 33$&$ 23$&$ -(\bar r \cdot r)\,r^2\,s$\\
  $33$&$ 24$&$ {\textstyle{1\over 4}}\,\bar \alpha_-\,\bar \alpha_+\,(\bar r \cdot r)\,r^2\,s$&$ 33$&$ 25$&$
    {\textstyle{1\over 2}}\,\bar \alpha_-\,(\bar r \cdot r)^2\,s$&$ 33$&$ 26$&$ -{\textstyle{1\over 2}}\,\bar \alpha_+\,(\bar r \cdot r)^2\,
    s$\\ $33$&$ 27$&$ -\bar r^2\,(\bar r \cdot r)\,s$&$ 33$&$ 28$&$
   {\textstyle{1\over 16}}\,\bar \alpha_-\,\alpha_-\,\bar \alpha_+\,\alpha_+\,(\bar r \cdot r)\,s$&$ 33$&$ 33$&$
   (\bar \alpha_-\,r^2\,s /2$\\ $33$&$ 36$&$ {\textstyle{1\over 2}}\,\alpha_-\,\bar r^2\,s $&$ 33$&$ 39$&$
   {\textstyle{1\over 4}}\,\alpha_-\,\alpha_+\,(3\,\bar \delta - 1)\,(\bar r \cdot r)\,s$&$ 33$&$ 40$&$
   {\textstyle{1\over 4}}\,\bar \alpha_-\,\bar \alpha_+\,(3\,\delta - 1)\,(\bar r \cdot r)\,s$\\
  $33$&$ 41$&$ -{\textstyle{1\over 4}}\,(\bar \delta\,\delta - 1)\,(\gamma_- - 3)\,(\bar r \cdot r)\,s$&$ 34$&$
   1$&$ (\bar r \cdot r)\,v^2$&$ 34$&$ 2$&$ -(\bar r \cdot r)\,v^2$\\
  $34$&$ 4$&$ -{\textstyle{1\over 4}}\,\bar \alpha_+\,\alpha_+\,(\bar r \cdot r)^2\,s$&$ 34$&$ 5$&$
   -{\textstyle{1\over 2}}\,\alpha_+\,\bar r^2\,(\bar r \cdot r)\,s$&$ 34$&$ 6$&$
   -{\textstyle{1\over 2}}\,\bar \alpha_+\,(\bar r \cdot r)\,r^2\,s$\\ $34$&$ 7$&$ -(\bar r \cdot r)^2\,s$&$
   34$&$ 8$&$ -{\textstyle{1\over 4}}\,\bar \alpha_-\,\alpha_-\,(\bar r \cdot r)^2\,s$&$ 34$&$ 9$&$
   {\textstyle{1\over 2}}\,\alpha_-\,\bar r^2\,(\bar r \cdot r)\,s$\\
  $34$&$ 10$&$ {\textstyle{1\over 2}}\,\bar \alpha_-\,(\bar r \cdot r)\,r^2\,s$&$ 34$&$ 11$&$
   -(\bar r \cdot r)^2\,s$&$ 34$&$ 12$&$
   {\textstyle{1\over 4}}\,(4\,v^2 - \alpha_-\,\bar \alpha_+\,(\bar r \cdot r)^2\,s)$\\
  $34$&$ 13$&$ -{\textstyle{1\over 2}}\,\alpha_-\,\bar r^2\,(\bar r \cdot r)\,s$&$ 34$&$ 14$&$
   {\textstyle{1\over 2}}\,\bar \alpha_+\,(\bar r \cdot r)\,r^2\,s$&$ 34$&$ 15$&$ (\bar r \cdot r)^2\,s$\\
  $34$&$ 16$&$ -{\textstyle{1\over 4}}\,\bar \alpha_-\,\alpha_+\,(\bar r \cdot r)^2\,s$&$ 34$&$ 17$&$
   {\textstyle{1\over 2}}\,\alpha_+\,\bar r^2\,(\bar r \cdot r)\,s$&$ 34$&$ 18$&$
   -{\textstyle{1\over 2}}\,\bar \alpha_-\,(\bar r \cdot r)\,r^2\,s$\\ $34$&$ 19$&$ (\bar r \cdot r)^2\,s$&$ 34$&$
   28$&$ {\textstyle{1\over 4}}\,\bar \alpha_-\,\alpha_+\,(\bar r \cdot r)\,s$&$ 34$&$ 32$&$
   -{\textstyle{1\over 2}}\,\alpha_+\,\bar r^2\,s$\\ $34$&$ 35$&$ {\textstyle{1\over 2}}\,\bar \alpha_-\,r^2\,s $&$ 34$&$
   39$&$ -\alpha_+\,(\bar r \cdot r)\,s$&$ 34$&$ 40$&$ \bar \alpha_-\,(\bar r \cdot r)\,s$\\
  $34$&$ 41$&$ (\bar \delta\,\delta + 1)\,(\bar r \cdot r)\,s$&$ 35$&$ 1$&$
   \alpha_+\,\bar r^2\,v^2 /2$&$ 35$&$ 2$&$ -{\textstyle{1\over 2}}\,\alpha_+\,\bar r^2\,v^2$\\
  $35$&$ 4$&$ -{\textstyle{1\over 8}}\,\bar \alpha_+\,\alpha_+^2\,\bar r^2\,(\bar r \cdot r)\,s$&$ 35$&$ 5$&$
   -{\textstyle{1\over 4}}\,\alpha_+^2\,(\bar r^2)^2\,s$&$ 35$&$ 6$&$
   -{\textstyle{1\over 4}}\,\bar \alpha_+\,\alpha_+\,\bar r^2\,r^2\,s$\\
  $35$&$ 7$&$ -{\textstyle{1\over 2}}\,\alpha_+\,\bar r^2\,(\bar r \cdot r)\,s$&$ 35$&$ 8$&$
   -{\textstyle{1\over 8}}\,\bar \alpha_-\,\alpha_-\,\alpha_+\,\bar r^2\,(\bar r \cdot r)\,s$&$ 35$&$ 9$&$
   {\textstyle{1\over 4}}\,\alpha_-\,\alpha_+\,(\bar r^2)^2\,s$
\end{tabular}
\caption{Continuation of Tab. V.}
\end{table}

\begin{table}[t]
\rescale
\renewcommand{\arraystretch}{1.03}
\begin{tabular}{|ll|c |ll|c |ll|c  }
$k$ & $n$ & $c_{k}^{(n)}$ \phantom{xxxxxxxxxxxxxxxxxxxxxxxx}&
$k$ & $n$ & $c_{k}^{(n)}$ \phantom{xxxxxxxxxxxxxxxxxxxxxxxx}&
$k$ & $n$ & $c_{k}^{(n)}$ \phantom{xxxxxxxxxxxxxxxxxxxxxxxx}\\ \hline

$35$&$ 10$&$ {\textstyle{1\over 4}}\,\bar \alpha_-\,\alpha_+\,\bar r^2\,r^2\,s$&$ 35$&$ 11$&$
   -{\textstyle{1\over 2}}\,\alpha_+\,\bar r^2\,(\bar r \cdot r)\,s$&$ 35$&$ 12$&$
   -{\textstyle{1\over 8}}\,\alpha_-\,\bar \alpha_+\,\alpha_+\,\bar r^2\,(\bar r \cdot r)\,s$\\
  $35$&$ 13$&$ -{\textstyle{1\over 4}}\,\alpha_-\,\alpha_+\,(\bar r^2)^2\,s$&$ 35$&$ 14$&$
   {\textstyle{1\over 4}}\,(\bar \alpha_+\,\alpha_+\,\bar r^2\,r^2\,s + 4\,v^2)$&$ 35$&$ 15$&$
   {\textstyle{1\over 2}}\,\alpha_+\,\bar r^2\,(\bar r \cdot r)\,s$\\
  $35$&$ 16$&$ -{\textstyle{1\over 8}}\,\bar \alpha_-\,\alpha_+^2\,\bar r^2\,(\bar r \cdot r)\,s$&$ 35$&$ 17$&$
   {\textstyle{1\over 4}}\,\alpha_+^2\,(\bar r^2)^2\,s$&$ 35$&$ 18$&$ -{\textstyle{1\over 4}}\,\bar \alpha_-\,\alpha_+\,
    \bar r^2\,r^2\,s$\\ $35$&$ 19$&$ {\textstyle{1\over 2}}\,\alpha_+\,\bar r^2\,(\bar r \cdot r)\,s$&$ 35$&$
   30$&$ {\textstyle{1\over 2}}\, \bar \alpha_-\,r^2\,s)$&$ 35$&$ 34$&$ -(\bar r \cdot r)\,s$\\
  $35$&$ 37$&$ {\textstyle{1\over 4}}\,\bar \alpha_-\,\alpha_+\,(\bar r \cdot r)\,s$&$ 35$&$ 39$&$
    {\textstyle{1\over 2}}\,\alpha_-\,\alpha_+\,\bar r^2\,s$&$ 35$&$ 40$&$ {\textstyle{1\over 2}}\,\bar \alpha_-\,\alpha_+\,\bar r^2\,
    s$\\ $35$&$ 41$&$ {\textstyle{1\over 2}}\,\alpha_+\,({\bar \delta}\,\delta - 1)\,\bar r^2\,s$&$ 36$&$ 1$&$
   -{\textstyle{1\over 2}}\,\bar \alpha_-\,r^2\,v^2$&$ 36$&$ 2$&$
   {\textstyle{1\over 2}}\, \bar \alpha_-\,r^2\,v^2$\\
  $36$&$ 4$&$ {\textstyle{1\over 8}}\,\bar \alpha_-\,\bar \alpha_+\,\alpha_+\,(\bar r \cdot r)\,r^2\,s$&$ 36$&$ 5$&$
   {\textstyle{1\over 4}}\,\bar \alpha_-\,\alpha_+\,\bar r^2\,r^2\,s$&$ 36$&$ 6$&$
   {\textstyle{1\over 4}}\,\bar \alpha_-\,\bar \alpha_+\,(r^2)^2\,s$\\
  $36$&$ 7$&$ {\textstyle{1\over 2}}\,\bar \alpha_-\,(\bar r \cdot r)\,r^2\,s$&$ 36$&$ 8$&$
   {\textstyle{1\over 8}}\,\bar \alpha_-^2\,\alpha_-\,(\bar r \cdot r)\,r^2\,s$&$ 36$&$ 9$&$
   -{\textstyle{1\over 4}}\,\bar \alpha_-\,\alpha_-\,\bar r^2\,r^2\,s$\\
  $36$&$ 10$&$ -{\textstyle{1\over 4}}\,\bar \alpha_-^2\,(r^2)^2\,s$&$ 36$&$ 11$&$
   {\textstyle{1\over 2}}\,\bar \alpha_-\,(\bar r \cdot r)\,r^2\,s$&$ 36$&$ 12$&$
   {\textstyle{1\over 8}}\,\bar \alpha_-\,\alpha_-\,\bar \alpha_+\,(\bar r \cdot r)\,r^2\,s$\\
  $36$&$ 13$&$ {\textstyle{1\over 4}}\,(\bar \alpha_-\,\alpha_-\,\bar r^2\,r^2\,s + 4\,v^2)$&$ 36$&$
   14$&$ -{\textstyle{1\over 4}}\,\bar \alpha_-\,\bar \alpha_+\,(r^2)^2\,s$&$ 36$&$ 15$&$
   -{\textstyle{1\over 2}}\,\bar \alpha_-\,(\bar r \cdot r)\,r^2\,s$\\
  $36$&$ 16$&$ {\textstyle{1\over 8}}\,\bar \alpha_-^2\,\alpha_+\,(\bar r \cdot r)\,r^2\,s$&$ 36$&$ 17$&$
   -{\textstyle{1\over 4}}\,\bar \alpha_-\,\alpha_+\,\bar r^2\,r^2\,s$&$ 36$&$ 18$&$
   {\textstyle{1\over 4}}\,\bar \alpha_-^2\,(r^2)^2\,s$\\
  $36$&$ 19$&$ -{\textstyle{1\over 2}}\,\bar \alpha_-\,(\bar r \cdot r)\,r^2\,s$&$ 36$&$ 29$&$
    -{\textstyle{1\over 2}}\,\alpha_+\,\bar r^2\,s$&$ 36$&$ 31$&$ {\textstyle{1\over 4}}\,\bar \alpha_-\,\alpha_+\,(\bar r \cdot r)\,
    s$\\ $36$&$ 36$&$ -(\bar r \cdot r)\,s$&$ 36$&$ 39$&$
    {\textstyle{1\over 2}}\,\bar \alpha_-\,\alpha_+\,r^2\,s$&$ 36$&$ 40$&$ {\textstyle{1\over 2}}\,\bar \alpha_-\,\bar \alpha_+\,r^2\,
    s$\\ $36$&$ 41$&$ -{\textstyle{1\over 2}}\,\bar \alpha_-\,({\bar \delta}\,\delta - 1)\,r^2\,s$&$ 37$&$
   1$&$ {\textstyle{1\over 4}}\,(\beta_+ + 1)\,(\bar r \cdot r)\,v^2$&$ 37$&$ 2$&$
   -{\textstyle{1\over 4}}\,(\beta_+ - 3)\,(\bar r \cdot r)\,v^2$\\ $37$&$ 3$&$ -(\bar r \cdot r)\,v^2$&$
   37$&$ 4$&$ -{\textstyle{1\over 16}}\,\bar \alpha_+\,\alpha_+\,(\beta_+ + 1)\,(\bar r \cdot r)^2\,s$&$ 37$&$ 5$&$
   -{\textstyle{1\over 8}}\,\alpha_+\,(\beta_+ + 1)\,\bar r^2\,(\bar r \cdot r)\,s$\\
  $37$&$ 6$&$ -{\textstyle{1\over 8}}\,\bar \alpha_+\,(\beta_+ + 1)\,(\bar r \cdot r)\,r^2\,s$&$ 37$&$ 7$&$
   -{\textstyle{1\over 4}}\,(\beta_+ + 1)\,(\bar r \cdot r)^2\,s$&$ 37$&$ 8$&$
   -{\textstyle{1\over 16}}\,\bar \alpha_-\,\alpha_-\,(\beta_+ + 1)\,(\bar r \cdot r)^2\,s$\\
  $37$&$ 9$&$ {\textstyle{1\over 8}}\,\alpha_-\,(\beta_+ + 1)\,\bar r^2\,(\bar r \cdot r)\,s$&$ 37$&$ 10$&$
   {\textstyle{1\over 8}}\,\bar \alpha_-\,(\beta_+ + 1)\,(\bar r \cdot r)\,r^2\,s$&$ 37$&$ 11$&$
   -{\textstyle{1\over 4}}\,(\beta_+ + 1)\,(\bar r \cdot r)^2\,s$\\
  $37$&$ 12$&$ -{\textstyle{1\over 16}}\,\alpha_-\,\bar \alpha_+\,(\beta_+ - 3)\,(\bar r \cdot r)^2\,s$&$ 37$&$
   13$&$ -{\textstyle{1\over 8}}\,\alpha_-\,(\beta_+ - 3)\,\bar r^2\,(\bar r \cdot r)\,s$&$ 37$&$ 14$&$
   {\textstyle{1\over 8}}\,\bar \alpha_+\,(\beta_+ - 3)\,(\bar r \cdot r)\,r^2\,s$\\
  $37$&$ 15$&$ {\textstyle{1\over 4}}\,((\beta_+ + 1)\,(\bar r \cdot r)^2 - 4\,\bar r^2\,r^2)\,s$&$
   37$&$ 16$&$ -{\textstyle{1\over 16}}\,\bar \alpha_-\,\alpha_+\,(\beta_+ - 3)\,(\bar r \cdot r)^2\,s$&$ 37$&$
   17$&$ {\textstyle{1\over 8}}\,\alpha_+\,(\beta_+ - 3)\,\bar r^2\,(\bar r \cdot r)\,s$\\
  $37$&$ 18$&$ -{\textstyle{1\over 8}}\,\bar \alpha_-\,(\beta_+ - 3)\,(\bar r \cdot r)\,r^2\,s$&$ 37$&$ 19$&$
   {\textstyle{1\over 4}}\,(\beta_+ - 3)\,(\bar r \cdot r)^2\,s$&$ 37$&$ 20$&$
   {\textstyle{1\over 4}}\,\alpha_-\,\alpha_+\,\bar r^2\,(\bar r \cdot r)\,s$\\
  $37$&$ 21$&$ {\textstyle{1\over 2}}\,\alpha_-\,(\bar r \cdot r)^2\,s$&$ 37$&$ 22$&$
   -{\textstyle{1\over 2}}\,\alpha_+\,(\bar r \cdot r)^2\,s$&$ 37$&$ 23$&$ -(\bar r \cdot r)\,r^2\,s$\\
  $37$&$ 24$&$ {\textstyle{1\over 4}}\,\bar \alpha_-\,\bar \alpha_+\,(\bar r \cdot r)\,r^2\,s$&$ 37$&$ 25$&$
    {\textstyle{1\over 2}}\,\bar \alpha_-\,(\bar r \cdot r)^2\,s$&$ 37$&$ 26$&$ -{\textstyle{1\over 2}}\,\bar \alpha_+\,(\bar r \cdot r)^2\,
    s$\\ $37$&$ 27$&$ -\bar r^2\,(\bar r \cdot r)\,s$&$ 37$&$ 28$&$
   {\textstyle{1\over 16}}\,\bar \alpha_-\,\alpha_-\,\bar \alpha_+\,\alpha_+\,(\bar r \cdot r)\,s$&$ 37$&$ 33$&$
   {\textstyle{1\over 2}}\, \bar \alpha_-\,r^2\,s$\\
  $37$&$ 38$&$ -{\textstyle{1\over 2}}\,\alpha_+\,\bar r^2\,s$&$ 37$&$ 39$&$
   {\textstyle{1\over 4}}\,\alpha_-\,\alpha_+\,(3\,{\bar \delta} - 1)\,(\bar r \cdot r)\,s$&$ 37$&$ 40$&$
   {\textstyle{1\over 4}}\,\bar \alpha_-\,\bar \alpha_+\,(3\,\delta + 1)\,(\bar r \cdot r)\,s$\\
  $37$&$ 41$&$ {\textstyle{1\over 4}}\,(\beta_+ - 3)\,({\bar \delta}\,\delta + 1)\,(\bar r \cdot r)\,s$&$ 38$&$ 1$&$
   (\bar r \cdot r)\,v^2$&$ 38$&$ 2$&$ -(\bar r \cdot r)\,v^2$\\
  $38$&$ 4$&$ -{\textstyle{1\over 4}}\,\bar \alpha_+\,\alpha_+\,(\bar r \cdot r)^2\,s$&$ 38$&$ 5$&$
   -{\textstyle{1\over 2}}\,\alpha_+\,\bar r^2\,(\bar r \cdot r)\,s$&$ 38$&$ 6$&$
   -{\textstyle{1\over 2}}\,\bar \alpha_+\,(\bar r \cdot r)\,r^2\,s$\\ $38$&$ 7$&$ -(\bar r \cdot r)^2\,s$&$
   38$&$ 8$&$ -{\textstyle{1\over 4}}\,\bar \alpha_-\,\alpha_-\,(\bar r \cdot r)^2\,s$&$ 38$&$ 9$&$
   {\textstyle{1\over 2}}\,\alpha_-\,\bar r^2\,(\bar r \cdot r)\,s$\\
  $38$&$ 10$&$ {\textstyle{1\over 2}}\,\bar \alpha_-\,(\bar r \cdot r)\,r^2\,s$&$ 38$&$ 11$&$
   (-(\bar r \cdot r)^2)\,s$&$ 38$&$ 12$&$ -{\textstyle{1\over 4}}\,\alpha_-\,\bar \alpha_+\,(\bar r \cdot r)^2\,s$\\
  $38$&$ 13$&$ -{\textstyle{1\over 2}}\,\alpha_-\,\bar r^2\,(\bar r \cdot r)\,s$&$ 38$&$ 14$&$
   {\textstyle{1\over 2}}\,\bar \alpha_+\,(\bar r \cdot r)\,r^2\,s$&$ 38$&$ 15$&$ (\bar r \cdot r)^2\,s$\\
  $38$&$ 16$&$ {\textstyle{1\over 4}}\,(4\,v^2 - \bar \alpha_-\,\alpha_+\,(\bar r \cdot r)^2\,s)$&$ 38$&$ 17$&$
   {\textstyle{1\over 2}}\,\alpha_+\,\bar r^2\,(\bar r \cdot r)\,s$&$ 38$&$ 18$&$
   -{\textstyle{1\over 2}}\,\bar \alpha_-\,(\bar r \cdot r)\,r^2\,s$\\ $38$&$ 19$&$ (\bar r \cdot r)^2\,s$&$ 38$&$
   28$&$ {\textstyle{1\over 4}}\,\alpha_-\,\bar \alpha_+\,(\bar r \cdot r)\,s$&$ 38$&$ 31$&$
   {\textstyle{1\over 2}}\, \alpha_-\,\bar r^2\,s)$\\
  $38$&$ 37$&$ -{\textstyle{1\over 2}}\,\bar \alpha_+\,r^2\,s$&$ 38$&$ 39$&$ \alpha_-\,(\bar r \cdot r)\,s$&$
   38$&$ 40$&$ -\bar \alpha_+\,(\bar r \cdot r)\,s$\\
  $38$&$ 41$&$ ({\bar \delta}\,\delta + 1)\,(\bar r \cdot r)\,s$&$ 39$&$ 1$&$
   -{\textstyle{1\over 2}}\,\alpha_-\,\bar r^2\,v^2$&$ 39$&$ 2$&$
   {\textstyle{1\over 2}}\, \alpha_-\,\bar r^2\,v^2)$\\
  $39$&$ 4$&$ {\textstyle{1\over 8}}\,\alpha_-\,\bar \alpha_+\,\alpha_+\,\bar r^2\,(\bar r \cdot r)\,s$&$ 39$&$ 5$&$
    {\textstyle{1\over 4}}\,\alpha_-\,\alpha_+\,(\bar r^2)^2\,s$&$ 39$&$ 6$&$ {\textstyle{1\over 4}}\,\alpha_-\,\bar \alpha_+\,\bar r^2\,
    r^2\,s$\\ $39$&$ 7$&$ {\textstyle{1\over 2}}\,\alpha_-\,\bar r^2\,(\bar r \cdot r)\,s$&$ 39$&$ 8$&$
   {\textstyle{1\over 8}}\,\bar \alpha_-\,\alpha_-^2\,\bar r^2\,(\bar r \cdot r)\,s$&$ 39$&$ 9$&$
   -{\textstyle{1\over 4}}\,\alpha_-^2\,(\bar r^2)^2\,s$\\
  $39$&$ 10$&$ -{\textstyle{1\over 4}}\,\bar \alpha_-\,\alpha_-\,\bar r^2\,r^2\,s$&$ 39$&$ 11$&$
   {\textstyle{1\over 2}}\,\alpha_-\,\bar r^2\,(\bar r \cdot r)\,s$&$ 39$&$ 12$&$
   {\textstyle{1\over 8}}\,\alpha_-^2\,\bar \alpha_+\,\bar r^2\,(\bar r \cdot r)\,s$\\
  $39$&$ 13$&$ {\textstyle{1\over 4}}\,\alpha_-^2\,(\bar r^2)^2\,s$&$ 39$&$ 14$&$
   -{\textstyle{1\over 4}}\,\alpha_-\,\bar \alpha_+\,\bar r^2\,r^2\,s$&$ 39$&$ 15$&$
   -{\textstyle{1\over 2}}\,\alpha_-\,\bar r^2\,(\bar r \cdot r)\,s$\\
  $39$&$ 16$&$ {\textstyle{1\over 8}}\,\bar \alpha_-\,\alpha_-\,\alpha_+\,\bar r^2\,(\bar r \cdot r)\,s$&$ 39$&$ 17$&$
   -{\textstyle{1\over 4}}\,\alpha_-\,\alpha_+\,(\bar r^2)^2\,s$&$ 39$&$ 18$&$
   {\textstyle{1\over 4}}\,(\bar \alpha_-\,\alpha_-\,\bar r^2\,r^2\,s + 4\,v^2)$\\
  $39$&$ 19$&$ -{\textstyle{1\over 2}}\,\alpha_-\,\bar r^2\,(\bar r \cdot r)\,s$&$ 39$&$ 30$&$
   -{\textstyle{1\over 2}}\,\bar \alpha_+\,r^2\,s$&$ 39$&$ 33$&$ -(\bar r \cdot r)\,s$\\
  $39$&$ 35$&$ {\textstyle{1\over 4}}\,\alpha_-\,\bar \alpha_+\,(\bar r \cdot r)\,s$&$ 39$&$ 39$&$
    {\textstyle{1\over 2}}\,\alpha_-\,\alpha_+\,\bar r^2\,s$&$ 39$&$ 40$&$ {\textstyle{1\over 2}}\,\alpha_-\,\bar \alpha_+\,\bar r^2\,
    s$\\ $39$&$ 41$&$ -{\textstyle{1\over 2}}\,\alpha_-\,({\bar \delta}\,\delta - 1)\,\bar r^2\,s$&$ 40$&$
   1$&$ {\textstyle{1\over 2}}\, \bar \alpha_+\,r^2\,v^2$&$ 40$&$ 2$&$
   -{\textstyle{1\over 2}}\,\bar \alpha_+\,r^2\,v^2$\\
  $40$&$ 4$&$ -{\textstyle{1\over 8}}\,\bar \alpha_+^2\,\alpha_+\,(\bar r \cdot r)\,r^2\,s$&$ 40$&$ 5$&$
   -{\textstyle{1\over 4}}\,\bar \alpha_+\,\alpha_+\,\bar r^2\,r^2\,s$&$ 40$&$ 6$&$
   -{\textstyle{1\over 4}}\,\bar \alpha_+^2\,(r^2)^2\,s$\\
  $40$&$ 7$&$ -{\textstyle{1\over 2}}\,\bar \alpha_+\,(\bar r \cdot r)\,r^2\,s$&$ 40$&$ 8$&$
   -{\textstyle{1\over 8}}\,\bar \alpha_-\,\alpha_-\,\bar \alpha_+\,(\bar r \cdot r)\,r^2\,s$&$ 40$&$ 9$&$
   {\textstyle{1\over 4}}\,\alpha_-\,\bar \alpha_+\,\bar r^2\,r^2\,s$\\
  $40$&$ 10$&$ {\textstyle{1\over 4}}\,\bar \alpha_-\,\bar \alpha_+\,(r^2)^2\,s$&$ 40$&$ 11$&$
   -{\textstyle{1\over 2}}\,\bar \alpha_+\,(\bar r \cdot r)\,r^2\,s$&$ 40$&$ 12$&$
   -{\textstyle{1\over 8}}\,\alpha_-\,\bar \alpha_+^2\,(\bar r \cdot r)\,r^2\,s$\\
  $40$&$ 13$&$ -{\textstyle{1\over 4}}\,\alpha_-\,\bar \alpha_+\,\bar r^2\,r^2\,s$&$ 40$&$ 14$&$
    {\textstyle{1\over 4}}\,\bar \alpha_+^2\,(r^2)^2\,s$&$ 40$&$ 15$&$ {\textstyle{1\over 2}}\,\bar \alpha_+\,(\bar r \cdot r)\,r^2\,
    s$\\ $40$&$ 16$&$ -{\textstyle{1\over 8}}\,\bar \alpha_-\,\bar \alpha_+\,\alpha_+\,(\bar r \cdot r)\,r^2\,s$&$
   40$&$ 17$&$ {\textstyle{1\over 4}}\,(\bar \alpha_+\,\alpha_+\,\bar r^2\,r^2\,s + 4\,v^2)$&$ 40$&$
   18$&$ -{\textstyle{1\over 4}}\,\bar \alpha_-\,\bar \alpha_+\,(r^2)^2\,s$\\
  $40$&$ 19$&$ {\textstyle{1\over 2}}\,\bar \alpha_+\,(\bar r \cdot r)\,r^2\,s$&$ 40$&$ 29$&$
   {\textstyle{1\over 2}}\, \alpha_-\,\bar r^2\,s$&$ 40$&$ 32$&$
    {\textstyle{1\over 4}}\,\alpha_-\,\bar \alpha_+\,(\bar r \cdot r)\,s$\\ $40$&$ 38$&$ -(\bar r \cdot r)\,s$&$ 40$&$ 39$&$
    {\textstyle{1\over 2}}\,\alpha_-\,\bar \alpha_+\,r^2\,s$&$ 40$&$ 40$&$ {\textstyle{1\over 2}}\,\bar \alpha_-\,\bar \alpha_+\,r^2\,
    s$\\ $40$&$ 41$&$ {\textstyle{1\over 2}}\,\bar \alpha_+\,({\bar \delta}\,\delta - 1)\,r^2\,s$&$ 41$&$ 1$&$
   -{\textstyle{1\over 4}}\,(\beta_- - 1)\,(\bar r \cdot r)\,v^2$&$ 41$&$ 2$&$
   {\textstyle{1\over 4}}\,(\beta_- + 3)\,(\bar r \cdot r)\,v^2$\\ $41$&$ 3$&$ -(\bar r \cdot r)\,v^2$&$ 41$&$
   4$&$ {\textstyle{1\over 16}}\,\bar \alpha_+\,\alpha_+\,(\beta_- - 1)\,(\bar r \cdot r)^2\,s$&$ 41$&$ 5$&$
   {\textstyle{1\over 8}}\,\alpha_+\,(\beta_- - 1)\,\bar r^2\,(\bar r \cdot r)\,s$\\
  $41$&$ 6$&$ {\textstyle{1\over 8}}\,\bar \alpha_+\,(\beta_- - 1)\,(\bar r \cdot r)\,r^2\,s$&$ 41$&$ 7$&$
   {\textstyle{1\over 4}}\,(\beta_- - 1)\,(\bar r \cdot r)^2\,s$&$ 41$&$ 8$&$
   {\textstyle{1\over 16}}\,\bar \alpha_-\,\alpha_-\,(\beta_- - 1)\,(\bar r \cdot r)^2\,s$\\
  $41$&$ 9$&$ -{\textstyle{1\over 8}}\,\alpha_-\,(\beta_- - 1)\,\bar r^2\,(\bar r \cdot r)\,s$&$ 41$&$ 10$&$
   -{\textstyle{1\over 8}}\,\bar \alpha_-\,(\beta_- - 1)\,(\bar r \cdot r)\,r^2\,s$&$ 41$&$ 11$&$
   {\textstyle{1\over 4}}\,(\beta_- - 1)\,(\bar r \cdot r)^2\,s$\\
  $41$&$ 12$&$ {\textstyle{1\over 16}}\,\alpha_-\,\bar \alpha_+\,(\beta_- + 3)\,(\bar r \cdot r)^2\,s$&$ 41$&$ 13$&$
   {\textstyle{1\over 8}}\,\alpha_-\,(\beta_- + 3)\,\bar r^2\,(\bar r \cdot r)\,s$&$ 41$&$ 14$&$
   -{\textstyle{1\over 8}}\,\bar \alpha_+\,(\beta_- + 3)\,(\bar r \cdot r)\,r^2\,s$\\
  $41$&$ 15$&$ -{\textstyle{1\over 4}}\,(\beta_- + 3)\,(\bar r \cdot r)^2\,s$&$ 41$&$ 16$&$
   {\textstyle{1\over 16}}\,\bar \alpha_-\,\alpha_+\,(\beta_- + 3)\,(\bar r \cdot r)^2\,s$&$ 41$&$ 17$&$
   -{\textstyle{1\over 8}}\,\alpha_+\,(\beta_- + 3)\,\bar r^2\,(\bar r \cdot r)\,s$\\
  $41$&$ 18$&$ {\textstyle{1\over 8}}\,\bar \alpha_-\,(\beta_- + 3)\,(\bar r \cdot r)\,r^2\,s$&$ 41$&$ 19$&$
   -{\textstyle{1\over 4}}\,((\beta_- - 1)\,(\bar r \cdot r)^2 + 4\,\bar r^2\,r^2)\,s$&$ 41$&$
   20$&$ {\textstyle{1\over 4}}\,\alpha_-\,\alpha_+\,\bar r^2\,(\bar r \cdot r)\,s$\\
  $41$&$ 21$&$ {\textstyle{1\over 2}}\,\alpha_-\,(\bar r \cdot r)^2\,s$&$ 41$&$ 22$&$
   -{\textstyle{1\over 2}}\,\alpha_+\,(\bar r \cdot r)^2\,s$&$ 41$&$ 23$&$ -(\bar r \cdot r)\,r^2\,s$\\
  $41$&$ 24$&$ {\textstyle{1\over 4}}\,\bar \alpha_-\,\bar \alpha_+\,(\bar r \cdot r)\,r^2\,s$&$ 41$&$ 25$&$
    {\textstyle{1\over 2}}\,\bar \alpha_-\,(\bar r \cdot r)^2\,s$&$ 41$&$ 26$&$ -{\textstyle{1\over 2}}\,\bar \alpha_+\,(\bar r \cdot r)^2\,
    s$\\ $41$&$ 27$&$ -\bar r^2\,(\bar r \cdot r)\,s$&$ 41$&$ 28$&$
   {\textstyle{1\over 16}}\,\bar \alpha_-\,\alpha_-\,\bar \alpha_+\,\alpha_+\,(\bar r \cdot r)\,s$&$ 41$&$ 34$&$
   -{\textstyle{1\over 2}}\,\bar \alpha_+\,r^2\,s$\\ $41$&$ 36$&$
   {\textstyle{1\over 2}}\, \alpha_-\,\bar r^2\,s)$&$ 41$&$ 39$&$
   {\textstyle{1\over 4}}\,\alpha_-\,\alpha_+\,(3\,{\bar \delta} + 1)\,(\bar r \cdot r)\,s$&$ 41$&$ 40$&$
   {\textstyle{1\over 4}}\,\bar \alpha_-\,\bar \alpha_+\,(3\,\delta - 1)\,(\bar r \cdot r)\,s$\\
  $41$&$ 41$&$ -{\textstyle{1\over 4}}\,(\beta_- + 3)\,({\bar \delta}\,\delta + 1)\,(\bar r \cdot r)\,s$&$ $&$
   $&$ $&$ $&$ $&$ $

\end{tabular}
\caption{Continuation of Tab. V.}
\end{table}

\end{widetext}

\clearpage

\bibliography{mybib}
\bibliographystyle{h-physrev}
\end{document}